\RequirePackage{fix-cm}
\documentclass[smallextended]{svjour3}       
\usepackage[12pt]{extsizes}
\usepackage{geometry}
\smartqed  
\usepackage{graphicx}
\usepackage[usestackEOL]{stackengine} 
\usepackage{latexsym}
\usepackage{amsmath}
\usepackage{amssymb}
\usepackage{color}

\DeclareMathOperator{\D}{\overset{\leftrightarrow}{D}}

\begin{document}

\title{Wave-particle interactions in quantum plasmas
}


\author{Amar P. Misra         \and
        Gert Brodin 
}


\institute{Amar P. Misra \at
              Department of Mathematics, Siksha Bhavana, Visva-Bharati University, Santiniketan-731 235,  India \\
              Tel.: +91-94333-68843\\
              \email{apmisra@visva-bharati.ac.in}           
           \and
           Gert Brodin \at
              Department of Physics, Ume{\aa} University, SE-901 87 Ume{\aa}, Sweden\\
              \email{gert.brodin@physics.umu.se}
}


\maketitle

\begin{abstract}
Wave-particle interaction (WPI) is one of the most fundamental processes in plasma physics in which one most prominent example is the Landau damping. Owing to its excellent energy-exchange mechanism, the WPI has gained increasing interest not only from theoretical points of view but also its many important applications including plasma heating and plasma acceleration.  In this review work, we present theoretical backgrounds of linear and nonlinear wave-particle interactions in quantum plasmas. Specifically, we  focus on the wave-particle interactions for homogeneous plasma waves (i.e., waves with infinite extent rather than a localized pulse) as well as for propagating electrostatic waves in the weak and strong quantum regimes  to demonstrate the modifications of several classical features  including those associated with resonant and trapped  particles.         Finally, the future perspectives of WPI in quantum plasmas are presented.   
\end{abstract}
\section{Introduction}\label{intro}
 Quantum plasmas have been a topic of important research for nearly sixty years due to their frequent occurrence and potential applications in many astrophysical plasmas (e.g., in the interiors of giant planets like Jupiter, brown and white dwarf stars, and outer crust of neutron stars), in laboratory devices via the compression of matter with lasers, x-rays or ion beams [e.g., the Lawrence Livermore National Laboratory, the Z-machine at Sandia National Laboratory, the Omega laser at the University of Rochester, the European free electron lasers FLASH and X-FEL in Germany and the Linac Coherent Light Source (LCLS) in Stanford], inertial confinement fusion (ICF) plasmas (during the initial phase), quark-gluon plasmas, solid-state plasmas, in semiconductor electron-hole plasmas, as well as in nanoplasmonics which is concerned with the interactions of quantum electrons in metallic nanostructures and electromagnetic radiation \cite{haug2009,shukla2010}. Quantum plasmas usually consist of different charged particles (e.g., electrons, positrons and protons) in which at least one component is  a fermion. In dense plasma environments, the number density of electrons/positrons is extremely high and hence they become degenerate and obey the Fermi-Dirac statistics.
 We briefly state under what conditions quantum effects start playing a role as follows: 
 \par 
 \textit{Firstly}, according to the Pauli's exclusion principle, there is at most one fermion in each quantum state, and each occupies a volume $h^3$ in phase space. So, the volume per each quantum state of a fermion in real space is $V_s=h^3/(2\pi mk_BT)^{3/2}\sim h^3/p^3$, i.e., the ratio of the volume in phase space   and the volume in momentum space. So, if $n$ is the number of particles per unit volume in phase space, the ratio becomes $nh^3/(2\pi mk_BT)^{3/2}\sim (T_F/T)^{3/2}$, i.e., a parameter $\chi=T_F/T$ must be defined to measure the degree of degeneracy of a particle, or to what extent the Pauli's principle has to be considered. Since $\chi$ can be expressed as $\chi=(1/2)(3\pi^2)^{2/3}(n\lambda_B^3)^{2/3}$, where $\lambda_B=\hbar/mv_t$ is the thermal de Broglie wavelength for a fermion (By default, a particle or a fermion means an electron as the same principle applies for other fermions), the quantum degeneracy effect  becomes important when $n\lambda_B^3\geq1$ and so $\chi\geq1$. Here, $h=2\pi\hbar$ is the Planck's constant, $m$ is the electron mass, $T$ is the electron temperature, $k_B$ is the Boltzmann constant, $v_t=\sqrt{k_BT/m}$ is the electron thermal velocity and $T_F$ is the Fermi temperature. 
 \par
\begin{figure}
 \begin{center}
 \includegraphics[scale=0.25]{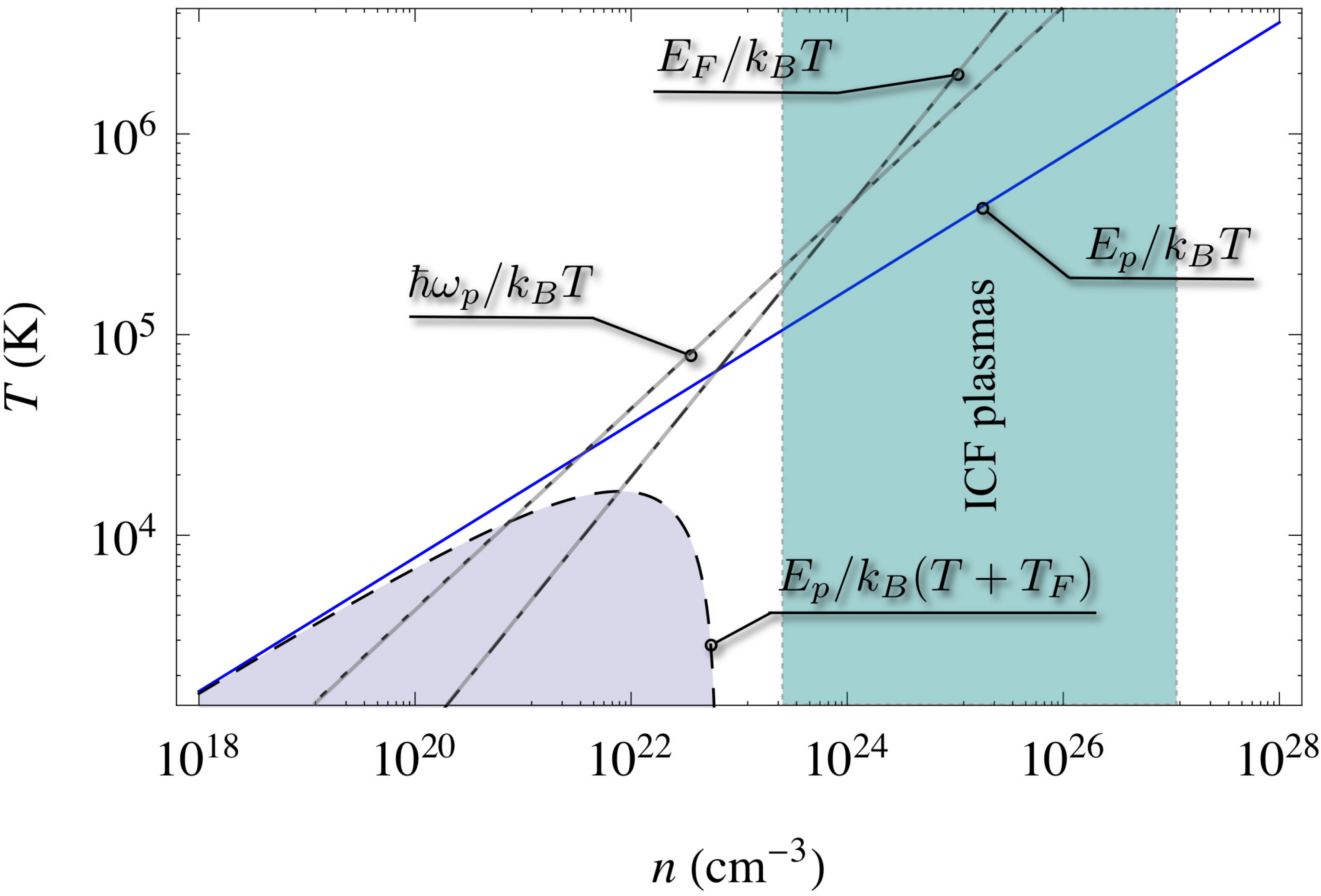}
 \caption{Classical and quantum plasma regimes are shown in the temperature-density parameter space. The figure is reproduced from Ref. \cite{zamanian2010}. Each curve represents where the parameter ratios shown are each equal to the unity.   The quantum effects become important below either of the lines 
 $E_{F}/k_{B}T=1$ and $\hbar \omega _{p}/k_{B}T=1$. While the ratio $E_{p}/k_{B}T$ denotes the strong coupling parameter in the moderate density regime, $E_{p}/k_{B}(T+T_{F})$ represents that in the high density regime. Here, $E_F=k_BT_F$ is the Fermi energy and     $E_{p}=e^{2}n_{0}^{1/3}/4\pi \varepsilon _{0}$ is the potential energy due to the nearest neighbor.  The blue shaded part is the region for collisional plasmas. }
 \label{fig-cl-q-regime}
 \end{center}
 \end{figure}

 \textit{Secondly},  since the wave function of a particle (electron) with momentum $mv_t$ has the wave length    $\lambda_B$,   the quantum effect is expected to be important at  length scale $k^{-1}\sim \lambda_B$. So, for collective oscillations an important length scale could be such that $(k\lambda_D)^{-1}\sim\lambda_B/\lambda_D\sim \hbar\omega_p/mv_t^2\geq1$, where $\lambda_D=v_t/\omega_p$ is the particle's Debye length and $\omega_{p}=\sqrt{ ne^2/\varepsilon_0m}$ is the electron plasma frequency in which $e$ is the elementary charge and $\varepsilon_0$ is the permittivity of free space. So, another parameter of interest can be defined   as $H=\hbar\omega_p/mv_t^2$.  For high densities or degenerate electrons $(T_F>T)$, $v_t$ is to be replaced by the Fermi velocity $v_F=\sqrt{k_BT_F/m}$. Thus, for low density plasmas, $H$ scales as $H\sim n^{1/2}/T$, and for high densities $H\sim n^{-1/6}$. 
 A schematic diagram showing the classical and quantum plasma regimes is shown in Fig. \ref{fig-cl-q-regime}
\par 
 When  plasma particles undergo electrostatic, electromagnetic and quantum forces (e.g., those associated with the statistical pressure, particle spin and exchange-correlation), they oscillate to generate high- or low-frequency electrostatic or electromagnetic waves. Interactions of these waves with fermions provide many interesting and important phenomena occurring in laboratory, space and astrophysical plasmas, as well as in many other environments as mentioned before.
For examples, the wave-particle interactions  (WPIs) play key roles in the excitation and damping of collective modes, diffusion in velocity space, i.e., thermalization, heating and acceleration of charged particles; transport of particles, momentum and energy.  In  \textit{laboratory}, the WPIs can become useful in many important applications including beat wave acceleration, plasma heating in magnetically confined fusion plasmas, edge transport reduction due to magnetic perturbation on multi-scale perturbations, and plasma  absorption of laser radiation in inertial fusion experiments  \cite{antoniazzi2005}. The WPIs in \textit{space plasmas} occur within the time scales of the plasma gyroperiod or the plasma oscillation period. They can play crucial roles, e.g., in the dynamics of energetic charged particles in the Earth's Van Allen radiation belts, in the formation of magnetopause boundary layers, precipitation of particles resulting into the formation of auroras, and transport of wave energy from one region to another \cite{tsurutani1997}. In \textit{ astrophysics}, the WPI can result into the radiation emission of hard $x$-rays and gamma rays, the stimulated Raman and Brillouin scattering as well as the acceleration of charged particles in relativistic regimes \cite{koch2006}.  Furthermore, the WPIs  in noble-metal nanoparticles can lead to surface-plasmon resonances which have potential applications in nanoscale optics and electronics \cite{thakkar2018}.  The studies of WPI typically include (i) Coherent WPI, resonance, trapping (ii) Chaos, quasilinear theory, and (iii) Weak and strong turbulence theory. In this article, we will, however, give emphasis   on the wave-particle resonance and associated wave damping in relativistic and nonrelativistic quantum plasmas with and without the effects of spin.  
\par 
One most prominent example of WPIs is the Landau damping. It is the damping of collective modes of oscillations in plasmas without any collision between charged particles. There are, in fact, two approaches to understand the physics of Landau damping: One approach considers the Landau damping in terms of dephasing of charged particles and the other considers Landau damping as the result of energy exchange between waves and particles due to resonance. In this review work, we, however, focus on the second approach. In $1946$, Lev Davidovich Landau first predicted and published a result on plasma oscillations \cite{landau1965}. He found that there would be exponential decay of coherent oscillations, i.e., Langmuir wave can suffer damping due to wave-particle interactions. He deduced this effect from a mathematical point of view while solving a Vlasov-Poisson system without its physical explanations. Although correct, the Landau's derivation was not meticulous from mathematical points of view and later resulted in several conceptual controversies. A number of works were devoted to resolve these issues. To mention few, in $1949$, Bohm and Gross pointed out that the Landau damping results into energy transfer from oscillating coherent field to its resonant particles \cite{bohm1949}. Later, in $1955$,  Van Kampen \cite{kapmen1955} and in $1959$,  Case \cite{case1959} proved that wave damping can be seen to occur with Fourier transforms and showed that the linearized Vlasov and Poisson equations have a continuous spectrum of singular normal modes.  Even after its mathematical verification by  Van Kampen \cite{kapmen1955} and Case \cite{case1959}, and experimental observation by Malmberg and Wharton in $1964$ \cite{malmberg1964}, it took almost twenty years to accept the reality of Landau damping. 
\par 
The subject of wave-particle interactions is very extensive. So, some choices of topics have to be made that can be covered.  
The paper is organized as follows: In Sec. \ref{sec-wpi-linear-th}, the linear Landau damping of Langmuir waves is treated  starting from the simple classical case, extending the theory to the quantum regime using the Wigner equation, and then  finally, covering also the effects of a relativistic background distribution. We start   Sec. \ref{sec-nonl-th-spinless}  by reviewing the nonlinear influence on Langmuir wave damping in homogeneous plasmas  considering the dynamics in both the weak and strong quantum regimes.  Next, we continue with nonlinear generalizations for localized pulse propagation. In particular, we  study the nonlinear wave-particle interactions both for ion-acoustic   and Langmuir pulses. In the end of Sec. \ref{sec-nonl-th-spinless}, we discuss the wave-particle interactions induced by the electron spin  properties. Finally,  the review ends with a summary and concluding discussion in Sec. \ref{summary}. 
\section{Wave-particle interactions: Linear theory} \label{sec-wpi-linear-th}
Wave-particle interaction is a process in which an exchange of energy takes place between waves and particles in a plasma.  Such an interaction leads to many interesting phenomena including the scattering and acceleration of particles as well as the growth or damping of waves. The growth (damping) of a wave amplitude  occurs depending on whether the wave gains (loses) energy from (to) the particles. In the following sections \ref{sec-basic} to \ref{sec-relat}, we will mainly focus on wave damping   as first described by Landau.  In Sec. \ref{sec-basic}, we   discuss the basic concept of Landau damping, the Landau's mathematical treatment to obtain the linear dispersion relation and the damping rate from the Vlasov-Poisson system. The concept of anti-damping or instability is also discussed with some illustrations. We also state the quantum kinetic equations for the description of electrostatic collective oscillations in quantum plasmas. Furthermore, the Landau damping of electrostatic waves in nonrelativistic and relativistic quantum plasmas with different background distributions of electrons is discussed  in Secs. \ref{sec-nonrelat} and \ref{sec-relat}.
 \subsection{Basic concept  of Landau damping}\label{sec-basic}
Before we begin with different aspects of wave-particles interactions, especially those of Landau damping, it is pertinent to introduce the basic concept of Landau damping. 
\subsubsection{\textit{Plasma oscillation and wave-particle interaction}}
We consider an electrically quasi-neutral plasma in equilibrium which consists of mobile electrons and  stationary positive ions forming only the background plasma. If electrons are displaced from their equilibrium position, a charge separation occurs and an electric field is created which acts as a restoring force to bring back electrons into their equilibrium position. However, due to their inertia, electrons accelerate towards the equilibrium position and overshoot it in the same way as an oscillating spring  
does. Thus, a standing wave (Langmuir oscillation) is generated with constant frequency $\omega_p$.  Note that in the Langmuir oscillation, the individual motion of electrons is not considered. Next,   we consider a random motion of electrons (e.g., due to their thermal velocity) with a given velocity distribution for the equilibrium state, and determine under what conditions a wave mode with a wave frequency $\omega$ and wave number $k$ exists.  We assume that the oscillating electrons produce electric fields of the following plane wave form with the phase velocity $v_p=\omega/k$:
\begin{equation}
E(x,t)=E_0\exp[i(kx-\omega t)],
\end{equation}
where $E_0$ is a constant amplitude of the wave.
The oscillating electrons, in turn, interact with the wave electric field they produce leading to the emergence of wave-particle interactions. As a result, the characteristics of particles and hence the field producing the forces are changed. In the wave-particle interaction, since the exchange of energy takes place between waves and particles, either growth (instability) or decay (damping) of the wave amplitude can occur. So, it is reasonable to assume $\omega$ as complex, i.e., $\omega=\omega_r+i\gamma$, so that  
\begin{equation}
E(x,t)=E_0\exp[i(kx-\omega t)]=E_0\exp[i(kx-\omega_r t)]\exp[\gamma t].
\end{equation}
Clearly, the electrostatic oscillation is damped if $\gamma<0$, otherwise for $\gamma>0$ we have an instability or anti-damping. Since particles can have, in general, different velocities, a simple picture is that in a background velocity distribution,
\begin{itemize}
\item If more particles move slowly than the wave, particles gain energy from the wave or wave loses energy to the particles, and the wave gets damped. 
\item If more particles have velocities larger than the wave, wave gains energy from the particles and wave is said to be unstable or anti-damped. 
\end{itemize}
It follows that the slope of the particle's velocity distribution may become important. However, this picture may not be completely correct. In fact, particles with very different velocities (i.e., much larger or lower than the wave) may not interact with the wave and so, no damping or instability is to occur.  
\subsubsection{\textit{Landau's mathematical treatment: Classical results}}
The wave-particle interaction is truly a kinetic phenomena, and so it can not be described by the fluid theory. Landau's treatment of WPI was based on a Vlasov-Poisson system. In this treatment, the equations for electron plasma oscillations in one-dimension are
\begin{equation}
\frac{\partial f}{\partial t}+v\frac{\partial f}{\partial x}+\frac{e}{m}\frac{\partial \phi}{\partial x}\frac{\partial f}{\partial v}=0, \label{eq-vlas}
\end{equation}
\begin{equation}
\frac{\partial^2\phi}{\partial x^2}=\frac{e}{\varepsilon_0}\left(\int f dv-n_0\right), \label{eq-poiss}
\end{equation}
where for electrostatic oscillations $E(x,t)=-\partial\phi/\partial x$ is used.
\par 
We look for a small amplitude plane wave solution of Eqs. \eqref{eq-vlas} and \eqref{eq-poiss}, and accordingly we perturb $f$ and $\phi$ about their equilibrium states as
\begin{equation}
f(x,v,t)=f_0(v)+f_1(x,v,t),~~\phi(x,t)=\phi_1(x,t).
\end{equation}
Thus, instead of considering the Vlasov's expression as a double Fourier transform, i.e., 
\begin{equation}
f_1(x,v,t)=\frac{1}{2\pi}\int_{-\infty}^{\infty}\int_{-\infty}^{\infty}\tilde{f_1}(k,v,\omega)e^{i(kx-\omega t)}dk d\omega
\end{equation} 
and similar for $\phi_1$, we consider the Landau's approach in which the perturbations vary as a Fourier transform in the space domain $(-\infty<x<\infty)$ and a Laplace transform in the time domain $(0<t<\infty)$, i.e., 
\begin{eqnarray}
\tilde{f_1}(k,v,t)&&=\frac{1}{2\pi}\int_{-\infty}^{\infty}f_1(x,v,t)e^{ikx}dx,~
\tilde{\phi_1}(k,t)=\frac{1}{2\pi}\int_{-\infty}^{\infty}\phi_1(x,t)e^{ikx}dx, \label{eq-expan1} \\  
f_1(k,v,s)&&=\int_{0}^{\infty}\tilde{f_1}(k,v,t)e^{-st}dt,~
\phi_1(k,s)=\int_{0}^{\infty}\phi_1(k,t)e^{-st}dt. \label{eq-expan2}
\end{eqnarray}
Next, linearizing Eqs. \eqref{eq-vlas} and \eqref{eq-poiss} and using Eqs. \eqref{eq-expan1} and \eqref{eq-expan2} we obtain the following dispersion relation for Langmuir waves \cite{landau1965}.
\begin{equation}
D(\omega,k)\equiv1-\frac{\omega_p^2}{n_0k^2}\int_{-\infty}^{\infty}\frac{\partial f_0/\partial v}{v-\omega/k}dv=0. \label{eq-disp-cl1}
\end{equation}
From Eq. \eqref{eq-disp-cl1}, it is clear that the wave-particle resonance occurs when $v=v_\text{ph}\equiv\omega/k$, i.e., when the particle velocity approaches the wave phase velocity. This is called the Landau resonance.  A physical picture is that when this resonance condition is satisfied, the particles do not experience  a rapidly fluctuating electric field of the wave, i.e., almost a static electric field in the particle's rest frame, and so they can interact strongly with the wave. Particles near the resonance moving slightly slower (faster) than the wave get accelerated (decelerated) by the wave electric field to   move with the wave phase velocity, and hence gain energy from (lose energy to) the wave. 
\par 
In a collisionless electron-ion plasma with immobile ions and Maxwellian background distribution of electrons there are  
\begin{figure}
\begin{center}
\includegraphics[width=3.3in,height=3in,trim=0.0in 0in 0in 0in]{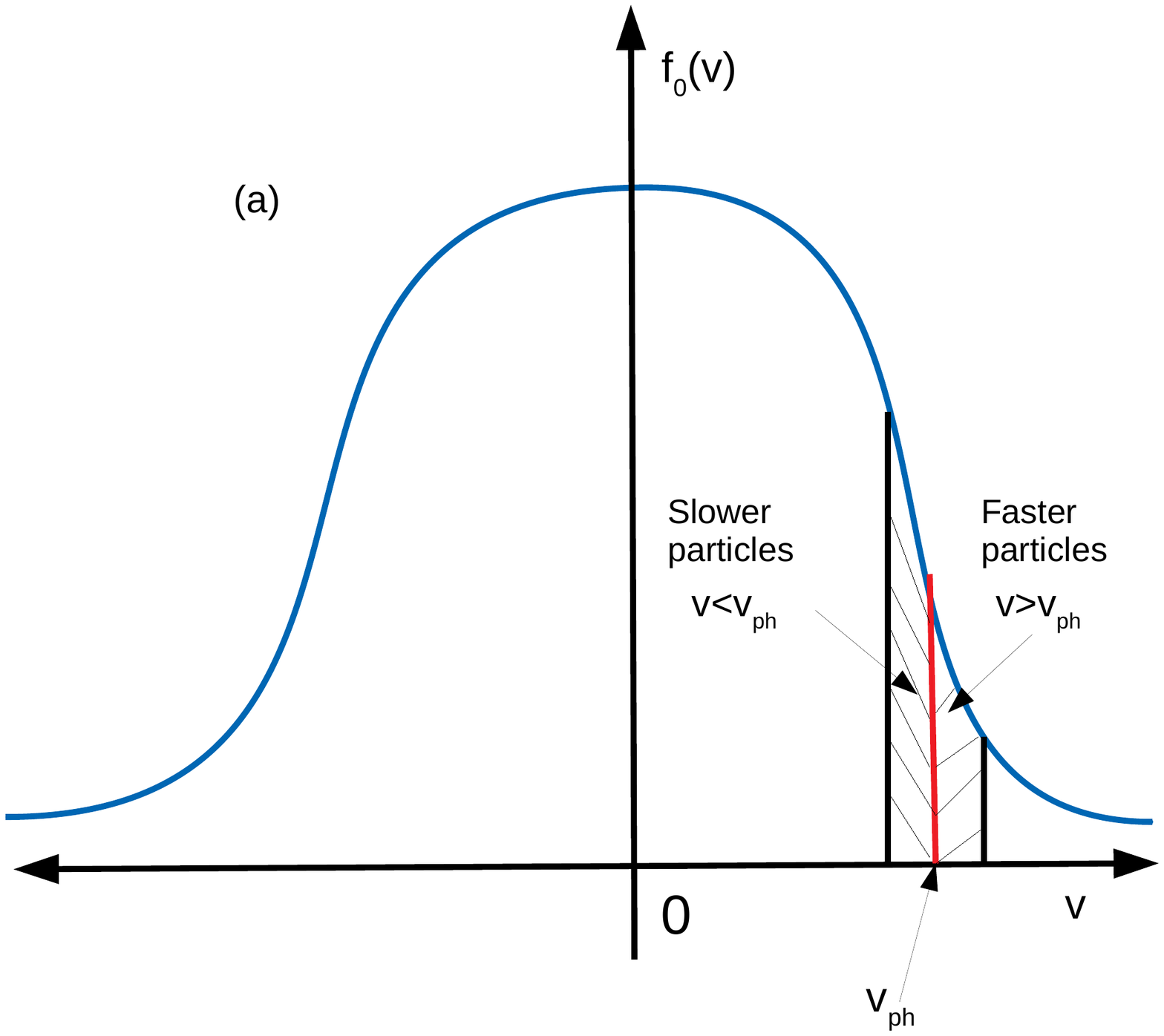}
\includegraphics[width=4in,height=3.5in,trim=0.0in 0in 0in 2in]{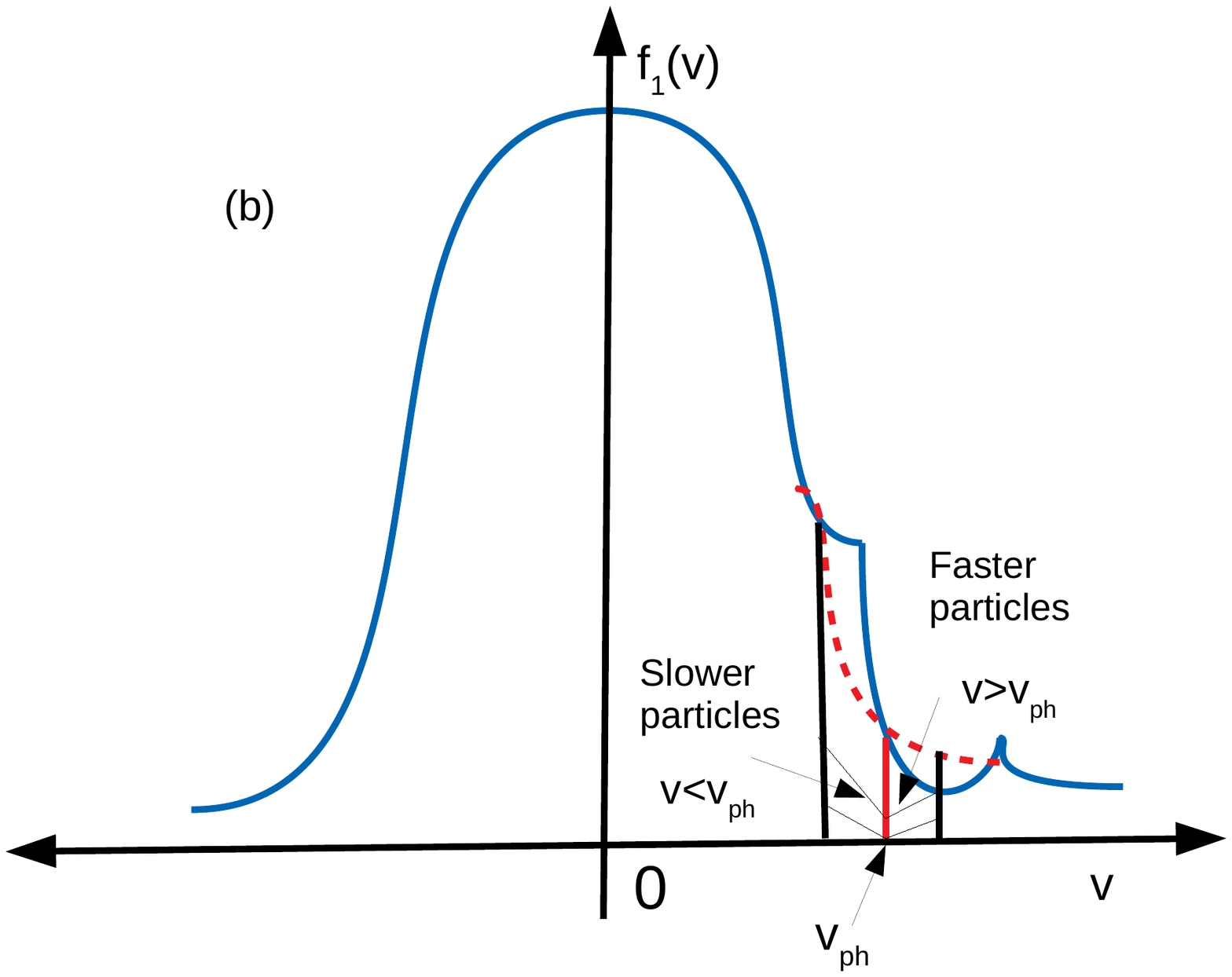}
\vspace{-100pt}
\caption{The Landau damping: (a) The initial distribution function $f_0(v)$ of electrons and (b) the perturbed distribution function $f_1(v)$  after an evolution due to the interaction of background electrons (dashed line) with the wave. The  resonance occurs in the region of a negative slope.  }
\label{fig:landau-dist}
\end{center}
\end{figure}
\begin{figure}
\begin{center}
\includegraphics[width=4in,height=3.5in,trim=0.0in 0in 0in 0in]{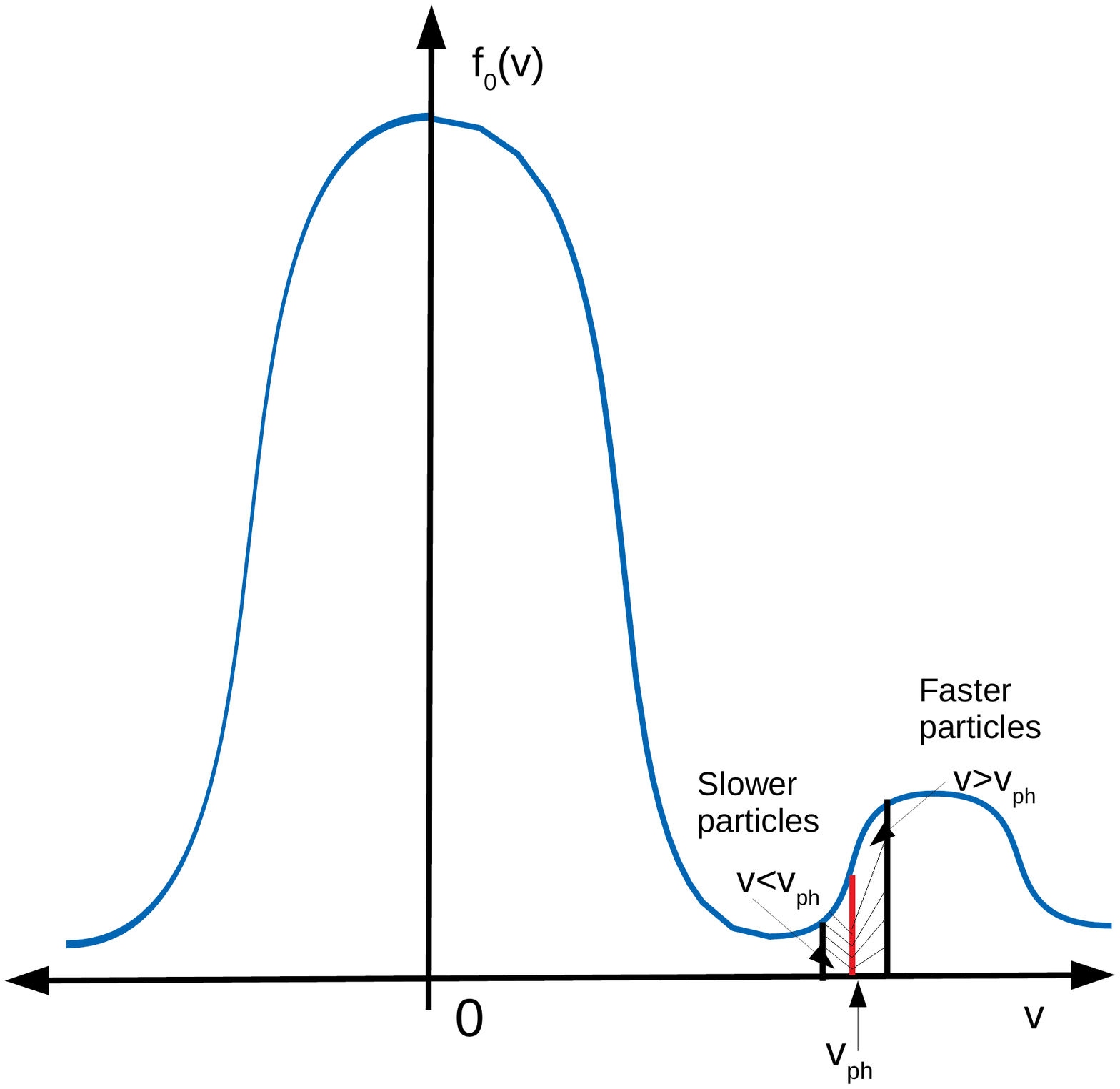}
\vspace{-50pt}
\caption{Landau anti-damping (instability): The bump-on-tail distribution function. The resonance occurs in the region of a positive slope.}
\label{fig:landau-bump-dist}
\end{center}
\end{figure}
more slower particles than the faster particles in the negative slope, and so the energy gained from the wave by the slower particles is more than that lost to the wave by the faster particles. As a result, a net wave damping occurs. In order to calculate the wave damping we consider $\omega=\omega_r+i\gamma$,  assume that the damping is weak, i.e., $|\gamma|\ll\omega_r$, and substitute it into the dispersion equation $D(\omega,k)=0$ to obtain
\begin{equation}
D(\omega,k)\equiv D_r(\omega_r,k)+iD_i(\omega_r,k)+i\gamma\frac{\partial D_r}{\partial\omega_r}=0. \label{eq-D}
\end{equation} 
After separating the real and imaginary parts, from Eq. \eqref{eq-D} we obtain
\begin{equation}
D(\omega_r,k)\equiv1-\frac{\omega_p^2}{n_0k^2}{\cal P}\int_{-\infty}^{\infty}\frac{\partial f_0/\partial v}{v-\omega_r/k}dv=0,   \label{eq-DR}
\end{equation}
\begin{equation}
\gamma=-\frac{D_i(\omega_r,k)}{\partial D_r/\partial \omega_r} ~\rm{with}~D_i(\omega_r,k)=-\frac{\pi\omega_p^2}{n_0k^2}\left[\frac{\partial f_0(v)}{\partial v}\right]_{v=\omega_r/k}. \label{eq-gamma-lin}
\end{equation}
For an one-dimensional Maxwellian background distribution of electrons we have
\begin{equation}
f_0(v)=\frac{n_0}{\sqrt{2\pi}v_t}\exp\left(-v^2/2v_t^2\right). \label{eq-dist-max}
\end{equation}  
  Figure \ref{fig:landau-dist}   shows the Landau damping  in which subplot (a) is for an initial distribution of thermal electrons with some narrow regions centered at the resonant velocity $v=v_\text{ph}$ showing the more  slower particles than the fast particles and  the subplot (b)  shows a perturbed distribution function, i.e., after an evolution  due to the interaction of the background distribution of electrons (dashed line) with the wave. Since particles with the velocity $v\approx v_p$ are trapped in the wave, this interaction results in the flattening of the distribution function $f_1(v)$ around the phase velocity (solid line). However, $f_1(v)$ contains the same number of particles which gain the total energy at the expense of the wave.   On the other hand, in a non-Maxwellian plasma,  if in some region of the phase space, the particle's distribution has more particles at higher velocities than those with lower velocities [see Fig. \ref{fig:landau-bump-dist}], then the wave will gain energy from the particles leading to what is known as ``bump-on-tail" instability  \cite{sarkar2015} or inverse Landau damping or Cherenkov instability. Thus, a beam of fast electrons having velocities much higher than their thermal speed will cause Langmuir waves to grow as there are available free energy of electrons. This kind of instability plays an important role, e.g., in solar radio bursts \cite{thejappa2018}. 
\par 
Next, for the Maxwellian background distribution of electrons [Eq. \eqref{eq-dist-max}], the classical dispersion relation for Langmuir waves and the Landau damping rate can be obtained from Eqs. \eqref{eq-DR} and \eqref{eq-gamma-lin}. In order that the  Langmuir waves are not   strongly  damped we must have $v_\text{ph}>v_t$. So, in the non-resonance region  assuming $v_\text{ph}\gg v$ and keeping terms up to the second-order of $v/v_\text{ph}$ in the binomial expansion of $(v-v_\text{ph})^{-2}$, we obtain 
\begin{equation}
\omega^2=\omega_p^2+3k^2v_t^2\omega_p^2/\omega^2.
\end{equation}
 If the thermal correction is small then replacing $\omega$ by $\omega_p$ (since $\omega=\omega_p$ for $v_t=0$) we obtain the following dispersion relation for Langmuir waves.
\begin{equation}
\omega^2=\omega_p^2+3k^2v_t^2. \label{eq-disp-cl}
\end{equation} 
The Landau damping rate is obtained from Eq. \eqref{eq-gamma-lin} as 
\begin{equation}
\gamma=-\frac{\sqrt{\pi}}{2\sqrt{2}}\left(\frac{\omega_r}{kv_t}\right)^3\omega_r\exp\left(-\frac{v_p^2}{2v_t^2}\right).\label{eq-gam0}
\end{equation}
In the expression of $\gamma$, one can approximate $\omega_r/k$ by $\omega_p/k$ and retain the thermal correction in the exponent to obtain
\begin{equation}
\gamma=-\frac{\sqrt{\pi}e^{-3/2}}{2\sqrt{2}}\frac{\omega_p}{(k\lambda_D)^3} \exp\left(-\frac{1}{2k^2\lambda_D^2}\right).\label{eq-gam1}
\end{equation}
 An alternative expression of $\gamma$ can be obtained from the relation
${\gamma}/{\omega_r}=- ({1}/{2})D_i(\omega_r,k)$ as 
\begin{equation}
\frac{\gamma}{\omega_r}=-\sqrt{\frac{\pi}{8}}\frac{\omega_p^2\omega_r}{k^3v_{t}^3} \exp\left(-\frac{\omega_r^2}{2k^2v_t^2}\right).\label{eq-gam2}
\end{equation}
The  expression \eqref{eq-gam2}  agrees with Eq. \eqref{eq-gam1} if one approximates   $\omega_r\sim\omega_p$.
\par 
Thus, from Eq. \eqref{eq-gam1} it follows that  since $\gamma<0$, there is indeed a collisionless damping of Langmuir waves. It is also evident that the damping becomes important for $k\lambda_D\sim o(1)$ and small for $k\lambda_D<1$.
 \subsection{Landau damping in nonrelativistic quantum plasmas}\label{sec-nonrelat}
 In the preceding section \ref{sec-basic}, we have discussed the basic concept of wave-particle interactions and, in particular, the Landau damping in classical plasmas, i.e., using the  linearized  Vlasov-Poisson system which predicts  the wave damping due to the phase velocity resonance only. However,   in the quantum regime, a new resonance mechanism enters the picture, and  we will see that the resonance velocity is modified by the  particle's dispersion. 
 \par 
 In contrast to a classical system  where the description of plasma particles is given in terms of a distribution function   $f(\mathbf{r},\mathbf{v},t)$ in $(6+1)$-dimensional phase space such that $f(\mathbf{r},\mathbf{v},t)d^3rd^3v$ gives  the number of particles in a volume element of phase space, a quantum state is described by  a wave-function $\psi$ of just one half of the phase space coordinates, either $\mathbf{r}$ or $\mathbf{v}$. In fact, the Heisenberg uncertainty principle, $drdv\geq\hbar/2$ does not provide any information about the particles in a phase space volume element. In this way, the Wigner formalism is introduced. The advantage of the Wigner function (which is not a probability density function as it can take negative values) in the formulation of quantum mechanics is that a classical Boltzmann's description can be recovered in the limit of $\hbar\rightarrow0$ where the uncertainty principle has no role. The Wigner function has numerous applications in  plasma physics, semiconductor physics, quantum optics, quantum chemistry, and quantum computing.  
\par 
The  electrostatic plasma collective oscillations in an electron-ion plasma with immobile ions can be described by the  quantum analog of the Vlasov-Poisson system, i.e.,  the three-dimensional Wigner-Poisson system, given by,
 \begin{equation}
\frac{\partial f}{\partial t}+{\bf v}\cdot\nabla_{\bf r} f+\frac{iem^3}{(2\pi)^3\hbar^4}\int\int d^3{\bf r}' d^3{\bf v}' e^{im({\bf v}-{\bf v}')\cdot {\bf r}'/\hbar} \left[\phi\left({\bf r}+\frac{{\bf r}'}{2},t\right)-\phi\left({\bf r}-\frac{{\bf r}'}{2},t\right)\right]f({\bf r},{\bf v}', t)=0, \label{wigner-3d}
\end{equation}
\begin{equation}
\nabla ^{2}\phi =\frac{e}{\varepsilon_0}\left( \int fd^3v-n_{0}\right),  \label{poisson-3d}
\end{equation}
or, in one-dimension, 
 \begin{equation}
\frac{\partial f}{\partial t}+v\frac{\partial f}{\partial x}+\frac{i e m}{2\pi\hbar^2}\int\int dx_0 dv_0 e^{im(v-v_0)x_0/\hbar} \left[\phi\left(x+\frac{x_0}{2}\right)-\phi\left(x-\frac{x_0}{2}\right)\right]f(x,v_0, t)=0, \label{wigner-1d}
\end{equation}
\begin{equation}
\frac{\partial^2\phi}{\partial x^2}=\frac{e}{\varepsilon_0}\left( \int fdv-n_{0}\right),  \label{poisson-1d}
\end{equation}
where $f$ is the Wigner distribution function,   $\phi $ is
the self-consistent electrostatic potential,  and   $n_{0}$ is the background number density of electrons and ions.
\par 
In the weak quantum limit, i.e.,  $H\equiv\hbar /mv_0L_0<1$, where $v_0$ and $L_0$ are, respectively, the characteristic velocity and length scales of oscillations, the integrand in the Wigner evolution equation  
 can be Taylor expanded to retain terms up to ${\cal{O}}(\hbar^2)$. Thus, in one-dimensional geometry, we obtain the following semi-classical Vlasov equation.
 \begin{equation}
\frac{\partial f}{\partial t}+v\frac{\partial f}{\partial x}+\frac{e}{m}\frac{\partial \phi}{\partial x} \frac{\partial f}{\partial v}-\frac{e\hbar^2 }{24 m^3} \frac{\partial^3 \phi}{\partial x^3} \frac{\partial^3 f}{\partial v^3} +{\cal{O}}(\hbar^4)=0.  \label{wigner-1d-semicl}
\end{equation}
Note that the Vlasov equation can be recovered from Eq. \eqref{wigner-1d-semicl} in the limit $\hbar\rightarrow0$.
 \par 
 We consider the propagation of electrostatic waves in a non-relativistic, collisionless and unmagnetized quantum plasma. The basic equations for the electron dynamics are the   Wigner-Moyal equation \eqref{wigner-3d} and the Poisson equation \eqref{poisson-3d}.  In order to obtain the linear dispersion relation for such  waves, we linearize  Eqs. \eqref{wigner-3d} and \eqref{poisson-3d} by separating $f$ and $\phi$ into their equilibrium and perturbation parts, i.e., $f({\bf r},{\bf v},t)= f_0(v) +f_1({\bf r},{\bf v},t)$ and 
$\phi({\bf r},t)=\phi_1({\bf r},t)$, and assume the perturbations to be of the form $\sim\exp(i{\bf k}\cdot{\bf r}-i\omega t)$, i.e., plane waves with frequency $\omega$ and wave vector $\mathbf{k}$. Thus, we obtain the following dispersion relation.
\begin{equation}
D(\omega,k)\equiv 1- \frac{\omega_{p}^2}{n_{0}}\int_{-\infty}^{\infty} \frac{f_0(v)}{(\omega-{\bf k}\cdot{\bf v})^2-{k^2v_q^2}}d^3v=0, \label{eq-disp-quantum}  
\end{equation}
where      $v_q=\hbar k/2m$ is the velocity associated with the plasmon quanta.
From Eq. \eqref{eq-disp-quantum}   some modifications to the classical dispersion relation can be noted.
\begin{itemize}
\item  The  dielectric function differs from the classical one in two ways: one with the background distribution and the other with the resonance condition.
\item The background distribution is either corresponding to the Fermi-Dirac statistics or Maxwell-Boltzmann statistics depending on particles are fully/partially degenerate or nondegenerate. The resonant velocity is other than the phase velocity, given by, $\omega-{\bf k}\cdot{\bf v}=\pm v_q $ or    $v^{\rm{res}}_{\pm}=\omega/k\pm v_q$  in one-dimensional geometry with ${\bf k}=k\hat{x}$.  
\item The modification of the resonant velocity is due to  the   quantum effects associated with the particle's dispersion. 
\item Of the two resonant velocities  $v^{\rm{res}}_{\pm}$, the lower one ($v^{\rm{res}}_{-}$) is of particular interest as it causes the wave damping more easily.
\item The expressions for the dispersion relation and the Landau damping rate will vary depending on the choice of the background distribution of electrons.  
\item The equilibrium distribution is always three-dimensional. So, even  in one-dimensional geometry one must consider the three-dimensional distribution function (Wigner) $f_0$, however,   projected on the $v_x$-axis,  i.e., $F_0(v) =\int\int f_0({\bf v})dv_ydv_z$ where $v^2=v_x^2+v_y^2+v_z^2$. 
\end{itemize} 
\par 
In order to find the expressions for the dispersion relation and the Landau damping rate, we first assume that  
the wave damping is small and the wave frequency is complex, i.e., $\omega=\omega_r+i\gamma$. Then    the time asymptotic solution for $\omega$ can be obtained by solving the dispersion equation $D(\omega,k)\equiv D_r(\omega_r,k)+iD_i(\omega_r,k)+i\gamma(\partial D_r(\omega_r,k)/\partial \omega_r)=0$, and separating the real and imaginary parts as
 \begin{equation}
D_r(\omega_r,k)\equiv 1- \frac{\omega_{p}^2}{n_{0}k^2}{\cal P}\int \frac{F_0(v)}{(v-\omega_r/k)^2-v_q^2}dv=0, \label{eq-Dr-quantum}  
\end{equation}
 and the Landau damping rate, given by,
 \begin{equation}
 \gamma=-\frac{D_i(\omega_r,k)}{\partial D_r/\partial \omega_r}, \label{eq-gam-quantum}
 \end{equation}
 where
 \begin{equation}
 D_i= -\frac{1}{2}\frac{\pi\omega_{p}^2}{v_qn_{0}k^2}\left[ F_0(v^{\rm{res}}_{+})-F_0(v^{\rm{res}}_{-}))  \right].  
 \end{equation}
 The linear dispersion properties and the damping rate can be analyzed for different electrostatic  waves with different background distributions of plasmas. Below we will discuss a few cases of interest.  
 \par 
 {\bf Case-I:} We consider the one-dimensional propagation of Langmuir waves  in the weak quantum regime in which the Langmuir wavelength  is larger than the thermal de Broglie wavelength of electrons,     i.e., $\lambda_B\equiv\hbar k/mv_t<1$.  This gives   $H\equiv\hbar\omega_p/mv_t^2<1$.      In this regime with $T\gg T_F$, the background distribution $f_0(v)$ of electrons can be considered to be the Maxwellian [Eq. \eqref{eq-dist-max}].
  In the semi-classical limit $\hbar k/mv_t<1$, Eqs. \eqref{wigner-1d} and \eqref{poisson-1d} can be  Fourier analyzed  to obtain the following  dispersion law and the Landau damping rate, given by,   \cite{chatterjee2016} 
\begin{equation}
 1-\frac{\omega_p^2}{n_0k^2}{\cal P}\int \frac{G(v)+\left(\hbar^2k^2/24 m^2\right) G^{\prime\prime}(v)}{v-\omega_r/k}dv=0  \label{eq-disp-sem-cl0}
\end{equation}
 \begin{equation}
 \gamma=\pi k \left[G\left(\frac{\omega_r}{k}\right)+\left(\hbar^2k^2/24 m^2\right) G^{\prime\prime}\left(\frac{\omega_r}{k}\right) \right] 
  \Big/ {\cal P}\int \frac{G(v)+\left(\hbar^2k^2/24 m^2\right) G^{\prime\prime}(v)}{\left(v-\omega_r/k\right)^2}dv,\label{eq-gam-sem-cl}
 \end{equation}
 
where $G(v)={\partial f_0 (v)}/{\partial v}$,   the prime in $G$ denotes derivative with respect to $v$,  and  ${\cal P}$ denotes the Cauchy Principal value. 
\par
\begin{figure*}[ht]
\centering
\includegraphics[height=2.2in,width=6.5in]{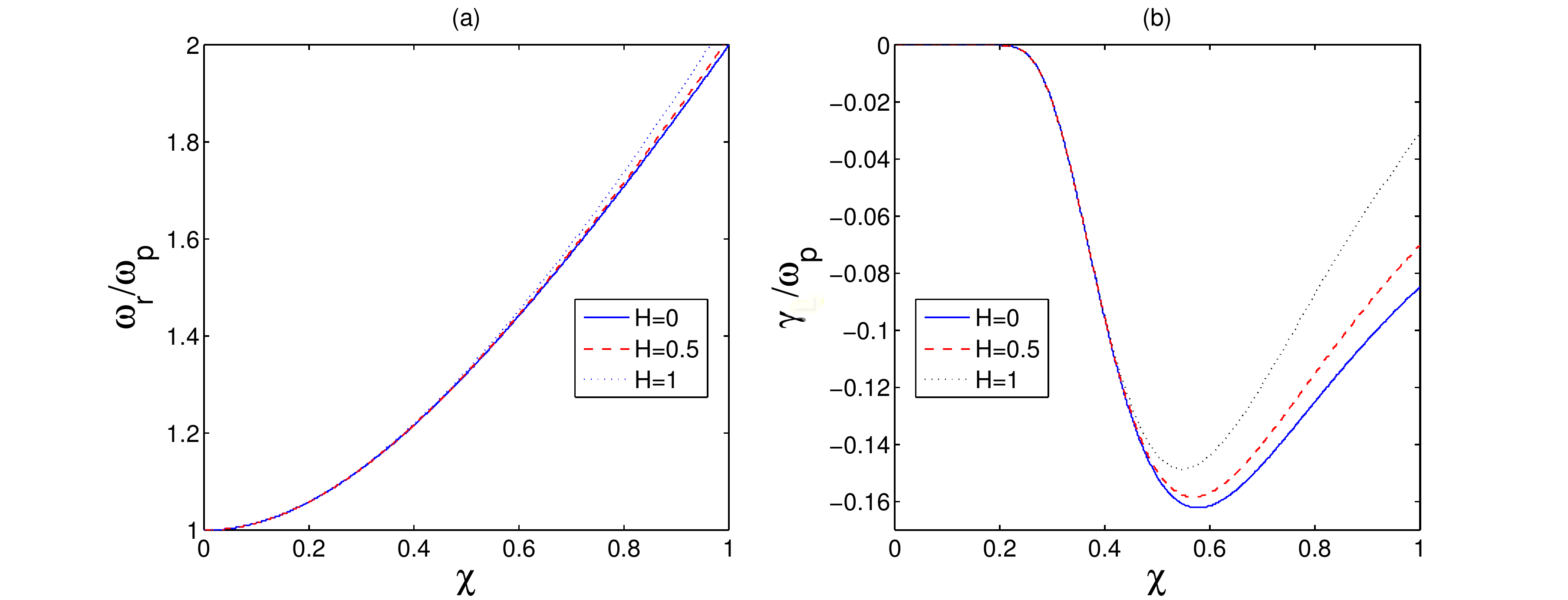}
\caption{  The wave dispersion [panel (a)] and the damping rate [panel (b)]  are shown (reproduced from Ref. \cite{chatterjee2016}) against $\chi\equiv k\lambda_D$ for different values of $H$ as in the legends.  }
\label{fig1}
\end{figure*}

 
In the region of small wave number, i.e.,  $k^2\lambda_D^2\ll1$,  and  the smallness of thermal corrections, the dispersion    relation and the Landau damping rate for Langmuir waves can be reduced.  Thus, from Eq.  \eqref{eq-disp-sem-cl0} we have
\begin{equation}
\omega_r^2 = \omega_p^2+ 3k^2v_t^2 +\frac{\hbar^2k^4}{4m^2}. \label{eq-disp-weak-Q}   
\end{equation}
In comparison with the classical dispersion relation [Eq. \eqref{eq-disp-cl}], we find that an additional term $\propto\hbar^2$ appears in Eq. \eqref{eq-disp-weak-Q}  due to the quantum particle dispersion.  The latter enhances the wave frequency and thus modifies the dispersion properties of Langmuir waves in quantum plasmas.  
Also, in the limit $\chi^2\equiv k^2\lambda_D^2\ll1$, the Landau damping rate [Eq.\eqref{eq-gam-sem-cl}]  reduces to
\begin{equation}
\gamma=- \sqrt{\frac{\pi}{8}} \frac{\omega_pe^{-3/2}}{k^3\lambda_D^3} \left(1+\frac{1}{24}H^2-\frac{1}{8}H^2k^2\lambda_D^2\right)\exp\left(-\frac{1}{2k^2\lambda_D^2}\right).   \label{eq-gam-weak-q1}
\end{equation}
An equivalent expression of $\gamma$ can be obtained from  
${\gamma}/{\omega_r}=- ({1}/{2})D_i(\omega_r,k)$ as 
\begin{equation}
\frac{\gamma}{\omega_r}=-\sqrt{\frac{\pi}{8}}\frac{\omega_p^2\omega_r}{k^3v_{t}^3}\left[1-\frac{1}{6}\frac{v_q^2}{v_t^2}\left(3-\frac{\omega_r^2}{k^2v_t^2}\right)\right] \exp\left(-\frac{\omega_r^2}{2k^2v_t^2}\right).\label{eq-gam-weak-q2}
\end{equation}
Equations \eqref{eq-gam-weak-q1} and \eqref{eq-gam-weak-q2} agree when one approximates $\omega_r\sim\omega_p$ in  Eq. \eqref{eq-gam-weak-q2} in the limit of small thermal and quantum corrections.  Comparing Eq. \eqref{eq-gam-weak-q2} with Eq. \eqref{eq-gam2}  we find that the magnitude of the Landau damping rate is increased due to the quantum effect. 
  The dispersion properties [Eq. \eqref{eq-disp-weak-Q}] and the Landau damping rate [Eq. \eqref{eq-gam-weak-q1}] are analyzed  by the influence of the quantum parameter $H$ as shown in Fig.  \ref{fig1}. Note that    different values of $H$  correspond to different plasma environments that are represented by the plasma number density $n_0$ and the  temperature $T$. For example, $H=0.5$ corresponds to the regime where $T=10^6$ K, $T_F/T=0.3$ and $ n_0=7\times10^{23}$ cm$^{-3}$, and $H=1$  corresponds to that where $T=6\times10^5$ K, $T_F/T=0.7$ and $ n_0=10^{24}$ cm$^{-3}$.  It is found that both the real part of the wave frequency and  the absolute value of the damping rate decreases with increasing values of $H$ in $0\lesssim H\lesssim1$.    Two subregions  $0\lesssim\chi\lesssim0.6$ and $0.6<\chi\lesssim1$ are found to exist, in one of which  $|\gamma_L|$ increases, whereas in the other it decreases.   It is concluded that in the wave-particle interaction the quantum effect influences the wave to lose energy to the particles more slowly than predicted in the classical theory.  
 \par 
 {\bf Case-II:} We consider a fully degenerate  plasma, i.e., a zero-temperature Fermi gas with the following background distribution of electrons.  
\begin{equation}
f_{0}(\mathbf{v})=\left\{ 
\begin{array}{cc}
{2m^{3}}/{(2\pi \hbar )^{3}}, & |\mathbf{v}|\leq v_{F} \\ 
0, & |\mathbf{v}|>v_{F},%
\end{array}%
\right.   \label{eq-dist-FD}
\end{equation}%
where $v_{F}=\sqrt{2E_{F}/m}$ is the electron Fermi velocity
and $E_{F}=\hbar ^{2}\left( 3\pi ^{2}n_{0}\right) ^{2/3}/2m$ is the Fermi
energy. Performing the velocity integral on the $v_yv_z$ plane, i.e., perpendicular to the $v_{x}$-axis and using the cylindrical coordinates in $v_{y}$ and $v_{z}$, we obtain  (replacing $v_{x}$ by $v$) 
\begin{equation}
 F_{0}(v)= \left\lbrace\begin{array}{cc} 
\left[{2\pi m^3}/{(2\pi\hbar)^3}\right](v_F^2-v^2),&|{ v}|\leq v_F \\
 0,&\rm{otherwise}. 
 \end{array}\right.  \label{eq-dist-FD-1d}
\end{equation}
We note that the  distribution function \eqref{eq-dist-FD}, which is flat topped in three dimensions, becomes parabolic in one dimension.  So, there is a possibility that the resonant velocity $v^{\rm{res}}_{\pm}$   falls in the negative slope of the distribution function $F_0(v)$ for which the wave damping occurs.  In order to assess it we must require an expression for the dispersion relation. The dispersion equation \eqref{eq-Dr-quantum} after evaluating the principal value integral using Eq. \eqref{eq-dist-FD-1d}  reduces to \cite{eliasson2010,misra2017}
\begin{equation}
1+\frac{3 \omega_p^2}{4 k^2 v_F^2} \left(2-\sum_{j=\pm1}\frac{j}{2v_q v_F%
}\left\lbrace v_F^2 - \left(v_p+jv_q \right)^2 \right\rbrace \log \left\vert%
\frac{v_p+jv_q-v_F}{v_p+jv_q+v_F}\right\vert\right) =0.
\label{eq-disp-FD1}
\end{equation}
\par 
Equation \eqref{eq-disp-FD1} can be analyzed numerically  to ascertain whether the resonant velocity $v^{\rm{res}}_{\pm}$ remains smaller than $v_F$ in some domain of $k\lambda_F$ for which the Landau damping can occur. Here, $\lambda_F=v_F/\omega_p$ is the Fermi wavelength, $v^{\rm{res}}_{\pm}/v_F=v_p/v_F\pm Hk\lambda_F/2$ and $H=\hbar\omega_p/m v_F^2$ (in degenerate plasmas $v_t$ is to be replaced by $v_F$). 
   In order that the quantum effects to be important and the Langmuir wavelength is not much larger than the de Broglie wavelength, we must have  $\hbar k/mv_F\equiv k\lambda_B=H k\lambda_F\lesssim1$, i.e., $k\lambda_F<1/H$. The smaller values of  $H~(\lesssim1)$ corresponds to high density regimes. 
 From the analysis of the dispersion relation \eqref{eq-disp-FD1} as in    Ref. \cite{misra2017}  it can be moted that there is a small  regime of the Langmuir wavelength    $0<k\lambda_F\lesssim1$ for which   $v^{\rm{res}}_{\pm}<v_F$ is satisfied.   For example, when $H\sim1$,  the relation $v^{\rm{res}}_{-}<v_F$ holds for    $0.9\lesssim k\lambda_F  \lesssim1$. As the values of $H$ decrease from $H=1$, the resonant velocities shift towards higher values of $k\lambda_F>1$. So, the Landau damping due to the one-plasmon resonance may occur for $k\lambda_F>1$ and $H\lesssim1$.  Consequently, such damping does not occur in the regime of $H<1$ with $0\lesssim k\lambda_F<1$ \cite{misra2017}, and so is  in the semi-classical limit $\hbar k/mv_F\ll1$. In the latter, the dispersion equation \eqref{eq-disp-FD1} reduces to
 \begin{equation}
 1+\frac{3 \omega_p^2}{2 k^2 v_F^2} \left(1-\frac{v_p}{v_F}  \log \left\vert\frac{v_p+v_F}{v_p-v_F}\right\vert\right) =0.
\label{eq-disp-sem-cl}
 \end{equation}
The term `semi-classical' is used because, the dispersion relation \eqref{eq-disp-sem-cl} can also be derived from the one-dimensional Vlasov-Poisson equation using the background distribution of electrons given by Eq. \eqref{eq-dist-FD}.
Since for $H<1$ and in the regime $0\lesssim k\lambda_F<1$, the Landau damping does not occur, the logarithmic functions in Eq. \eqref{eq-disp-FD1} can be expanded for small wave numbers. Thus, retaining the terms involving $k$ up to $k^4$, one obtains \cite{eliasson2010} 
\begin{equation}
\omega^2=\omega_p^2+\frac{3}{5}k^2v_F^2+(1+\alpha)\frac{\hbar^2k^4}{4m^2},\label{eq-disp-Lang-FD}
\end{equation}
where $\alpha=(48/175)m^2v_F^4/\hbar^2\omega_p^2$ is  a correction term which becomes smaller than unity in low-density plasmas, e.g., metals and semiconductors. However, it can be larger than unity in highly dense environments. A term similar to  $\alpha$ was also obtained and discussed by Ferrell in his work on the characteristics of electron plasma oscillations in metals \cite{ferrell1957}.  
\par 
It is to be noted that a critical value  $k_{\rm{cr}}$ of the wave number $k$ and hence the corresponding critical wave frequency $\omega_{\rm{cr}}$  exist such that the Landau damping occurs for $k>k_{\rm{cr}}$ and $\omega>\omega_{\rm{cr}}$. The critical values can be obtained from the resonance condition $v=\omega/k-\hbar k/2m\leq v_F$, i.e., 
\begin{equation}
\omega_{\rm{cr}}\equiv k_{\rm{cr}}v_F+\hbar k_{\rm{cr}}^2/2m, \label{eq-omeg-cr} 
\end{equation}
 and the following reduced equation for $k_\text{cr}$ [after substituting $\omega=\omega_{\rm{cr}}$ in Eq. \eqref{eq-disp-FD1}] \cite{eliasson2010}
\begin{equation}
1+\frac{3\omega_p^2}{4k_{\rm{cr}}^2v_F^2}\left[2-\left(2+\frac{\hbar k_{\rm{cr}}}{m v_F}\right)\log\left(1+\frac{2mv_F}{\hbar k_{\rm{cr}}}\right)  \right]=0. \label{eq-k-cr}
\end{equation}
An approximate expression for the Landau damping rate can be obtained by using the formalism $\gamma/\omega_r=-(1/2)D_i(\omega_r,k)$    and  noting that $v_F^3=3\pi^2n_0\hbar^3/m^3$ as
\begin{equation}
\frac{\gamma}{\omega_r}=-\frac{3\pi}{4}\frac{\omega_r\omega_p^2}{k^3v_F^3}.
\end{equation}
 It follows that the Landau damping rate $|\gamma|$ gets reduced at higher values of $k$. The regions for the existence of damped and undamped waves are discussed in Ref. \cite{eliasson2010}.   
\par 
 \textbf{Case-III:}   
 Following the work of Melrose and Mushtaq \cite{melrose2010}, we consider the background distribution of electrons as  a three-dimensional Fermi gas  with arbitrary degeneracy. In this case, the linear dielectric function will be the same as Eq. \eqref{eq-disp-quantum}, however, the three-dimensional background distribution of electrons is given by 
   \begin{equation} 
 f_0(v)\equiv 2\bar {n}(v)\left(\frac{m}{2\pi\hbar}\right)^3=2\left(\frac{m}{2\pi\hbar}\right)^3\left[1+\exp\left(\frac{\varepsilon-\mu_0}{k_BT}\right)\right]^{-1},\label{eq-dist-3d}
\end{equation}
 where $\bar{n}(v)$ is the occupation number of electrons, $\varepsilon=(m/2)(v_x^2+v_y^2+v_z^2)$ is the kinetic energy and  $\mu_0$ is the equilibrium chemical potential  related to the equilibrium number density $n_0$ which satisfies the following charge neutrality condition.
\begin{equation} \label{eq-mu0}
n_0=\int f_0(v) d^3v. 
\end{equation}
The parameter $\mu_0/k_BT$ determines the level of degeneracy of electrons.  It can take from large negative values to large positive values as one enters the regions from nondegenerate to fully degenerate plasmas. Thus, $\xi_0\equiv \exp(\mu_0/k_BT)\ll1$ in the nondegenerate limit   $T\gg T_{F}$      for which one can recover the Maxwellian distribution and   $\xi_0\gg1$   in the fully degenerate limit $(T\ll T_{F})$. The case with $T\gtrsim T_{F}$ such that $\xi_0~(<0)$ is of moderate value, corresponds to  a partially degenerate plasma.  Noting that $d^3v=v^2\sin\theta dvd\theta d\phi$ with $0<v<\infty$, $0\leq \theta\leq\pi$ and $0\leq\phi\leq 2\pi$, and integrating over $\theta$ and $\phi$, we obtain from Eq. \eqref{eq-disp-quantum} the following expression for the electron susceptibility $[D(\omega,k)\equiv 1-\chi(\omega,k)]$ \cite{melrose2010}
\begin{equation}
\chi(\omega,k)=\frac{4\pi e^2m}{\varepsilon_0(2\pi\hbar)^3}\frac{1}{2k\Delta}\int d\varepsilon\bar{n}(\varepsilon)\left[\log \Big|\frac{\omega-a\sqrt{\varepsilon}+\Delta}{\omega+a\sqrt{\varepsilon}+\Delta}\Big|-\log \Big|\frac{\omega-a\sqrt{\varepsilon}-\Delta}{\omega+a\sqrt{\varepsilon}-\Delta}\Big|\right], \label{eq-chi}
\end{equation}
where $\Delta=kv_q$ and $a=\sqrt{(2/m)}k$.
\par 
Equation \eqref{eq-mu0} is rewritten as
\begin{equation}
n_0=\frac{2m^3}{(2\pi\hbar)^3}\sqrt{\frac{2}{m}}\frac{4\pi}{m}\int^{\infty}_0\sqrt{\varepsilon}\bar{n}(\varepsilon)d\varepsilon.
\end{equation}
Next, using in it the distribution function  \eqref{eq-dist-3d} and noting that
\begin{equation}
Li_{\nu}(-z)=-\frac{1}{\Gamma(\nu)}\int_0^{\infty}\frac{s^{\nu-1}}{1+z^{-1}e^s}ds,~\nu>0;~~Li_{\nu}(z)=\sum_{n=1}^{\infty}\frac{z^n}{n^{\nu}},~|z|<1,\label{eq-Li}
\end{equation}
 we obtain  
 \begin{equation}
 n_0=-\frac{2(2\pi)^{3/2}m^3v_t^3}{(2\pi\hbar)^3}Li_{3/2}(-\xi_0),   \label{eq-no-ad}
 \end{equation}
 or, using  the expression for the Fermi temperature   $k_BT_F=(\hbar^2/2m)(3\pi^2n_0)^{2/3}$, we   write
 \begin{equation}
 -Li_{3/2}(-\xi_0)=\frac{4}{3\sqrt{\pi}}\left(\frac{T_F}{T}\right)^{3/2},
 \end{equation}
 Such an expression of $T_F$ can also be obtained from Eq. \eqref{eq-no-ad} in the limit of $\xi_0\gg1$ so that $\mu_0\approx k_BT_F$ and $Li_\nu(-\xi_0)\approx -(\mu_0/k_BT)^\nu/\Gamma(\nu+1)$.
The expression for $n_0$ [Eq. \eqref{eq-no-ad}] is applicable for arbitrary degeneracy of electrons, and using it one can obtain the total number density as 
\begin{equation}
n=n_0\frac{Li_{3/2}(-\xi)}{Li_{3/2}(-\xi_0)}, 
\end{equation} 
where $\xi=\exp(\mu/k_BT)$. 
\par 
A power series expansion in $\xi_0$ of the expression of $\bar{n}$, i.e., $\bar{n}(\varepsilon)=\left[1+\xi_0^{-1}\exp(\varepsilon/k_BT)\right]^{-1}$ can be made in the limit of $\xi_0\ll1$ to give
\begin{equation}
\bar{n}(\varepsilon)=\sum_{s=1}^{\infty}(-1)^{s-1}\xi_0^{s}\exp(-s\varepsilon/k_BT). \label{eq-nbar-small-xi}
\end{equation} 
 The  expression for $\bar{n}(\varepsilon)$ [Eq. \eqref{eq-nbar-small-xi}] can be inserted in  Eq. \eqref{eq-chi} to yield  
 \begin{equation}
 D(\omega,k)=1-\frac{2\pi^{3/2}e^2m^2v_t^2}{\varepsilon_0(2\pi\hbar)^3k\Delta}\sum_{s=1}^{\infty}(-1)^{s-1}\frac{\xi_0^s}{s}\left[Z\left(\sqrt{\frac{s}{k_BT}}y_+\right)-Z\left(\sqrt{\frac{s}{k_BT}}y_-\right)\right], \label{eq-D-Zpm}
 \end{equation}
where $y_{\pm}=(\omega\pm\Delta)/a$ and $Z$ is the plasma dispersion function, given by,
\begin{equation}
Z(\zeta)=\frac{1}{\sqrt{\pi}}\int_{-\infty}^{\infty}\frac{e^{-t^2}}{t-\zeta}dt.
\end{equation}
\par 
Since the plasma dispersion function can have real and imaginary parts, and also $\zeta$ can be large or small, three   cases may be of interest: the case where the Landau resonance contributes; the high-  $(Y_{\pm}\equiv y_{\pm}/\sqrt{k_BT}=(\omega\pm\Delta)/\sqrt{2}kv_t\gg1)$ and low-frequency $(Y_{\pm}\ll1)$ limits according to when $\zeta$ is large or small. The low-frequency limit is disregarded to this study as we will simplify the dispersion relation for high-frequency Langmuir waves and associated Landau damping.  
\par 
In the limit of $Y_{\pm}\gg1$, only the real part of $Z(\zeta)$, where   
\begin{equation}
Z(\zeta)=i\sqrt{\pi}e^{-\zeta^2}-\zeta^{-1}\left[1+(1/2\zeta^2)+(3/4\zeta^4)+\cdots\right],~|\zeta|\gg1,
\end{equation}
is of particular interest.  So, one obtains
\begin{equation}
Z\left(\sqrt{\frac{s}{k_BT}}y_+\right)-Z\left(\sqrt{\frac{s}{k_BT}}y_-\right)=\sqrt{\frac{2}{s}}\frac{2\Delta k v_t}{\omega^2-\Delta^2}\left[1+\frac{k^2v_t^2}{s}\frac{3\omega^2+\Delta^2}{(\omega^2-\Delta^2)^2}+\cdots\right]. \label{eq-Zpm}
\end{equation}
Thus, using the expression \eqref{eq-Zpm}, Eq. \eqref{eq-D-Zpm} reduces to 
\begin{equation}
  D(\omega,k)=1-\frac{\omega_p^2}{\omega^2-\Delta^2}-\omega_p^2k^2v_t^2\frac{3\omega^2+\Delta^2}{(\omega^2-\Delta^2)^3} G, \label{eq-D-Zpm-redu}
\end{equation}
where $G=Li_{5/2}(-\xi_0)/Li_{3/2}(-\xi_0)\approx \sqrt{1+(T_F/5T)^2}$ in which an interpolation has been made by assuming that  
$G\rightarrow1$ for $T\gg T_F$ (non-degenerate limit) and $G\rightarrow v_F^2/5v_t^2$ for $T\ll T_F$ (completely degenerate limit). Although, Eq. \eqref{eq-D-Zpm-redu} is obtained using the power series expansion of $\bar{n}(\varepsilon)$ in the limit of $\xi_0\ll1$, an alternative derivation [for details see Eq. (9) of \cite{melrose2010}] suggests that the dielectric function \eqref{eq-D-Zpm-redu} is applicable for arbitrary degeneracy of electrons. Thus, from Eq. \eqref{eq-D-Zpm-redu}, we obtain the following dispersion relation for Langmuir waves in plasmas with arbitrary degeneracy.
\begin{equation}
\omega^2=\omega_p^2+\Delta^2+\omega_p^2k^2v_t^2\frac{3\omega^2+\Delta^2}{(\omega^2-\Delta^2)^2} G.\label{eq-Disp-redu1}
\end{equation}  
In absence of any thermal flow, we have $\omega^2=\omega_p^2+\Delta^2$. So, if the thermal correction is small, this expression of $\omega^2$ can be substituted in Eq. \eqref{eq-Disp-redu1} to yield
  \begin{equation}
\omega^2=\omega_p^2+3k^2v_t^2G+\frac{\hbar^2k^4}{4m^2},\label{eq-Disp-redu2}
\end{equation} 
where we have retained the terms involving $k$ up to   ${\cal O}(k^4)$.   In the nondegenerate limit  $G\rightarrow1$,   Eq. \eqref{eq-Disp-redu2} reduces to the known dispersion relation for Langmuir waves [\textit{cf.} Eq. \eqref{eq-disp-weak-Q}], i.e.,
\begin{equation}
 \omega^2=\omega_p^2+3k^2v_t^2+\frac{\hbar^2k^4}{4m^2}.
 \end{equation}
On the other hand, in the fully degenerate limit, i.e.,  $G\rightarrow v_F^2/5v_t^2$, Eq. \eqref{eq-Disp-redu2} gives  
\begin{equation}
\omega^2=\omega_p^2+(3/5)k^2v_F^2+\frac{\hbar^2k^4}{4m^2},
\end{equation}
 which agrees with  Eq. \eqref{eq-disp-Lang-FD}  obtained before except an additional factor $(1+\alpha)$ to the term $\propto k^4$. Such a disagreement may be due to an approximation made in the derivation of Eq. \eqref{eq-D-Zpm-redu} in the nondegenerate limit   $\xi_0\ll1$.   Melrose and Mushtaq \cite{melrose2010} made an interpolation formula between the nondegenerate and fully degenerate limits to obtain the following dispersion relation for Langmuir waves  with arbitrary degeneracy.
 \begin{equation}
 \omega^2=\omega_p^2+3k^2\left[v_t^4+\left(\frac{v_F^2}{5}\right)^2\right]^{1/2}+\frac{\hbar^2k^4}{4m^2}.\label{eq-Disp-redu3}
\end{equation}
\par    
From Eq. \eqref{eq-D-Zpm}, the imaginary part can be obtained as 
\begin{equation}
\Im D(\omega,k)=\sqrt{\frac{\pi}{2}}\frac{\omega_p^2}{kv_t}\frac{1}{2\Delta}\left[Li_1\left(-\xi_0 e^{-Y_-^2}\right)-Li_1\left(-\xi_0 e^{-Y_+^2}\right)\right]\Big/Li_{3/2}(-\xi_0),
\end{equation}  
which, after using the relation  $Li_1(z)=\sum_{n=1}^{\infty}z^n/n=-\log(1-z)$, reduces to
\begin{equation}
\Im D(\omega,k)=\sqrt{\frac{\pi}{2}}\frac{\omega_p^2}{kv_t}\frac{1}{2\Delta}\frac{1}{Li_{3/2}(-\xi_0)}\log \left(\frac{1+\xi_0 e^{-Y_+^2}}{1+\xi_0 e^{-Y_-^2}}\right).
\end{equation} 
Thus, for arbitrary degeneracy, the Landau damping rate of Langmuir waves can be obtained by using either $\gamma/\omega_r=-(1/2)D_i(\omega_r,k)$ or $\gamma=-{D_i(\omega_r,k)}/\left(\partial D_r/\partial \omega_r\right)$   and noting that $\partial D_r/\partial \omega_r \approx 2\omega_p^2/\omega_r^3$ for small $k$, i.e., 
 \begin{equation}
\gamma=-\sqrt{\frac{\pi}{2}}\frac{\omega_r^3}{kv_t}\frac{1}{4\Delta}\frac{1}{Li_{3/2}(-\xi_0)}\log \left(\frac{1+\xi_0 e^{-Y_+^2}}{1+\xi_0 e^{-Y_-^2}}\right). \label{eq-gam-arbD}
\end{equation} 
From Eq. \eqref{eq-gam-arbD} it can be assessed that the Landau damping in degenerate plasmas becomes smaller than that in non-degenerate plasmas. 
An alternative derivation of the dielectric function for Langmuir waves in one-dimensional geometry in arbitrary degenerate plasmas can be found in Ref. \cite{rightley2016}.   
\par
\textbf{Case-IV:} So far we have studied the dispersion properties and the Landau damping rates of high-frequency Langmuir waves as described in Cases I to III. Here, we study those for   low-frequency electron-acoustic waves (EAWs) in a  partially degenerate plasma with two-temperature (low with the suffix `$l$' and high with the suffix `$h$') electrons and stationary ions. Such partially degenerate plasmas, where the background distribution of  electrons deviate from   thermodynamic equilibrium,  can appear in the context of laser produced plasmas or ion-beam driven plasmas   \cite{hau-riege2011,gibbon2005}. 
Similar to the previous cases, our starting point is the  Wigner-Moyal and the Poisson system [Eqs. \eqref{wigner-3d} and \eqref{poisson-3d}] which are rewritten for $\alpha$-species electrons as
\begin{equation}
\begin{split}
\frac{\partial f_\alpha}{\partial t}+{\bf v}\cdot\nabla_{\bf r} f_\alpha+\frac{iem^3}{(2\pi)^3\hbar^4}\int\int d^3{\bf r}' d^3{\bf v}' e^{im({\bf v}-{\bf v}')\cdot {\bf r}'/\hbar} &
  \left[\phi\left({\bf r}+\frac{{\bf r}'}{2},t\right)-\phi\left({\bf r}-\frac{{\bf r}'}{2},t\right)\right]\\
  &\times f_\alpha({\bf r},{\bf v}', t)=0, \label{wigner-3d-alpha}
\end{split}
\end{equation}
and the Poisson equation
\begin{equation}
 \nabla^2 \phi=  \frac{e}{\varepsilon_0}\left( \sum_{\alpha=l,h} \int f_\alpha d^3v-n_{0}\right), \label{pois-3d-alpha}
\end{equation} 
where $n_0$ is the unperturbed number density of stationary ions.
 For the one-dimensional propagation of EAWs   along the $x$-axis, the background   distribution function for   electrons is the projected Fermi-Dirac distribution  (writing $v_x$ as $v$).
 \begin{equation} 
\begin{split}
 f_\alpha^{(0)}(v)&=\int\int f^{3D}_\alpha({\bf v})dv_ydv_z\\
 &=2\left(\frac{m}{2\pi\hbar}\right)^3\int\int\left[1+\exp\left(\frac{\varepsilon-\mu_\alpha}{k_BT_{\alpha}}\right)\right]^{-1}dv_ydv_z\\
 &=\frac{3}{4} \frac{n_{\alpha0}}{v_{F\alpha}}\frac{T_\alpha}{T_{F\alpha}}\log\left[1+\exp\left(-\frac{\frac{1}{2}mv^2-\mu_\alpha}{k_BT_{\alpha}}\right)\right], \label{eq-distb-fn}
  \end{split}
\end{equation}
where   $\mu_\alpha$ is the equilibrium chemical potential which satisfies the following charge neutrality condition.
\begin{equation} \label{eq-mu1}
n_0=\sum_\alpha\int f_\alpha^{(0)}(v) dv. 
\end{equation}
As said before, in the nondegenerate limit $(T_\alpha\gg T_{F\alpha})$, the parameter $\xi_\alpha=\mu_\alpha/k_BT_\alpha$ is large and negative, while it is large and positive in the fully degenerate limit $(T_\alpha\ll T_{F\alpha})$. We, however, consider the case of $T_\alpha\gtrsim T_{F\alpha}$ such that $\xi_\alpha~(<0)$ is of moderate value. 
 Here, we note that there are certain parameter restrictions  imposed by the Pauli's exclusion principle as we cannot have a phase space density of the background Wigner function exceeding $2(m/2\pi\hbar)^{3}$. As a result, the parameters for the high- and low-temperature  electron  distributions cannot be chosen independently. The strictest criterion appears for $\varepsilon=0$ leading to
\begin{equation}
\left[{1+\exp \left(-\frac{\mu _{l}}{k_{B}T_{l}}\right)}\right]^{-1}+\left[{1+\exp \left(-\frac{\mu
_{h}}{k_{B}T_{h}}\right)}\right]^{-1}\leq 1. \label{eq-mu2}
\end{equation}
For a partially degenerate low-temperature distribution (i.e., $k_{B}T_{l}\sim E_{F}$, $\mu _{l}\sim 2E_{F}$), this condition is typically fulfilled if the high-temperature distribution is not too far from the 
classical Maxwell-Boltzmann regime such that $-\mu _{h}/k_{B}T_{h}\gtrsim 3$. 
\par
Fourier  analyzing   Eqs. \eqref{wigner-3d-alpha} and \eqref{pois-3d-alpha} by considering $f_\alpha(x,v,t)= f_\alpha^{(0)} +f_\alpha^{(1)}$ and 
$\phi(x,t)=\phi^{(1)}$, and assuming the perturbations to be of the form  $\sim\exp(ikx-i\omega t)$,    we obtain the following dispersion relation.   
\begin{equation}
D(\omega,k)\equiv 1-\sum_{\alpha=l,h}\frac{\omega_{p\alpha}^2}{n_{\alpha_0}k^2}\int_{-\infty}^{\infty} \frac{f_\alpha^{(0)}(v)}{(v-\omega/k)^2-{v_q^2}}dv=0, \label{eq-disp}  
\end{equation}
where   $\omega_{p\alpha}=\sqrt{ n_{\alpha0}e^2/\varepsilon_0 m}$ is the plasma frequency for $\alpha$-species electrons.
Assuming the  wave damping to be small with  $\omega=\omega_r+i\gamma$, we obtain from    $D(\omega,k)\equiv D_r(\omega_r,k)+iD_i(\omega_r,k)+i\gamma[\partial D_r(\omega_r,k)/\partial \omega_r]=0$ the dielectric function 
 \begin{equation}
D_r(\omega_r,k)\equiv 1-\sum_{\alpha=l,h}\frac{\omega_{p\alpha}^2}{n_{\alpha0}k^2}{\cal P}\int \frac{f_\alpha^{(0)}(v)}{(v-\omega_r/k)^2-v_q^2}dv=0, \label{eq-Dr}  
\end{equation}
 and the Landau damping rate 
 \begin{equation}
 \gamma=-\frac{D_i(\omega_r,k)}{\partial D_r/\partial \omega_r}, \label{eq-damp}
 \end{equation}
 where
 \begin{equation}
 D_i= -2\pi v_q\sum_{\alpha=l,h}\frac{\omega_{p\alpha}^2}{n_{\alpha0}k^2}\left[ f_\alpha^{(0)}(v^\text{res}_{+})-f_\alpha^{(0)}(v^\text{res}_{-})  \right],  \label{eq-Di}
 \end{equation}
in which  $v^\text{res}_{\pm}=\omega_r/k\pm  v_q$  denotes the plasmon resonance velocity.  
\par
 Next, substituting the distribution function \eqref{eq-distb-fn} into Eq. \eqref{eq-Dr} and evaluating the integrals in two different regimes $|v|<\omega_r/k\pm v_q$ and $|v|>\omega_r/k\pm v_q$, i.e., 
 \begin{equation} 
 \begin{split}
{\cal P}\int_{-\infty}^{\infty}&=\lim_{\epsilon\rightarrow0+}\left[\int_{-\infty}^{-(\lambda\pm v_q)}+\int_{-(\lambda\pm v_q)}^{(\lambda\pm v_q)-\epsilon}+\int_{(\lambda\pm v_q)+\epsilon}^{\infty}\right]\\
&=\lim_{\epsilon\rightarrow0+}\left[\int_{-(\lambda\pm v_q)}^{(\lambda\pm v_q)-\epsilon}+2\int_{(\lambda\pm v_q)+\epsilon}^{\infty}\right], \label{eq-int}
\end{split}
\end{equation}  
and noting that the exponential function in the distribution function [Eq. \eqref{eq-distb-fn}] is small in the partially degenerate regime, we obtain \cite{misra2021}
 \begin{equation}
 D_r(\omega_r,k)\equiv 1-\frac{3}{4}\sum_{\alpha=l,h}\omega_{p\alpha}^2e^{\xi_\alpha}\left[2\sqrt{2}\frac{v_{t\alpha}/v_{F\alpha}}{\omega_r^2-k^2v_q^2}+\frac{\sqrt{2\pi}v_{t\alpha}-\omega_r/k}{k^2v_{t\alpha}^2v_{F\alpha}}   \right]=0. \label{eq-Dr1}
 \end{equation}
Equation \eqref{eq-Dr1} describes both the high-frequency and relatively low-frequency branches of electrostatic waves.  In the limit of $\omega_r/k\pm v_q\gg v_{th}$, the high-frequency Langmuir wave (LW) mode can be obtained from Eq. \eqref{eq-Dr1} by considering the first and the second terms as
\begin{equation}
\omega_r^2=k^2v_q^2+\frac{3}{\sqrt{2}}\sum_{\alpha=l,h}\omega_{p\alpha}^2\frac{v_{t\alpha}}{v_{F\alpha}}e^{\xi_\alpha}, \label{eq-LW}
\end{equation}
where $v_{t\alpha}=\sqrt{2k_BT_\alpha/m}$ is the thermal velocity and $v_{F\alpha}=\sqrt{2k_BT_{F\alpha}/m}$ the Fermi velocity of $\alpha$-species electrons.
On the other hand, for $\omega_r/k\pm v_q\ll v_{tl}$, the first and the third terms of Eq.  \eqref{eq-Dr1} can be combined to yield the following dispersion relation for the   EAW mode.
\begin{equation}
\omega_r=k\left(\sqrt{2\pi}\sum_{\alpha=l,h}\frac{\omega_{p\alpha}^2e^{\xi\alpha}}{v_{t\alpha} v_{F\alpha}}-\frac{4}{3}k^2 \right)\Big/\sum_{\alpha=l,h}\frac{\omega_{p\alpha}^2e^{\xi\alpha}}{v^2_{t\alpha} v_{F\alpha}}. \label{eq-EAW}
\end{equation}
Furthermore, Eq. \eqref{eq-mu1} for the  equilibrium chemical potentials    reduces to
 \begin{equation}
\frac{3}{4}\sqrt{\pi}\sum_{\alpha=l,h}\left(\frac{T_\alpha}{T_{F\alpha}}\right)^{3/2}n_{\alpha0}e^{\xi_\alpha}=n_0.
\end{equation}
Since the dispersion relations \eqref{eq-LW} and \eqref{eq-EAW} are obtained by assuming the smallness of the exponential function in the distribution function \eqref{eq-distb-fn}, the classical results of LWs and EAWs cannot be recovered directly from Eqs.  \eqref{eq-LW} and \eqref{eq-EAW}.  The reason is that the Fermi-Dirac distribution approaches the Maxwell-Boltzmann distribution in the limit of low density or high temperature, i.e., when the integrand in Eq. \eqref{eq-distb-fn} is much smaller than the unity. 
\par
\begin{figure*}! 
\includegraphics[scale=0.5]{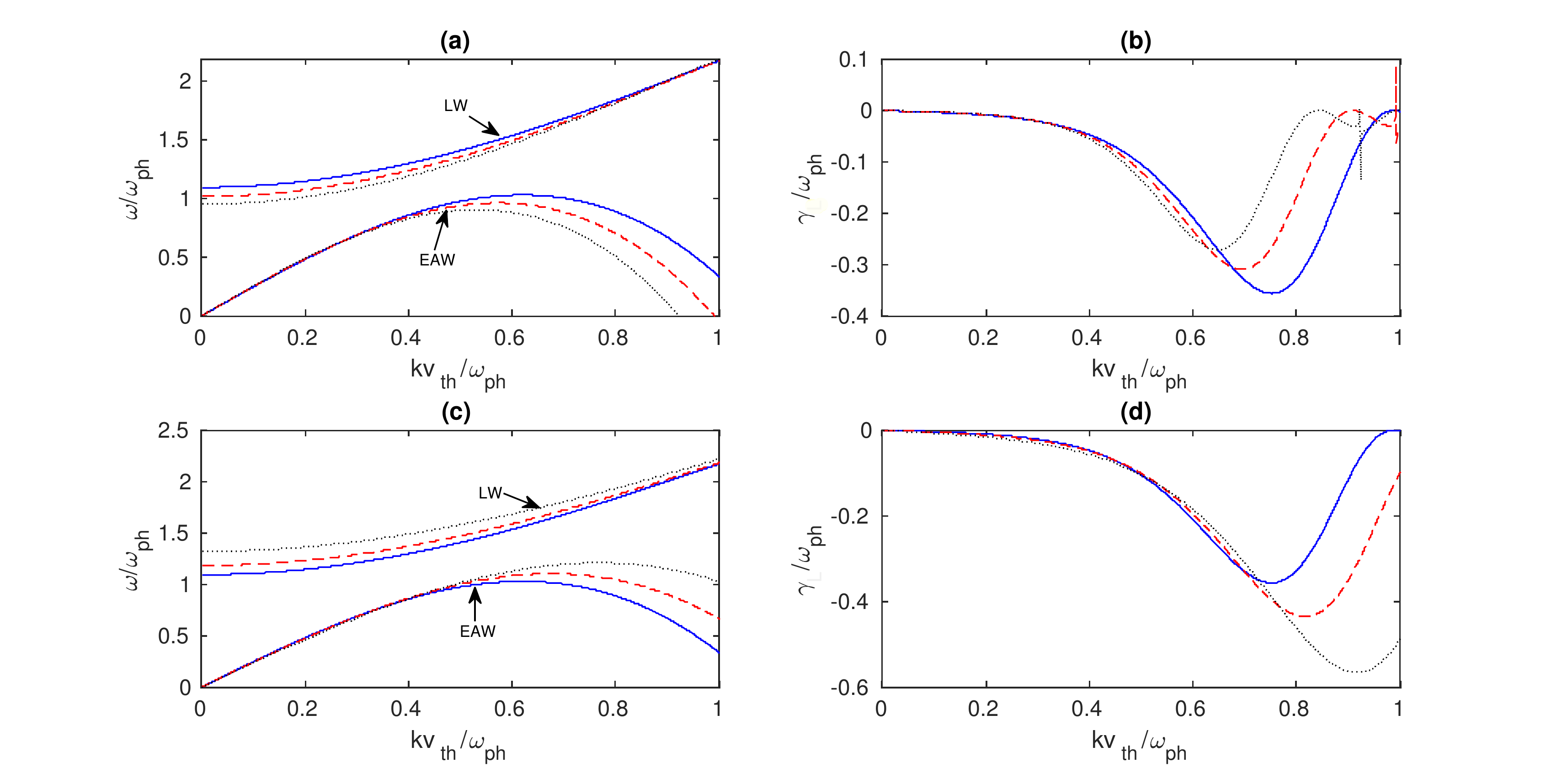}
\caption{The profiles of the dispersion curves  of  Langmuir waves (LWs) and electron-acoustic waves (EAWs) [Subplots (a) and (c)],   given by Eqs. \eqref{eq-LW} and \eqref{eq-EAW}, and the Landau damping rates for EAWs [Subplots (b) and (d)],   given by  Eq. \eqref{eq-damp},     are shown. The figure is reproduced from Ref. \cite{misra2021}. For subplots (a) and (b), the fixed parameters are $n_{l0}=2\times10^{24}$ cm$^{-3}$,  $T_l=10^6$ K and $T_h=2.5 T_l$; the solid, dashed and dotted lines, respectively, correspond to  $n_{h0}=10n_{l0}$, $11n_{l0}$ and $12n_{l0}$.  For subplots (c) and (d), the fixed parameters are $n_{l0}=2\times10^{24}$ cm$^{-3}$, $n_{h0}=10n_{l0}$ and  $T_l=10^6$ K; the solid, dashed and dotted lines, respectively, correspond to  $T_h=2.5 T_l$, $2.7 T_l$ and $3 T_l$. }
\label{fig:disp-curve}
\end{figure*}

      
From Eq. \eqref{eq-EAW} we note that the EAW has the properties similar to the ion-acoustic waves in an electron-ion plasma.  
For typical plasma parameters $n_{l0}=2\times10^{24}$ cm$^{-3}$, $n_{h0}=10-12n_{l0}$,  $T_l=10^6$ K and $T_h=2.5-3 T_l$, and with $kv_{th}/\omega_{ph}\equiv k\lambda_h\ll1$,       Eq. \eqref{eq-EAW} reduces to $\omega_r\approx 2.5kv_{th}$. This predicts the phase velocity of EAWs   a bit higher than predicted in the classical theory \cite{holloway1991} $(\omega_r\approx 1.31kv_{th})$. The profiles of the dispersion curves and the Landau damping rates are shown in  Fig. \ref{fig:disp-curve}.    It is found  that depending on the values of the ratios $T=T_l/T_h$ or  $N=n_{l0}/n_{h0}$,    a critical value $K_c$ of $K\equiv k\lambda_h$ exists,  beyond which the EAW frequency  can turn over, going to zero, and then assume negative values.   Such a distinctive nature of the wave frequency does not appear in classical plasmas where the high- and low-frequency branches  form a thumb-like curve \cite{holloway1991} and it may be due to the finite temperature degeneracy of background electrons.    A considerable regime of $K$ where the  Landau damping rate of EAWs remains weak    is found to be $0\lesssim K\lesssim 0.5$ or $\hbar k/mv_{th}\lesssim0.2$.
 \subsection{Landau damping  in relativistic quantum plasmas}\label{sec-relat}
 In high-enrgy density plasmas, especially in the laser-based  inertial  fusion plasma experiments and  laser-based plasma compression schemes,   electrons become highly relativistic due to laser-driven ponderomotive force. So, it is required to consider a relativistic quantum kinetic model for the description of wave-particle interactions  in high-energy density plasmas. A theoretical study along this line was made by Zhu and Ji \cite{zhu2012}. According to their work, we consider the relativistic quantum kinetic model which is established by the covariant Wigner function and Dirac equation.  The covariant form of one-particle Wigner function is defined as
 \begin{equation}
 f^\mu(x,p)=\frac{1}{(2\pi\hbar)^4}\int d^4y\exp(-ip^\mu y_\mu/\hbar) \Big\langle \bar{\psi}\left(x+\frac{1}{2}y\right)\gamma^\mu\psi \left(x-\frac{1}{2}y\right) \Big\rangle. \label{eq-Wign-rel}
 \end{equation}
Here, $p$ is the relativistic particle momentum,    $\gamma^\mu$ with $\mu=0,1,2,3$ are the   matrices which can be expressed  in terms of $2\times2$   sub-matrices of the Pauli matrices and the $2\times2$  identity matrix, and are such that $\lbrace \gamma^\mu, \gamma^\nu \rbrace=2\eta^{\mu\nu}I_4$, where $\eta$ is the Minkowski metric with metric signature $(+ - - -)$,  $I_4$ is the $4\times4$ identity matrix, and the bracketed expression $\lbrace a, b \rbrace=ab+ba$  denotes the anti-commutator. The angular brackets  $\langle...\rangle$ in Eq. \eqref{eq-Wign-rel} denote  a quantum statistical average, i.e., $\langle...\rangle=\rm{Tr}\lbrace ...\rho\rbrace$ with $\rho$ denoting the density operator that characterizes the statistical state of the system  and $\psi$  (with $\bar{\psi}$ denoting its conjugate)   the wave function  satisfying the following Dirac equation.
\begin{equation}
\left[\gamma^\mu\left(i\hbar\partial_\mu+\frac{e}{c}A_\mu(x)\right)-mc\right]\psi(x)=0, \label{eq-Dirac}
\end{equation}  
where $A_\mu=(-\phi,cA)$ is the Minkowski four-vector operator of the electromagnetic potential which satisfies the D'Alembert's equation in covariant form
\begin{equation}
\square \langle A^\mu(x)\rangle=-\frac{e}{\varepsilon_0mc}\int d^4p p^\mu f(x,p).\label{eq-A-DA}
\end{equation}  
Here, $\square\equiv \partial^\mu\partial_\mu$ is the D'Alembert operator and $f(x,p)$ is defined by
 \begin{equation}
 f(x,p)=\frac{1}{(2\pi\hbar)^4}\int d^4y\exp(-ip^\mu y_\mu/\hbar) \Big\langle \bar{\psi}\left(x+\frac{1}{2}y\right) \psi\left(x-\frac{1}{2}y\right) \Big\rangle. \label{eq-Wign-rel2}
 \end{equation}
Next, the evolution equation for $f^\mu(x,p)$ can be obtained by taking the derivative of Eq. \eqref{eq-Wign-rel} and using the Dirac equation \eqref{eq-Dirac} as
\begin{equation}
\begin{split}
\partial_\mu f^{\mu}(x,p)+\frac{ie}{\hbar c}\frac{1}{(2\pi\hbar)^4}\int\int d^4yd^4p^\prime \exp\left[-i(p-p^\prime)\cdot y/\hbar\right] & \left[  A_\mu\left(x+\frac{1}{2}y\right)- A_\mu\left(x-\frac{1}{2}y\right)\right]\\
&\times f^\mu(x,p^\prime)=0, \label{eq-Wign-rel3}
\end{split}
\end{equation}
 where $p\cdot y=p^\mu y_\mu$. The Dirac current operator $\bar{\psi}\left(x+{y}/{2}\right)\gamma^\mu\psi \left(x-{y}/{2}\right)$ can be decomposed into a convective part and the   current due to the spin  and magnetic moment of particles.  In the weakly relativistic limit, the spin and magnetic moment contributions can be neglected. So, under this approximation one obtains from Eq. \eqref{eq-Wign-rel} using the Lorenz gauge condition, i.e., $\partial_\mu A^\mu(x)=0$ as (See for details Ref. \cite{zhu2012})
 \begin{equation}
 f^\mu(x,p)\approx \frac{1}{mc}\left(p^\mu+\frac{e}{c}A^\mu(x)\right)f(x,p),
 \end{equation}
 and Eqs. \eqref{eq-A-DA} and \eqref{eq-Wign-rel3} reduce to
 \begin{equation}
\square \langle A^\mu(x)\rangle=-\frac{e}{\varepsilon_0mc}\int d^4p p^\mu f(x,p)-\frac{4\pi e^2}{mc^2}\langle A^\mu(x)\rangle\int d^4pf(x,p),\label{eq-A-DA1}
\end{equation} 
 \begin{equation}
 \begin{split}
&\left(p^\mu+\frac{e}{c}A^\mu(x)\right)\partial_\mu f^{\mu}(x,p)-\frac{ie}{\hbar c}\frac{1}{(2\pi\hbar)^4}\int\int d^4yd^4p^\prime \exp\left[-i(p-p^\prime)\cdot y/\hbar\right] \\
&\times\left[A_\mu\left(x+\frac{1}{2}y\right)- A_\mu\left(x-\frac{1}{2}y\right)\right]\left(p^\mu+\frac{e}{c}A^\mu(x)\right)f^\mu(x,p^\prime)=0. \label{eq-Wign-rel4}
\end{split}
\end{equation}
Next,  Fourier analyzing Eqs. \eqref{eq-A-DA1} and \eqref{eq-Wign-rel4} by assuming $f(x,p)=f_0(p)+f_1(x,p)$ and $\langle A^\mu (x)\rangle=O+A_1^\mu (x)$, and using the Lorenz gauge condition $k_\mu A_1^\mu(k)=0$, we obtain the following dispersion relation.
\begin{equation}
\rm{Det}\left[\left(\Omega_p^2/c^2-k_\nu k^\nu\right)\eta^{\mu\sigma}+\omega_p^2K^{\mu\sigma}/c^2\right]=0, \label{eq-Disp-Rel}
\end{equation} 
where $\Omega_p^2=(\omega_p^2/n_0)\int d^4pf_0(p)$  is the relativistic quantum plasma frequency and $K^{\mu\sigma}$ is the dielectric permittivity tensor, given by,
\begin{equation}
K^{\mu\sigma}=\frac{1}{\hbar n_0}\int d^4p\frac{p^\mu p^\sigma}{k_\nu p^\nu}\left[f_0\left(p+\frac{1}{2}\hbar k\right)-f_0\left(p-\frac{1}{2}\hbar k \right)\right]. \label{eq-Kmusig}
\end{equation} 
In a frame where the equilibrium four-velocity component is $(1,0,0,0)$ and the wave propagates along the third-axis, Eq. \eqref{eq-Disp-Rel} reduces to
\begin{equation}
\left(\Omega_p^2-c^2k_\mu k^\mu-\omega_p^2K^{11}\right)^2\left[\left(\Omega_p^2-c^2k_\mu k^\mu+\omega_p^2K^{00}\right)\left(\Omega_p^2-c^2k_\mu k^\mu-\omega_p^2K^{33}\right)+\omega_p^4K^{30}/c^2\right]=0. \label{eq-Disp-Rel1}
\end{equation}
The first factor of Eq. \eqref{eq-Disp-Rel1} gives the transverse mode
\begin{equation}
\omega_T^2=\Omega_p^2+c^2k^2-\omega_p^2 K^{11}.\label{eq-disp-tr}
\end{equation}
On the other hand, the second factor of Eq. \eqref{eq-Disp-Rel1} after using the current conservation equation $k_\mu\Pi^{\mu\nu}(k)=0$, where  $\Pi^{\mu\nu}$ is the polarization tensor, i.e., $\Pi^{\mu\nu}(k)=-\left(\omega_p^2K^{\mu\nu}+\Omega_p^2\eta^{\mu\nu}\right)/4\pi$, gives the following longitudinal mode.
\begin{equation}
\omega_L^2=\Omega_p^2+c^2k^2+\omega_p^2 K^{00}-\frac{\omega}{kc}\omega_p^2K^{30}. \label{eq-disp-long}
\end{equation}
In a fully degenerate plasma   the   background distribution of electrons can be considered as
\begin{equation}
f_0(p)=\frac{2mc^3}{(2\pi\hbar)^3}\int d^4p^\prime\delta(p-p^\prime)2\theta(p^\prime_0c)\delta(p^{\prime2}c^2-m^2c^4)\theta(\varepsilon_F-p_0^{\prime}c), \label{eq-dist-rel}
\end{equation} 
where $\delta$ is the Dirac delta function, $\theta$ is the Heaviside step function, $\varepsilon_F$ is the electron Fermi energy and  $p_0^\prime$ is the classical momentum.
\par 
Substituting the distribution function [Eq. \eqref{eq-dist-rel}] into Eq. \eqref{eq-Kmusig} and assuming that $\hbar\omega\ll E_p\equiv (m^2c^4+p^2c^2)^{1/2}$, $\hbar k\ll p$ and $v_F\ll v_\text{ph}$,  we obtain reduced expressions for $K^{00}$, $K^{30}$, and $K^{11}$. Using these expressions of $K$'s and  considering the weakly relativistic limit $(p_F\ll mc)$,  we obtain from Eq. \eqref{eq-disp-long} 
\begin{equation}
\omega_L^2=\omega_p^2\left(1-\frac{1}{2}\beta_F^2\right)+\frac{3}{5}k^2v_F^2\left(1-\frac{3}{2}\beta_F^2\right)+\frac{\hbar^2k^4}{4m^2}\left(1-\frac{3}{2}\beta_F^2\right),~~(v_F\ll v_\text{ph}) \label{eq-disp-long1}
\end{equation}
where $\beta_F=v_F/c$. Similarly, one can obtain the dispersion relation for electromagnetic wave in a weakly relativistic quantum plasma as
\begin{equation}
\omega_T^2=\omega_p^2\left(1-\frac{1}{2}\beta_F^2\right)+c^2k^2+\frac{\hbar^2k^4}{4m^2}\left(1-\frac{3}{2}\beta_F^2\right). \label{eq-disp-tr1}
\end{equation}
In the limit of $\beta_F\rightarrow0$, Eq. \eqref{eq-disp-long1} reduces to the dispersion relation  of Langmuir waves in a nonrelativistic fully degenerate plasma as discussed in Case II. Also, in the classical limit $\hbar\rightarrow0$ together with the limit $\beta_F\rightarrow0$, we recover from Eq. \eqref{eq-disp-tr1} the classical dispersion relation of electromagnetic waves. Note that since in the weakly relativistic approximation, the phase velocity of electromagnetic waves is much higher than the Fermi velocity, the possibility of Landau damping is ruled out. So, we are interested only with the Langmuir waves. An expression for the Landau damping rate can be obtained by using $\gamma/\omega_r=-(1/2)D_i(\omega_r,k)$  as \cite{zhu2012}
\begin{equation}
\frac{\gamma}{\omega_r}=-\frac{3\pi}{4}\frac{\omega_p^2\omega_r}{k^3v_F^3}\left(1+\beta_F^2+\frac{1}{8}\frac{\hbar^2 k^2}{m^2c^2} \right), ~~(v_F>v_\text{ph}), \label{eq-damp-rel}
\end{equation} 
where $\omega_r\equiv \omega_L$.
\par 
A comparison of the dispersion properties and associated Landau damping of Langmuir waves can be made in classical plasmas, nonrelativistic quantum plasmas and relativistic quantum plasmas.   
Comparing the Landau damping rates for Langmuir waves in classical and fully degenerate plasmas we find that
\begin{equation}
\frac{\gamma_\mathrm{cl}}{\gamma_\mathrm{F}}\sim\pi\sqrt{2\pi}e^{-3/2}n_0\lambda_B^3\sim1.8n_0\lambda_B^3. \label{eq-gam-cl/F}
\end{equation}
 Equation \eqref{eq-gam-cl/F} shows that since for quantum plasmas $n_0\lambda_B^3\geq1$, the Landau damping rate in classical plasmas is higher than that in fully degenerate plasmas. On the other hand, the ratio of the damping rates in non-relativistic and relativistic quantum 
plasmas gives
\begin{equation}
\frac{\gamma_\mathrm{rq}}{\gamma_\mathrm{nq}}\sim 1+\beta_F^2+\frac{1}{8}\frac{\hbar^2 k^2}{m^2c^2}. \label{eq-gam-rq-nq}
\end{equation}
It follows from Eq.  \eqref{eq-gam-rq-nq}   that the Landau damping rate in relativistic regime is a bit higher than that in nonrelativistic regime due to the relativistic factor $\beta_F$ and the quantum recoil associated with the particle dispersion   $ \propto\hbar^2$.  Similarly,    the classical Landau damping  rate can also be shown to be higher than that   in non-relativistic quamtum plasmas with finite temperature degeneracy. The reason is that for degenerate plasmas, most of the electron  energy levels are filled up to the Fermi energy and number of free electrons to take part in the resonance is reduced.  As a result, the energy conversion between Langmuir waves and degenerate particles are not so effective as in classical plasmas. 
The dispersion relations and the Landau damping rates so  obtained in different plasmas with various background distributions are summarized in  Table \ref{table-disp}.  It is noted that depending on the quantum effect   weak or strong,   the background distribution changes from Boltzmann to Fermi-Dirac statistics.  
 \begin{center}
 \begin{table}
 \begin{tabular}{|c|c|c|c|}       
\hline                   
System & \vtop{\hbox{\strut Background} \hbox{\strut distribution} } & Dispersion relation  & \vtop{\hbox{\strut Landau damping} \hbox{\strut rate}}\\ \hline  
Classical plasma & Maxwellian & $\omega^2=\omega_p^2+3k^2v_t^2$ & \vtop{ \hbox{\strut $\frac{\gamma}{\omega}=-\sqrt{\frac{\pi}{8}}\frac{\omega_p^2\omega}{k^3v_{t}^3}$} \hbox{\strut $\times \exp\left(-\frac{\omega^2}{2k^2v_t^2}\right)$ }} \\ \hline
\vtop{\hbox{\strut Non-relativistic}\hbox{\strut quantum plasma}}  &  Maxwellian  & \vtop{\hbox{\strut $\omega^2=\omega_p^2+3k^2v_t^2$}\hbox{\strut $+\frac{\hbar^2k^4}{4m^2}$} } & \vtop{ \hbox{\strut $\frac{\gamma}{\omega}=-\sqrt{\frac{\pi}{8}}\frac{\omega_p^2\omega}{k^3v_{t}^3}\left[1-\frac{1}{6}\frac{v_q^2}{v_t^2}\left(3-\frac{\omega^2}{k^2v_t^2}\right)\right]$} \hbox{\strut $\times \exp\left(-\frac{\omega^2}{2k^2v_t^2}\right)$ }} \\  \hline
\vtop{\hbox{\strut Non-relativistic}\hbox{\strut quantum plasma}}  & \vtop{\hbox{\strut Fermi-Dirac}\hbox{\strut (Zero temperature) }}     & \vtop{\hbox{\strut $\omega^2=\omega_p^2+\frac{3}{5}k^2v_F^2$}\hbox{\strut $+(1+\alpha)\frac{\hbar^2k^4}{4m^2}$} } & $\frac{\gamma}{\omega}=-\frac{3\pi}{4}\frac{\omega\omega_p^2}{k^3v_F^3}$  \\  \hline
  \vtop{\hbox{\strut Non-relativistic}\hbox{\strut quantum plasma}}& \vtop{\hbox{\strut Fermi-Dirac}\hbox{\strut (Finite temperature)}}   & \vtop{\hbox{\strut $\omega^2=\omega_p^2+3k^2$}\hbox{\strut $\times \left[v_t^4+(v_F^2/5)^2\right]^{1/2}$} \hbox{$+\frac{\hbar^2k^4}{4m^2}$} } & \vtop{\hbox{\strut $\frac{\gamma}{\omega}=-\sqrt{\frac{\pi}{2}}\frac{\omega^2}{kv_t}\frac{(1/4\Delta)}{Li_{3/2}(-\xi)}$} \hbox{\strut $\times\log \left(\frac{1+\xi e^{-Y_+^2}}{1+\xi e^{-Y_-^2}}\right)$}}   \\ \hline 
\vtop{\hbox{\strut Relativistic}\hbox{\strut quantum plasma}}& \vtop{\hbox{\strut Fermi-Dirac}\hbox{\strut (Zero temperature)}}   & \vtop{\hbox{\strut $\omega^2=\omega_p^2\left(1-\frac{1} {2}\beta_F^2\right)$}\hbox{\strut $+\frac{3}{5}k^2v_F^2 \left(1-\frac{3} {2}\beta_F^2\right)$} \hbox{$+\frac{\hbar^2k^4}{4m^2}\left(1-\frac{3} {2}\beta_F^2\right)$} } & \vtop{ \hbox{\strut $\frac{\gamma}{\omega}=-\frac{3\pi}{4}\frac{\omega\omega_p^2}{k^3v_F^3}$ } \hbox{\strut $ \times\left( 1+\beta_F^2+\frac{1}{8}\frac{\hbar^2k^2}{m^2c^2}\right)$ } } \\ \hline 
\end{tabular}
\caption{Dispersion relations and Landau damping rates derived in classical, non-relativistic quantum plasmas, and relativistic quantum plasmas.} \label{table-disp}
\end{table}
\end{center}
\section{Wave-particle interactions: Nonlinear theory} \label{sec-nonl-th-spinless}
 So far we have discussed the linear theory of wave-particle interactions, especially the Landau   damping in classical and quantum regimes. We have seen that while the linearized Vlasov-Poisson system in classical and weak quantum (semiclassical) regimes gives the phase velocity resonance, the linearized Wigner-Moyal equation predicts the wave damping where the particle's resonant velocity is shifted from the phase velocity by a velocity $v_q=\hbar k/2m$ due to quantum effects. Going beyond the linear theory, we will see that while the phase velocity or group velocity  is  still the resonant velocity in the classical or  weak quantum  regime, in the strong quantum regime there appear some additional resonances with velocity shifts $nv_q$, $n=2,3,...$, called multi-plasmon resonances which can occur due to   simultaneous absorption (or emission) of multiple plasmon quanta \cite{brodin2017}.
 \par 
  On the other hand, it is known from the classical theory that for a homogeneous plasma wave, i.e., a wave with infinite extent rather than a localized pulse, the linear Landau damping   can  turn into   nonlinear bounce  oscillations with  the bounce frequency of trapped particles   $\omega _{B}=(ek^{2}\Phi /m)^{1/2}$ \cite{oneil1965,nicholson1983}, where $\Phi $ is the potential amplitude of the wave field. However, in quantum plasmas not only the modification of such classical behaviors  occurs  but also a complete suppression of the linear Landau resonance can be seen depending on which regime (weak or strong quantum) we consider. Our aim in Sec. \ref{sec-non-hom} is to demonstrate these phenomena, especially to show the existence of bounce-like oscillations even in absence of trapped particles    in the weak quantum regime and the emergence of nonlinear multi-plasmon resonance in the strong quantum regime.  Although, many of the more well-known aspects of Landau damping can be studied for an infinite plane wave,  there are still some rooms to generalize this setup to the more realistic case of localized   waves and wave packets where the phase velocity or group velocity resonances   enter the picture in the weak quantum regime together with the nonlinear multi-plasmon resonances in the strong quantum regime.
     We will also consider these nonlinear resonant wave-particle  interactions in Secs. \ref{sec-iaw-sem-cl}  to \ref{sec-nld-LW-multi}  on the assumption that the particle trapping time by  the wave, i.e., $\tilde{\omega}^{-1} _{B}$ is typically longer than the  time  by which the wave gets damped, i.e.,  $\gamma>\tilde{\omega}_B$, where $\gamma$ is the linear Landau damping rate.     
\subsection{Wave-particle interactions in the nonlinear homogeneous regime} \label{sec-non-hom}
   For a homogeneous plasma, particles to be trapped and the nonlinearities to be important, the bounce frequency must fulfill  $\omega _{B}$ $>\gamma$. As noted, the classical behaviors can be modified in the quantum regime; so, we consider  two  regimes, namely the weak quantum and the strong quantum regimes.  \textit{Firstly}, for the case of weak damping, when the linear resonance is located in the tail of the distribution, quantum effects can influence the dynamics in a regime that is seemingly classical. In particular, even if the conditions for the classical regime, $v_{F}\ll v_{t}$ and $\hbar k/mv_{t}\ll 1$, are both fulfilled, the nonlinear regime of wave-particle interaction may still be strongly modified by quantum effects. This phenomenon will be considered in the first subsection \ref{sec-wqr} below. \textit{Secondly}, we will consider a completely degenerate system, and focus on the strong quantum regime, in which case $\hbar k/mv_{F}\sim 1$. Here, we will be concerned with the case where linear wave-particle damping is suppressed completely, but where nonlinear wave-particle interaction is possible due to processes involving simultaneous absorption of multiple wave-quanta.
 \subsubsection{The weak quantum regime} \label{sec-wqr}
In this subsection, we consider a nearly classical case  with $v_{F}\ll v_{t}$
and $\hbar k/mv_{t}\ll 1$, and a resonance in the tail of the background
electron distribution. Due to the weak quantum condition, the linear
resonant velocity will be close to the phase velocity. To assure that the
damping is modest, i.e. $\gamma /\omega $ $\ll 1$, we will assume that $ 
\omega /kv_{t}<1$ with some margin. After an initial simplification, using the
above inequalities, the dynamical equation is solved numerically as in Ref. 
\cite{brodin2015}. The key steps in simplifying the full Wigner-Poisson
system are as follows:
\begin{enumerate}
\item Due to the resonance being located in the tail of the distribution,
the nonlinearity sets in for a modest amplitude. Specifically, we have $%
ek\Phi/m\omega v_{t}\ll 1$. Together with the condition $\hbar k/mv_{t}\ll 1$,
this means the linear Vlasov equation applies for most of the velocity space,
except close to the resonance, where the full equation (nonlinear Wigner
equation) must be solved.
\item The resonance region is defined by $[v_\mathrm{res}-\delta
v_\mathrm{res},v_\mathrm{res}+\delta v_\mathrm{res}]$, and defines the region where the full
nonlinear Wigner equation is solved (as opposed to the linearized Vlasov
equation). As long as $\delta v_\mathrm{res}$ fulfills $\hbar k/2m\ll \delta
v_\mathrm{res}\ll kv_{t}^{2}/\omega \,$, the results are insensitive to the exact
width of the resonance region $\delta v_\mathrm{res}.$
\item In the resonance region, the ansatz for the one-dimensional Wigner
function is a general periodic function, i.e., it can be written in the form 
\begin{equation}
g=g_{0}(\mathbf{v})+\delta g_{0}(\mathbf{v,}t)+\left[ \sum_{n=1}^{\infty
}g_{n}(\mathbf{v,}t)\exp i[n(kz-\omega t)]+\mathrm{c.c}\right]. 
\end{equation}
\item In a practical calculation scheme, the sum over harmonics needs to be
truncated. The number of harmonics to be required will vary, mainly depending on the
combined quantum and nonlinearity parameter \ $e\Phi /\hbar \omega
_{B}=m\omega _{B}/\hbar k^{2}$.
\end{enumerate}
\par
Based on the points 1 to 4, Ref. \cite{brodin2015} reduced the Wigner-Poisson
equations to the following normalized equations. 
\begin{equation}
\frac{\partial \hat{\Phi}(t)}{\partial t}=\frac{1}{\pi }\int_{\mathrm{res}%
}g_{1}dv,  \label{Norm-1}
\end{equation} 
\begin{eqnarray}
\frac{\partial g_{1}}{\partial t}+ivg_{1} &=&\hat{\Phi}\left[ 1+\frac{%
g_{0}(v+\delta v_{q})-g_{0}(v-\delta v_{q})}{2\delta v_{q}}\right] +\hat{\Phi%
}^{\ast }\left[ \frac{g_{2}(v+\delta v_{q})-g_{2}(v-\delta v_{q})}{2\delta
v_{q}}\right],   \label{Norm-2} \\
\frac{\partial g_{n}}{\partial t}+invg_{n} &=&\hat{\Phi}\left[ \frac{%
g_{n-1}^{\ast }(v+\delta v_{q})-g_{n-1}^{\ast }(v-\delta v_{q})}{2\delta
v_{q}}\right] +\hat{\Phi}^{\ast }\left[ \frac{g_{n+1}(v+\delta
v_{q})-g_{n+1}(v-\delta v_{q})}{2\delta v_{q}}\right],   \label{Norm-3}
\end{eqnarray} 
and
\begin{equation}
\frac{\partial g_{0}}{\partial t}=\hat{\Phi}\left[ \frac{g_{1}^{\ast
}(v+\delta v_{q})-g_{1}^{\ast }(v-\delta v_{q})}{2\delta v_{q}}\right] +\hat{%
\Phi}^{\ast }\left[ \frac{g_{1}(v+\delta v_{q})-g_{1}(v-\delta v_{q})}{%
2\delta v_{q}}\right],   \label{Norm-4}
\end{equation}%
where $n\geq 2$ in Eq. (\ref{Norm-3}). The physical quantities are normalized
according to:  the time $t\rightarrow\gamma t$,  the   velocity  $v_z\rightarrow
kv_{z}/\gamma$, the  potential  $\Phi\rightarrow ek^{2}\Phi
/\gamma^{2}m$ and the normalized harmonics of the Wigner function is
  $(kg_{n}/\gamma)\left[\partial G_{0}(v_{z})/\partial v_{z}\right]_{\omega
/k}$. Finally, the quantum velocity shift, occurring in the arguments of $g_{n}$, is given by $\delta v_{q}=hk^{2}/2m\gamma$.
\par
An advantage with the above system of Eqs. (\ref{Norm-1})--(\ref{Norm-4}), as
compared to the initial system, is that the time-steps larger than the inverse
plasma frequency can be used. Moreover, the equations only need to be solved
in a small part of velocity space, close to the resonance. Finally, the
spatial dependence is solved for analytically in Eqs. (\ref{Norm-1})--(\ref{Norm-4}%
), which further simplifies the numerics. A detailed numerical study of Eqs.
(\ref{Norm-1})--(\ref{Norm-4}) was presented in Ref. \cite{brodin2015}.
Here, we will only present the main features, which were the following:

\begin{enumerate}
\item \textbf{A quantum modification of the nonlinear bounce frequency.}
Specifically, the classical nonlinear bounce frequency $\omega _{B}$ is
replaced by a quantum correspondence $\tilde{\omega}_{B}$ with a lower value
given by $\tilde{\omega}_{B}=\omega _{B}/(1+\hbar k^{2}/2m\gamma)$.
Thus, when $\hbar k^{2}/2m\gamma\gg 1$ is fulfilled, there is a
substantial difference in the bounce frequency. Due to the smallness of the
linear damping, the decrease in the bounce frequency can be appreciable even in
the regime $\hbar k/mv_{t}\ll 1$.

\item \textbf{Quantum suppression of the nonlinear bounce oscillations.}
Classically, for $\omega _{B}>\gamma$, we have nonlinear bounce
oscillations. However, the corresponding quantum condition is $\tilde{\omega} 
_{B}>\gamma$. Thus, in the regime where $\tilde{\omega}_{B}<\gamma
<\omega _{B}$, classical theory will be thoroughly invalidated  as the
nonlinear oscillations will be completely suppressed, and instead linear
damping takes place. Apart from Ref. \cite{brodin2015}, this feature was
also observed in Ref. \cite{daligault2014}, using a slightly different
approach. 

\item \textbf{Bounce-like oscillations in the absence of trapped particles. }%
Classically nonlinear oscillations take place due   to particles being
trapped in the potential well of the plasma oscillations. In the regime 
where $\omega _{B}>\hbar k^{2}/2m$, however, the connection between
nonlinear oscillations and trapped particles completely disappears. In this
regime, there are no trapped particles, since the lowest energy state of the
potential well will be higher than the trapping potential. However, the
nonlinear bounce-like oscillations can still take place. In this case, the
energy oscillates between different harmonics of the perturbed distribution
function leading to bounce-like oscillations of the electrostatic
potential.
\end{enumerate}
\par
The above features taken together   show that the quantum behaviors can be seen in a regime
that on the surface is classical  as we have used $\hbar k/mv_{t}\ll 1$ and $%
v_{F}\ll v_{t}$. The main reason making this possible is that the
characteristic scale length in velocity space is very short for the case of
a resonance in the tail of the electron distribution. As a result, the sharp
localization in velocity space triggers a large quantum uncertainty in
physical space, and hence the system will be subject to quantum
modifications of the classical theory. While the above results refer to
quantum modifications when the resonance is located in the tail of the
distribution, we note that some of the above features also remain  when the
resonance lies close  to the bulk of the distribution  although  the plasma
temperature and density must fit into the typical quantum regime in that
case. Specifically, numerical results presented in Ref. \cite{suh1991} 
show  that the quantum suppression of nonlinear behaviors can be applicable also
in that case.

\subsubsection{The strong quantum regime} \label{sec-sqr}

In the strong quantum regime, we have short wavelengths $\hbar k\sim mv_{F}$ 
  and a high plasma density, i.e.,  $H\sim 1$.   Typically, there is no regime
resembling the classical bounce oscillations in that case. However, an
interesting effect occurs in the degenerate regime, when the resonance of Eq.
\eqref{eq-Dr-quantum} occurs at a velocity slightly larger than the Fermi velocity. In that
case, the linear wave-particle damping may be absent  and be replaced by a
nonlinear counterpart. To understand how this may happen, let us take a first
  look at the linear wave-particle resonance. As noticed in Eq.
\eqref{eq-Dr-quantum}, the quantum mechanical adjustment of the resonant velocity is given by

\begin{equation}
v^{\text{res}}=\frac{\omega }{k}\rightarrow v^{\text{res}}=\frac{\omega }{k}%
\pm \frac{\hbar k}{2m}.  \label{Eq-15}
\end{equation}%
Let us study the physical meaning of this result briefly. When a particle absorbs or
emits a wave quantum it can increase or decrease the momentum according to 
\begin{equation}
\hbar k_{1}\pm \hbar k=\hbar k_{2},  \label{Eq-16}
\end{equation}%
and at the same time the energy changes according to%
\begin{equation}
\hbar \omega _{1}\pm \hbar \omega =\hbar \omega _{2}.  \label{Eq-17}
\end{equation}%
Next, we identify $\hbar k_{1}/m$ (or equally well $\hbar k_{2}/m$) with the
resonant velocity $v^{\text{res}}$ and note that for small amplitude waves,
the frequencies and wave numbers $(\omega _{1,2},k_{1,2})$ obey the
free particle dispersion relation $\omega _{1,2}=\hbar k_{1,2}^{2}/2m$.
Using these relations, we can deduce that the energy momentum relations
[Eqs.~(\ref{Eq-16}) and (\ref{Eq-17})] imply the modification of the resonant
velocity as seen in Eq.~(\ref{Eq-15}). An interesting possibility,  which was
studied in Ref.~\cite{brodin2017}, is the simultaneous absorption (or
emission) of multiple wave quanta  rather than a single wave quantum at a
time. In that case, Eqs. (\ref{Eq-16}) and (\ref{Eq-17}) are replaced by 
\begin{equation}
\hbar k_{1}\pm n\hbar k=\hbar k_{2},  \label{Eq-18}
\end{equation}%
and 
\begin{equation}
\hbar \omega _{1}\pm n\hbar \omega =\hbar \omega _{2},  \label{Eq-19}
\end{equation}%
where $n=1,2,3,\ldots $ is an integer. Accordingly, performing the same
algebra as for the linear case, the resonant velocities now become 
\begin{equation}
v_{\pm n}^{\text{res}}=\frac{\omega }{k}\pm n\frac{\hbar k}{2m}.
\label{Eq-20}
\end{equation}%
When we pick the minus sign in Eq.~(\ref{Eq-20}), the resonant velocity for
absorbing multiple wave quanta can be considerably low provided the
wavelengths are short. As a consequence, in the case of Langmuir waves, the
damping rate due to absorption of multiple wave quanta can be much larger
than the standard linear damping rate. This is due to the larger number of
resonant particles in the former case.
\par
While the physical conditions are slightly different, mathematically things
are very similar to the weak quantum case  as we need to solve the
Vlasov-Poisson system. Importantly, in the weakly nonlinear regime,
wave-particle interaction due to  multi-plasmon damping is a slow process 
such that a perturbative approach can be applied. Moreover, the division of
velocity space into a resonance   and a nonresonance regions is still
possible. As a result, the system of equations   resembles that of the previous
section (Sec. \ref{sec-wqr}). Specifically, the simplified Wigner equation, after weakly nonlinear
approximations have been made, can be written as
\begin{equation} \label{eq:nonlinearIII}
\begin{split}
\partial _{t}f_{0}& =-\frac{ie}{\hbar }(\Phi _{1}\D_{1}f_{1}^{\ast }-\Phi
_{1}^{\ast }\D_{1}f_{1}+\Phi _{2}\D_{2}f_{2}^{\ast }-\Phi _{2}^{\ast }\D 
_{2}f_{2}),    \\
\partial _{t}{f_{1}}-i(\omega -kv_{z})f_{1}& =-\frac{ie}{\hbar }(\Phi _{1}\D 
_{1}(F_{0}+f_{0})-\Phi _{1}^{\ast }\D_{1}f_{2}+\Phi _{2}\D_{2}f_{1}^{\ast
}-\Phi _{2}^{\ast }\D_{2}f_{3}),    \\
\partial _{t}f_{n}-in(\omega -kv_{z})f_{n}& =-\frac{ie}{\hbar }(\Phi _{1}\D 
_{1}f_{n-1}-\Phi _{1}^{\ast }\D_{1}f_{n+1}+\Phi _{2}\D_{2}f_{n-2}^{\ast
}-\Phi _{2}^{\ast }\D_{2}f_{n+2}), \quad (n>1),  
\end{split}
\end{equation}
where  $f_{n}$ denotes the harmonics of the Wigner-function. Moreover, we use
the same quantum velocity shift $\mathbf{v}_{q}$ as before  and we have
introduced a velocity shift operator $\D_{n}$,  defined by, 
\begin{equation}
(\D_{n}f)(\mathbf{v})=f(\mathbf{v}+n\mathbf{v}_{q})-f(\mathbf{v}-n\mathbf{v}%
_{q}).
\end{equation}
\par
While the governing equations resemble Eqs. (\ref{Norm-2})--(\ref{Norm-4}),
an important difference is that  the harmonics of the electrostatic
potential must be included in the treatment. Moreover, before solving the
equations, further simplifications need to be done, which is slightly
different depending on whether the main damping is due to the resonance for $%
n=2$ (two-plasmon resonance) or for $n=3$ (three plasmon resonance) (see
Ref. \cite{brodin2017} for details). Finally,  the equations are solved numerically.
Interestingly,   the results for two-plasmon damping and three-plasmon damping
are similar. In both the cases,  since the damping mechanism is nonlinear, the
damping rate decays with the amplitude. Importantly, the numerical results
for the damping rate can be fitted to the following expression.

\begin{equation}
|\Phi (t)|=|\Phi (0)|/(1+t/t_{0})^{1/2},
\end{equation}%
where  $t_{0}$ is a characteristic damping time that scales as 
\begin{equation}
t_{0}\sim C(v_{q},\omega /k)\left\vert \frac{\hbar \omega }{e\Phi (0)}%
\right\vert ^{2}\frac{1}{\omega }.  \label{eq:dampingTime}
\end{equation}%
From the numerical results, it is found that the dimensionless coefficient $C$
varies in between $0.03-0.5$ as a function of the velocity shift and the phase
velocity  (see Ref. \cite{brodin2017} for details). While the magnitudes of the
two-plasmon   and the three-plamon damping rates are of
comparable magnitude, generally the damping due to the three-plasmon processes
occurs slightly faster. This follows from the fact that the resonance occurs
somewhat deeper into the bulk of the background electron distribution for
the three-plasmon resonance.
\par
While the effect of multi-plasmon damping is most pronounced for a
completely degenerate system  as the competing linear processes may vanish
completely, it can also be prominent at a finite temperature  as discussed
in some detail in Ref. \cite{brodin2017}.  Moreover, as should be clear from
the discussion leading up to Eq. (\ref{Eq-20}) that the damping mechanism is of
a very general nature. In principle, in case the wavelength is short enough
to make quantum effects important, the same type of multi-quanta damping
mechanism applies to all types of wave-modes  not just the plasmons.
     \subsection{Nonlinear Landau damping of ion-acoustic solitary waves in the weak quantum regime}\label{sec-iaw-sem-cl}
While the nonlinear wave-particle interaction  in homogeneous plasmas is of basic theoretical interest, in a practical context, wave-particle interaction typically competes with other nonlinear processes.  In particular, it is well known that the nonlinear propagation of small amplitude ion-acoustic waves (IAWs) in a plasma with warm electrons and cold ions is asymptotically governed by the Korteweg-de Vries (KdV) equation. The significant modification of this equation due to electron Landau damping was noted and studied by Ott and Sudan \cite{ott1969} on the assumption that particle's trapping time is much longer than that of Landau damping. The theory was later advanced by Vandam and Taniuti \cite{vandam1973} to take into account the ion Landau resonance under the consideration that the Landau damping is a far-field approximation of the Vlasov equation, i.e., a small amplitude long-wavelength wave  will damp after a long time.  The theory of Landau damping of IAWs was, however, further studied in the context of plasmas in the semiclassical or weak quantum regime by Barman and Misra \cite{barman2017}. According to their work, we consider the nonlinear propagation of ion-acoustic waves (IAWs) and the  wave-particle interaction  in an unmagnetized collisionless plasma  with weak quantum  effects, i.e., when   the typical  ion-acoustic    length scale  is larger than the thermal de Broglie wavelength.    In order to include the resonance effects both from quantum electrons and classical ions  we consider the semi-classical Vlasov equation for electrons, Vlasov equation for ions and the Poisson equation, given by,
  \begin{equation}
 \frac{\partial f_e}{\partial t}+v\frac{\partial f_e}{\partial x}+\frac{1}{m_0}\frac{\partial \phi}{\partial x}\frac{\partial f_e}{\partial v}-\frac{H^2}{24m_0^2}  \frac{\partial^3 \phi}{\partial x^3} \frac{\partial^3 f_e}{\partial v^3}=0, \label{nond-Vlasov-eqn-electron} 
\end{equation}
\begin{equation}
 \frac{\partial f_i}{\partial t}+v\frac{\partial f_i}{\partial x}-\frac{\partial \phi}{\partial x}\frac{\partial f_i}{\partial v}=0, \label{nond-Vlasov-eqn-ion} 
\end{equation}
\begin{equation}
 {\frac{\partial^2 \phi}{\partial x^2}}=-\sum_{j=e,i} \theta_{j} \int f_j dv, \label{nond-poisson-eq}
\end{equation} 
In Eqs. \eqref{nond-Vlasov-eqn-electron} to \eqref{nond-poisson-eq} we have normalized the physical quantities according to $v\rightarrow v/c_s$,  $\phi\rightarrow e\phi/k_BT_e$, $n_{j}\rightarrow n_{j}/n_{0}$, and $f_{j}\rightarrow f_{j}c_s/n_{0}$ where $c_s=\sqrt {k_BT_e/m_i}\equiv\omega_{pi}\lambda_D$ is the IAW speed with $\omega_{pi}=\sqrt{n_{0}e^2/\varepsilon_0m_i}$   denoting  the ion plasma frequency. Also,  $n_{0}$ is the equilibrium number density of electrons and ions, and $T_j$ is the thermodynamic temperature of electrons $(j=e)$ and ions $(j=i)$. The space and time variables are normalized by $\lambda_D$ and $\omega^{-1}_{pi}$ respectively. Furthermore,  $m_0=m/m_i$ is the electron to ion mass ratio,  $H=\hbar\omega_{p}/k_BT_e$   is the dimensionless  quantum parameter  denoting the ratio of the electron plasmon energy to the thermal energy and $\theta_j=\mp1$ for   $j=e~(i)$. 
 \par
\begin{figure}
\begin{center}
\includegraphics[scale=0.4]{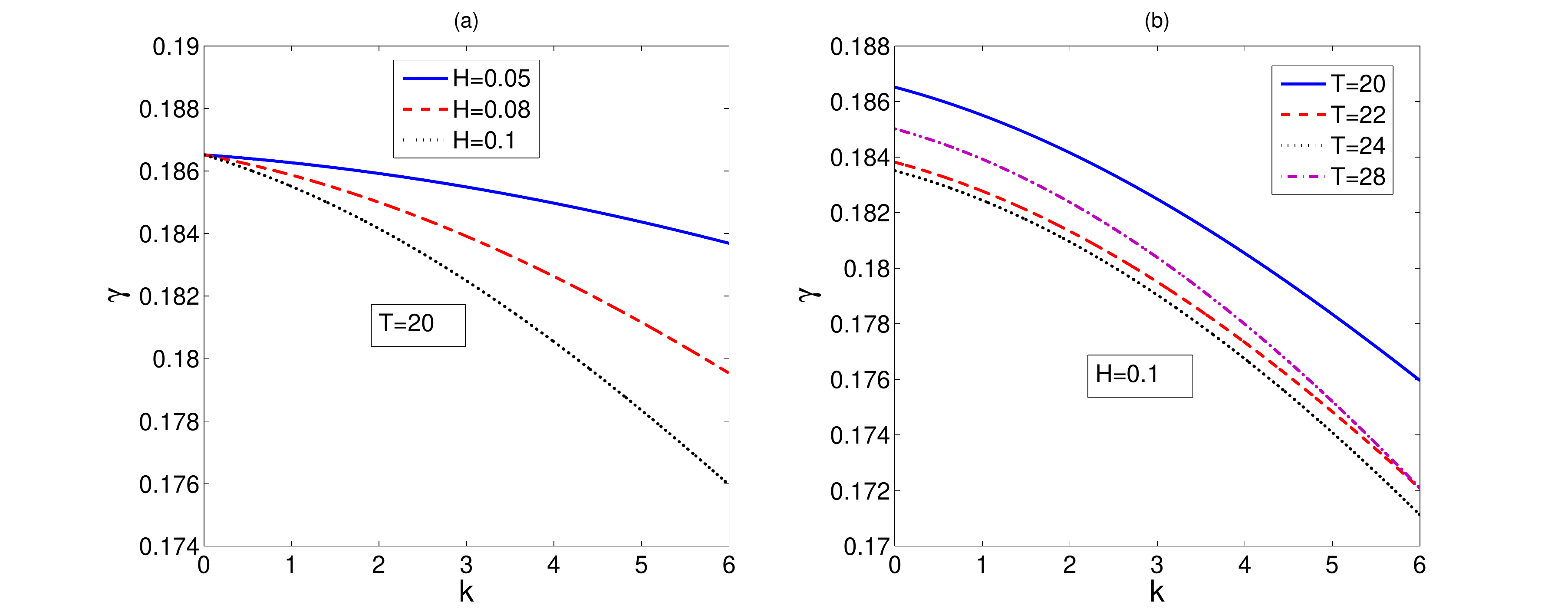}
\caption{The Landau damping rate $\gamma$ (normalized by $\omega_{pi}$) is plotted against the wave number  $k$ (normalized by $\lambda_{D}^{-1}$) in three different cases: (a) when $T$ is fixed and $H$ varies, (b) when $H$ is fixed and $T$ varies and (c) when the value of $T$  is relatively lower than that in the plots (a) and (b). The figure is reproduced from Ref. \cite{barman2017}. }
\label{fig:damping}
\end{center}
\end{figure}

 
 An evolution equation for the small  amplitude IAWs can be derived following Refs. \cite{vandam1973,barman2017}, i.e.,  using the   multi-scale asymptotic expansion  technique in which $\phi$ and $f_j$ are expanded in different powers of $\epsilon$, where $\epsilon~(\lesssim1)$ is a small positive scaling parameter  measuring the weakness of perturbations. In the weak quantum regime, the background distribution of electrons and ions [i.e.,  $f^{(0)}_{j}$, for $j=e, i$] can be assumed to be the Maxwellian.  Furthermore, different  expansions  for $f_{j}$  are to be considered  in the non-resonance $\left(|v-\omega/k|\gg o(\epsilon)\right)$ and resonance  $(v\approx\omega/k)$ regions.   Also,  in order to properly include  the contributions of resonant   particles,      the multi-scale Fourier-Laplace transforms for $f_{j}-f^{(0)}_{j}$ and $\phi$  are to be  employed.    A standard perturbation scheme with the stretched coordinates 
 $\xi=\epsilon^{1/2}x,~ \sigma=\epsilon^{1/2}t,~s=\epsilon^{3/2}x$  
  yields the following evolution equation for the first order potential perturbation of IAWs (for details   see Ref. \cite{barman2017}).
 \begin{equation}
\frac{\partial\phi}{\partial s}+\alpha\phi\frac{\partial\phi}{\partial\zeta}+\beta\frac{\partial^3\phi}{\partial\zeta^3}
+\gamma~\text{P}\int_{-\infty}^{\infty}(\zeta-\zeta')^{-1}\frac{\partial\phi(\zeta')}{\partial\zeta'}d\zeta'=0, \label{KdV}
\end{equation}
where $\zeta=\xi-v_p\sigma$ and   the coefficients are given by  $\alpha=b/a$, $\beta=1/a$ and $\gamma=c/a$ in which  $a$, $b$ and $c$ are simplified to
\begin{equation}
a=2v_p^{-2}\left(1+6v_p^{-2}T^{-1}\right)+v_p H^2k, \label{a-reduced}
\end{equation}
\begin{equation}
b=3v_p^{-4}+30T^{-1}v_p^{-6}-1, \label{b-reduced}
\end{equation}
\begin{equation}
c=\epsilon^{-1}\frac{v_p}{\sqrt{2\pi}}\left[m_0^{1/2}+T^{3/2}\exp\left(-\frac{Tv_p^2}{2}\right)\right].\label{c-reduced}
\end{equation}
Here, $T=T_e/T_i$ with $T_i$ denoting the ion temperature and the term $\propto\gamma$ appears due to the wave-particle resonance and  $v_p$ is the nonlinear wave phase speed $\lambda\equiv\omega/k$, given by,  
 \begin{equation}
v_p^2=\frac{1+\sqrt{1+(12/T)\left(1+{H^2k^2}/{12}\right)}}{2\left(1+{H^2k^2}/{12}\right)}.\label{lambda1}
\end{equation} 
 It is noted that the dispersion relation is  modified by the quantum correction  $\propto H$.
 In the limit of $T\gg1$ and   $H^2k^2/12\ll1$,   Eq. \eqref{lambda1} reduces to 
\begin{equation}
v_p\approx1+\frac{3}{2T}-\frac{H^2k^2}{24}. \label{lambda2}
\end{equation}
 It follows that  in contrast to the quantum fluid theory \cite{haas2003} or classical kinetic theory \cite{vandam1973}, the phase velocity  $v_p$  is no longer  a constant, i.e., the wave becomes dispersive due to the quantum effects. 
A careful analysis shows that   the wave speed $v_p$ always    decays with the wave number $k$. However, it can be increased or decreased depending on the values of $H$ and $T$.   The linear damping rate $\gamma$ [Fig. \ref{fig:damping}]  is also seen to decrease with increasing values of $k$ and  $H$.   However,     a critical value of $T\sim24$ exists below (above) which the value of $\gamma$ decreases (increases) with an increasing value  of $T$.   
\par 
It is pertinent to mention that in the derivation of the  KdV equation \eqref{KdV},   not only the  Landau damping (linear resonance) contributes to the wave dynamics,  there also appears a term involving the effects of particle trapping (nonlinear resonance). However, we have disregarded such term as the results with the trapping effects are similar to  the classical theory \cite{vandam1973}.  So,   we study  mainly  the linear Landau damping effect on the ion-acoustic solitary waves. We also note that each of the coefficients $a,~b$ and $c$ are modified by the quantum parameter $H$, in absence of which one recovers the classical results of Vandam \textit{et al.} \cite{vandam1973}.  
\par 
In order to study the effects of the linear Landau damping on the profile of ion-acoustic solitary waves we find an approximate solitary wave solution  of Eq. \eqref{KdV}  on the assumption that  the effect of the Landau damping ($\varpropto\gamma$) is small, i.e.,  $\gamma\ll\alpha ~(\gtrsim\beta)$, which holds when $T\gtrsim20$ and $H<1$, as \cite{barman2014}   
\begin{equation}
\phi=\Psi~\text{sech}^2\left[\left(\zeta-\frac {\alpha}3\int_{0}^{s}\Phi ds\right)/W\right]+o(\gamma),\label{soliton-sol}
\end{equation}
where  $\Psi=\Phi_0\left(1+{s}/{s_0}\right)^{-2}$ is the amplitude of the solitary wave solution of the  KdV equation \eqref{KdV}, and $\Phi=3U_0/\alpha$ is the corresponding amplitude,   $W=\left({12\beta}/{\Phi\alpha}\right)^{1/2}\equiv\sqrt{4\beta/U_0}$ is the width and $U_0=\Phi\alpha/3$ is the constant phase speed (normalized by $c_s$) of the solitary wave solution of the KdV equation in absence of the Landau damping (i.e., when $\gamma$ or $c=0$).   Also, $\Phi=\Phi_0$   at $s=0$ and $s_0$ is given by
\begin{equation}
s_0^{-1}=\frac{\gamma}{4}\sqrt{\frac{\alpha\Phi_0}{3\beta}}\text{P}\int_{-\infty}^{\infty}\int_{-\infty}^{\infty}\frac{\text{sech}^2z}{z-z'}\frac{\partial}{\partial z'}\left(\text{sech}^2z'\right)dzdz'.\label{tau'}
\end{equation}
It is found that the wave amplitude decays with time and the decay rate is  relatively  low  (compared to the classical result \cite{ott1969}) in the weak quantum regime.  
\par
Some estimates for the bounce freuencies of electrons and ions, as well as  a comparison of the contributions from the linear and nonlinear resonances on the wave damping can be made.  If $\omega_B= \sqrt{e\phi/m}/W$ denotes the bouncing frequency,   the electron trapping time  by a solitary pulse is $\sim\omega_{B}^{-1}$.   Since for small amplitude perturbations, the wave potential scales as  $\phi\sim\epsilon$, we have  $\omega_B\sim\omega_{p}\sqrt{\epsilon}\sim\omega_{pi}\epsilon^{-1/2}$  for $\epsilon\sim\sqrt{m/m_i}$. However, from Fig. \ref{fig:damping}  it can be estimated that the Landau damping rate, $\gamma\sim0.18\omega_{pi}$ for some values of $T,~H$ and $k$. So, the condition $\gamma<\omega_B$ holds  for electrons be trapped. On the other hand,   since for ions  $\omega_B\sim\omega_{pi}\sqrt{\epsilon}$, one has $\gamma>\omega_B$ and  ion trapping may be neglected.     For the nonlinear resonance  we find  \cite{vandam1973}
\begin{equation}
\int_{res.}f_{j} dv\approx-\frac{v_p}{\sqrt{2\pi}}\left[\exp\left(-\frac{1}{2}m_0v_p^2\right)+\epsilon T^{3/2}\exp\left(-\frac{1}{2}Tv_p^2\right)\right].\label{trap-estimate}
\end{equation}
Thus, from Eqs. \eqref{c-reduced} and \eqref{trap-estimate}, it is clear that the effects of the linear resonance is relatively higher than that of the nonlinear one (trapping). Nevertheless, ions may be reflected  by a solitary pulse and propagate as a precursor \cite{tiwari2016}.
\par
 Some important points are to be mentioned.  In the semiclassical regime  since   $T_e>T_F$ holds, the Pauli blocking is reduced  and the particles' collisions  can  influence the dynamics of IAWs. However, the inclusion of a collisional term in the semiclassical Vlasov equation is not so straightforward. If a small collisional effect (e.g., Coulomb collision) is introduced, the effective electron-electron collision frequency scales as $\nu_{ef}\sim \epsilon\omega_{p}\left(n_0\lambda_D^3\right)^{-1}\sim\epsilon^{-2}\omega_{pi}\left(n_0\lambda_D^3\right)^{-1}$. For moderate density plasmas with  $n_0\sim6\times10^{23}$ cm$^{-3}$ and $T_e\sim7\times10^6$ K, one can have $\left(n_0\lambda_D^3\right)^{-1}~(\sim0.13)> \epsilon~(\sim0.02)$   and $H\sim0.05$. Thus, $\nu_{ef}~(\gtrsim\epsilon^{-1}\omega_{pi})>\omega_{B}\sim\sqrt{\epsilon}\omega_{p}\sim\omega_{pi}\epsilon^{-1/2}$, and consequently, the trapping of electrons will be destroyed. Furthermore, depending on the values of $T$, $H$ and $k$, the Landau damping contribution $\propto\gamma$ can be even larger than the damping due to the collisional effects. In this way, one can safely neglect the collisional effects in the dynamics of IAWs.        
    \subsection{Nonlinear Landau damping of Langmuir wave envelopes in the weak quantum regime}\label{sec-nls-sem-cl}
In this section, we consider the resonant wave-particle interactions and amplitude modulation of Langmuir wave packets in the weak quantum regime. Here, instead of the phase velocity resonance as   in the case of IAWs (\textit{cf.} Sec. \ref{sec-iaw-sem-cl}),    the group velocity resonance   occurs and  contributes to the wave damping in the nonlinear regime. We note that the group velocity resonance can similarly be important as for the phase velocity, i.e., for  particle acceleration and transport of particle, momentum and energy. Also, due to this resonance,  the transformation of wave energy takes place from   high-frequency side bands to the low-frequency ones which may result into  the onset of weak or strong turbulence in nonlinear plasma media.
\par
   The modulational instability (MI) has been a well-known mechanism for the evolution of wave packets due to energy localization  in plasmas. It manifests the   exponential growth of a small plane wave perturbation in the medium. Such a gain leads to the amplification of the sidebands leading the uniform wave to break up into a train of oscillations. In this way, the MI acts as a precursor for the formation of bright or dark envelope solitons in dispersive plasma media. However, the wave envelopes can be damped due to the wave-particle interactions. In classical plasmas, Ichikawa \textit{et al.} \cite{ichikawa1974} first investigated the theory of Landau damping of Langmuir wave envelopes due to resonant particles having the group velocity of the wave  assuming that the typical time scale of oscillations is much longer than the bouncing period of particles trapped in the potential trough.   They  showed that the nonlinear wave-particle resonance leads to the modification of the nonlinear Schr{\"o}dinger (NLS) equation with a nonlocal nonlinearity. Further modifications of the nonlinearities and dispersion of the  NLS equation also appear due to the quantum particle's dispersion \cite{chatterjee2016}. To demonstrate it we consider the weak quantum regime, i.e., $\hbar k/mv_t<1$ (or $H=\hbar\omega_p/mv_t^2<1$, where $\omega_p=\sqrt{2 n_0e^2/\varepsilon_0m}$ is the electron-positron plasma oscillation frequency and $v_t=\sqrt{k_BT/m}$ is the thermal velocity of electrons and positrons) and the modulation of Langmuir wave envelopes  with the effects of the wave-particle resonance in an electron-positron-pair plasma. The results will be similar for electron-ion plasmas with stationary ions.  Here, we  assume that $T_e=T_p=T$ and $T> T_F$. So, in the weak quantum regimme,  the background distributions of electrons and positrons can be described by the   Maxwellian-Boltzmann distributions [\textit{cf.} Eq. \eqref{eq-dist-max}].      It has been shown that besides giving rise to the modification of the nonlinearity and dispersion, the Landau damping rate and the decay rate of the wave amplitude are greatly reduced by the quantum particle dispersion \cite{chatterjee2016}.   
\par
Similar to Sec. \ref{sec-iaw-sem-cl},  our basic equations are the semiclassical Vlasov equation for electrons and positrons and the Poisson equation.
  \begin{equation}
\frac{\partial f_{\alpha}}{\partial t}+v\frac{\partial f_{\alpha}}{\partial x}-\frac{e_\alpha}{m_\alpha}\frac{\partial \phi}{\partial x} \frac{\partial f_{\alpha}}{\partial v}+\frac{e_\alpha\hbar^2 }{24 m_\alpha^3} \frac{\partial^3 \phi}{\partial x^3} \frac{\partial^3 F_{\alpha}}{\partial v^3} +{\cal O}(H^4)=0,  \label{wigner-eq}
\end{equation}
\begin{equation}
 {\frac{\partial^2 \phi}{\partial x^2}}=- \sum \frac{e_\alpha}{\varepsilon_0} \int f_\alpha dv, \label{poisson-eq}
\end{equation}
where $e_\alpha=\mp1$ for electrons $(\alpha=e)$ and positrons $(\alpha=p)$ respectively, and $f_\alpha$ is the Wigner distribution function for $\alpha$-species particles. 
\par 
Introducing the multiple space-times scales with the stretched coordinates $x\rightarrow x+\epsilon ^{-1} \eta +\epsilon^{-2}\zeta,~ ~t\rightarrow t+\epsilon ^{-1}\sigma$, the expansions for $\phi$ and $f_\alpha$ in powers of a small positive number $\epsilon$ and using the Fourier-Laplace integrals (see for details, Refs. \cite{ichikawa1974,chatterjee2015,chatterjee2016}  we obtain   the following nonlinear Schr{\"o}dinger (NLS) equation for the small but finite  amplitude perturbation $\phi(\xi, \tau)$   \cite{chatterjee2016}.  
\begin{equation}
i\frac{\partial\phi}{\partial \tau}+P \frac{\partial^2\phi}{\partial \xi^2}+Q |\phi|^2 \phi +\frac{R }{\pi}{\cal P}\int\frac{ |\phi(\xi',\tau)|^2}{\xi-\xi'} d\xi' \phi+i\tilde{\gamma}\phi=0, \label{nls}
\end{equation}
where $\xi=\eta-v_g \sigma$ with $v_g$ denoting the group velocity of the envelope, and the coefficients of the group velocity dispersion $(P)$,   local cubic nonlinear $(Q)$ and nonlocal nonlinear $(R)$ terms are   simplified (in the limit of $\chi^2\equiv k^2\lambda_D^2\ll1$ with $\lambda_D=\left(\varepsilon_0k_B T/2 n_0 e^2\right)^{1/2}$ denoting the   plasma Debye length) to give \cite{chatterjee2016} 
\begin{equation}
P=\frac{3}{2}\frac{\omega_p}{k_d^2} \left[1- \frac{1}{2}(9-H^2) \chi^2+\frac{85}{8} H^2 \chi^4\right], \label{P-reduced}
\end{equation}
\begin{equation}
Q= -\frac{1}{2} \left( \frac{e}{k_BT}\right) ^2 \omega_p \chi^2 \left( 1-\frac{H^2}{4} \chi^2\right), \label{Q-reduced}
\end{equation}
\begin{equation}
R=\frac{3}{2} \left( \frac{e}{k_BT}\right) ^2 \left( \frac{\pi}{2}\right) ^{1/2} \omega_p\chi^3 \left( 1-\frac{13H^2}{24}\chi^2\right). \label{R-reduced} 
\end{equation}
The coefficients $P$, $Q$ and $R$ of the NLS equation \eqref{nls} are modified by the quantum parameter $H$ associated with the particle dispersion. The nonlocal term $\propto R$ appears due to the wave-particle resonance having the group velocity of the wave envelopes. This resonance contribution also modifies  the local nonlinear coefficient $Q$, which appears due to the carrier wave self-interactions.  The   damping coefficient $\tilde{\gamma}$   associated with the   phase velocity resonance in the linear regime is given by
\begin{equation}
\tilde{\gamma}=\frac{\theta (s)\gamma}{\epsilon^2}, \label{S}
\end{equation}
where $\theta(s)$ is unity for $s=0$ and vanishes otherwise. Clearly, if the linear damping rate is higher order than $\epsilon^2$, the contribution from the term $\propto \tilde{\gamma}$ is relative small compared to that of the nonlinear Landau damping $\propto R$.  
\par 
We focus in the regime of small $k$ and $H$. In particular, for  $ k^2\lambda_D^2\ll1$  and  the smallness of thermal corrections, the dispersion    relation and the Landau damping rate  are simplified to
\begin{equation}
{\omega_r}^2 = {\omega_p}^2\left( 1+ 3\chi^2 +\frac{1}{4}H^2 \chi^4\right), \label{dispersion-reduced}
\end{equation}
\begin{equation}
\gamma=- \sqrt{\frac{\pi}{8}} \frac{\omega_p}{\chi^3} \exp \left[ {-\frac{1}{2\chi^2}} \left(1+3 \chi^2 +\frac14H^2\chi^4 \right) \right]
\left[1+\frac{H^2}{24}- \frac{H^2 \chi^2}{2}\left(\chi^2+\frac14\right)   \right]. \label{landau-damping-reduced}
\end{equation}
Equations \eqref{dispersion-reduced} and \eqref{landau-damping-reduced} are similar to   Eqs. \eqref{eq-disp-weak-Q} and \eqref{eq-gam-weak-q1} derived in Case I of Sec. \ref{sec-nonrelat} for electron-ion plasmas and thus   the qualitative properties of the wave dispersion and the linear Landau damping rate will remain the same as for electron-ion   plasmas.  Although, the phase velocity is the resonant velocity in the linear regime, the  group velocity resonance occurs in the nonlinear propagation of Langmuir wave envelopes.   
\par
It is pertinent to examine the conservation laws   for the NLS equation \eqref{nls}. Although, the mass and momentum conservations hold,    the nonlocal nonlinear term $\propto R$   violates the energy conservation law  \cite{chatterjee2015,chatterjee2016,misra2017}. Since $R>0$ [Eq. \eqref{R-reduced}] for any values of $\chi$ and $H$ in the interval $(0~1]$ the time derivative of the energy integral $I_3=\int\left[|\partial_\xi\phi|^2-\left(Q/2P\right)|\phi|^4\right]d\xi$ is negative, i.e.,
 \begin{equation}
\frac{\partial I_3}{\partial\tau}=-\frac{R}{\pi}\int s^2|\hat{\phi} 
(s,\tau)|^2|\hat{\phi}(-s,\tau)|^2ds<0 ~~\mathrm{for}~ R>0.  \label{eq-energ-conser-q1}
\end{equation}
This implies that  an initial perturbation (e.g., in the form of a plane wave) will decay to zero with time, and hence a steady state solution of the NLS equation \eqref{nls} with $|I_3|<\infty$ may not be possible. While the sign of the nonlocal coefficient $R$ is important for determining  the conservation of energy, the sign of $PQ$ plays a key role for the frequency  up-shift or down-shift $(\Omega_r)$  and the rate of transfer of the wave energy $(\Gamma)$ to the particles.  It is found that the quantum parameter $H$ shifts the positive and negative regions of $PQ$ around the values of $\chi$ \cite{chatterjee2016}.
\par
\begin{figure*}[ht]
\centering
\includegraphics[height=2.5in,width=6.5in]{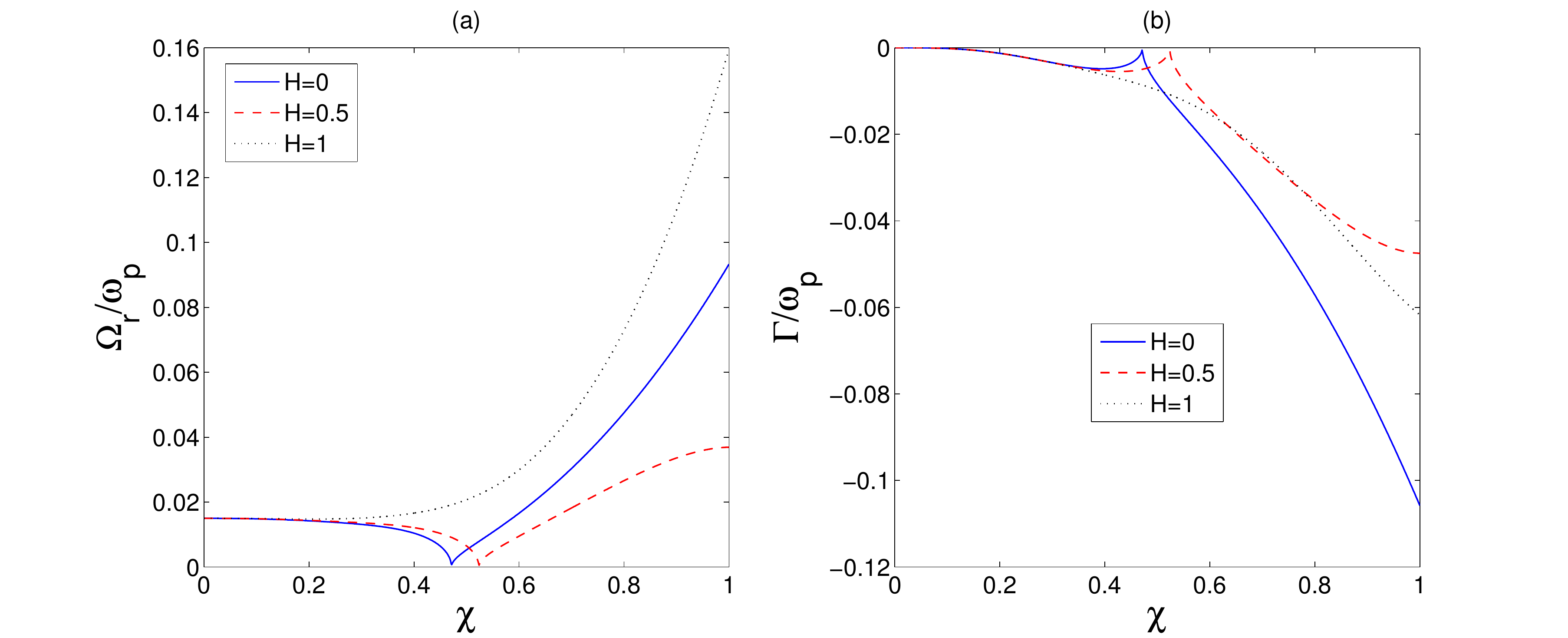}
\caption{The nondimensional frequency shift $\Omega_r/\omega_p$    and the energy transfer rate $\Gamma/\omega_p$   are plotted against the nondimensional carrier wave number $\chi\equiv k\lambda_D$ for different values of $H$ as in the legends and for a  fixed $\rho_0=K=0.1$. The figure is reproduced from Ref. \cite{chatterjee2016}.  }
\label{fig3}
\end{figure*}

 
A standard modulational instability analysis of a plane wave solution of  Eq. \eqref{nls}     of the   form  
\begin{equation}
\phi= \rho^{1/2}\exp\left(i \int ^\xi \frac{\sigma}{2P} d\xi\right), \label{sol-nls}
\end{equation}
where $\rho$ and $\sigma$ are real functions of $\xi$ and $\tau$, by means of a plane wave perturbation with   frequency $\Omega~(=\Omega_r+i\Gamma)$ and wave number $K$, reveals that the Langmuir wave packet is always unstable due to the presence of $R>0$ associated with the  group velocity resonance and is independent of the signs of  $P$ and $Q$. The key features of the instability analysis are as follows:
\begin{itemize}
\item In the  small amplitude limit with $\rho_0\ll |P/2Q|K^2$, where $\rho_0$ is the initial value of $\rho$, the frequency shift ($\Omega_r$) is related to the group velocity dispersion and   the imaginary part $\Gamma$ gives the nonlinear wave damping due to the group velocity resonance.  In the opposite limit, both $\Omega_r$ and $\Gamma$ can exist in the regions of $\chi$ and $H$ where $PQ<0$. However, their maximum values can be obtained in the region for  $PQ>0$.
\item Both $\Omega_r$ and $\Gamma$ can increase or decrease depending on the values of $\chi$ and $H$. However, they can vanish at a critical value of $\chi$ where the group velocity dispersion turns over from negative to positive values by the quantum effect. 
\end{itemize}
Some qualitative features of $\Omega_r$ and $\Gamma$ are presented in Fig. \ref{fig3} for different values of $H$ \cite{chatterjee2016}.
\par
From the energy conservation law, it is seen that the Langmuir wave energy decays with time due to the nonlocal term $\propto R$  of the NLS equation \eqref{nls} associated with the group velocity resonance. Following Ref. \cite{chatterjee2015},   approximate soliton solutions  of the NLS equation \eqref{nls}  when  the wave damping $(\propto R)$ is small can be presented in two different cases. For $PQ>0$, the solution can be written as
 \begin{equation}
\phi(\xi,\tau)=\sqrt{\phi_{0}(\xi,0)}\left(1-i\frac{\tau}{\tau_0}\right)^{-1/2}\text{sech}~{z}\exp(i\theta), \label{sol1-approx-nls}
\end{equation}
where  $z=(\xi-v_0\tau)/L$, $\theta= \left[v_0\xi+\left(\Omega_0-{v_0^2}/{2}\right)\tau\right]/2P$, with $v_0,~L,~\Omega_0,~\theta_1$ being constants, and $\tau_0$ is given by  
\begin{equation}
\tau_0^{-1}=\frac{\sqrt{2}R\phi_{0}(\xi,0)}{{\pi}^{3/2}\theta_1}\left[\frac{\cosh{(\pi\theta_1)}-1}{ \sinh{\left({\pi\theta_1}/{2}\right)}}\right]  {\cal{P}}  \int_{-\infty}^{\infty}\int_{-\infty}^{\infty}\left(\frac{\text{sech}^2 z'}{z-z'}\right) \text{sech}^2z \exp(i\theta_1z)dz dz'. 
\end{equation}
 \par 
On the other hand, for $PQ<0$, an approximate solitary wave solution of Eq. \eqref{nls}   is given  by \cite{chatterjee2015}
\begin{equation}
\phi={\phi_{0}(\xi,0)}\left(1-i\frac{\tau}{\tau_0}\right)^{-1/2}\text{tanh}{z}\exp(i\theta), \label{sol2-approx-nls}
\end{equation}
where $\theta$ and $\tau_0$ are given by 
\begin{equation}
\theta=\frac{1}{2P}\left[v_0\xi+ \left(2PQ\phi_0^2(\xi,0)-\frac{v_0^2}{2}\right)\tau\right], \label{theta-PQ<0}
\end{equation} 
\begin{equation}
\begin{split}
\tau_0^{-1}=\left(\frac{2}{\pi}\right)^{3/2}\frac{R\phi_{0}^2(\xi,0)\left[1-\cosh(\pi\theta_2)\right]}{\delta(\tau)(1-\cosh\pi\theta_2)+\theta_2 \sinh\left(\frac{\pi\theta_2}{2}\right)}{\cal{P}} &\int_{-\infty}^{\infty} \int_{-\infty}^{\infty} 
 \left(\frac{\tanh^2 z'}{z-z'}\right) \\
& \times \tanh^2z \exp(i\theta_2z)dz dz',
\end{split}
\end{equation}
with $\delta(\tau)$ denoting the Dirac delta function and $\theta_2$ a real constant.
\par 
The decay rate $| \left(1-i\tau/\tau_0\right)^{-1/2}|$ can be analyzed for both the cases of $PQ>0$ and $PQ<0$. It is found that the solitary wave amplitude decays with time   and   the rate is relatively low  (compared to the classical case)  due to the effects of the quantum particle dispersion. 
\subsection{Nonlinear Landau damping of electron-acoustic waves due to multi-plasmon resonances} 
As discussed in Sec. \ref{sec-sqr} that in the strong quantum regime, linear Landau resonance is   suppressed, however, nonlinear wave-particle interaction is possible due to   simultaneous absorption of multiple wave-quanta rather than a single wave quantum at a time. 
 The purpose of this section is to consider this phenomena for a low-frequency electrostatic wave and to show that deviating from the classical or semiclassical regime, not only the phase velocity resonance occurs, there also appear multi-plasmon resonances  in the nonlinear regime. As an illustration, we consider the nonlinear wave-particle interaction and  evolution of small-amplitude  electron-acoustic waves (EAWs) in a  partially degenerate plasma with two-temperature electrons and stationary ions.    To be brief,    in certain environments, e.g., in the interior of giant stars like white dwarfs, gas giants like Jupiter and in laser produced plasmas or ion-beam driven plasmas, the   background electrons deviating from the thermodynamic equilibrium    can  have a relatively high-temperature tail such that they can be grouped into two distinct components with different thermodynamic temperatures $T_l$ and $T_h$ for low  $(\alpha=l)$ and high-temperature $(\alpha=h)$ electrons such that $T_h>T_l$ and $T_\alpha \gtrsim T_{F\alpha}$, where $T_{F\alpha}$ denotes the    Fermi temperature of $\alpha$-species electrons. 
  Although, the theory is independent of the background distribution,    we consider  the Fermi-Dirac distribution at finite temperature $(T_\alpha\neq0)$. The plasmas with two groups of electrons cannot be fully degenerate from quantum mechanical points of view.  The other relevant details are given in  Ref. \cite{misra2021}.     
  From the linear theory of EAWs as in Sec. \ref{sec-nonrelat}, it can be  assumed that   a low-frequency mode with the dispersion $\omega \propto k$   exists. Also,  the Landau damping due to the linear plasmon resonance is weak and the wave damping occurs after a long time of propagation for which  the nonlinear evolution of EAWs can be described by the Korteweg-de Vries (KdV) equation. Furthermore, similar to classical plasmas, we assume that the  background high-temperature electrons is relatively densely populated compared to the low-temperature species, i.e.,  $n_{h0}> n_{l0}$. 
 Also, the EAWs are weakly dispersive such that  in the regime $\hbar k/m\lesssim v_{tl}$,   the Wigner-Moyal equation is still valid and some quantum effects due to the particle's dispersion become significant in  the  wave-particle interactions.    
\par 
The basic equations are the same as Eqs. \eqref{wigner-3d-alpha} and \eqref{pois-3d-alpha}, i.e., the Wigner-Moyal and Poisson system. Also, the background distributions of partially degenerate electrons are as given by Eq. \eqref{eq-distb-fn}.   The conditions for the equilibrium chemical potential $\mu_\alpha$ remain also the same as Eqs. \eqref{eq-mu1} and \eqref{eq-mu2} given in Case IV of Sec. \ref{sec-nonrelat}.
 Having known from  the linear theory  (Sec. \ref{sec-nonrelat})  that the  EAW has a cubic order dispersion and the Landau damping rate is small, we  derive an evolution equation for the weakly nonlinear EAWs in a degenerate plasma using the multiple-scale perturbation technique. Some special attention must be devoted to the higher order (in the amplitude) resonances that occur in the Wigner theory. 
In particular, 
due to the nonlinearities, we will have Landau resonances with resonant velocities that are shifted an amount $\pm n v_{q}$ in momentum space (See e.g., Ref. \cite{brodin2017}) compared to the resonant velocity of
classical theory (i.e., the phase velocity). Here, $n=1$ gives the velocity shift  already appeared in the linear theory  as described in Sec. \ref{sec-nonrelat}. Close to the resonant velocities, the Wigner
equation must be analyzed in more detail. In the quantum regime,    the classical   resonance velocity is changed 
according to 
\begin{equation}
v^{\text{res}}=\frac{\omega }{k}\rightarrow v^{\text{res}}_{\pm n}=\frac{\omega }{k} 
\pm n\frac{\hbar k}{2m}.  \label{eq-reson}
\end{equation}%
The details  of this modification is given in Sec. \ref{sec-sqr}.  
 \par 
Dividing the velocity space into the resonance and nonresonance regions, using the multiple scale expansion technique which involves the modified Gardner-Morikawa transformation  (i.e., $\xi=\epsilon^{1/2}x,~\sigma=\epsilon^{1/2}t,~\tau=\epsilon^{3/2} t$, where $\epsilon>0$ is a small scaling parameter) the  perturbation expansions for the Wigner function  $f_\alpha$ and the potential $\phi$, and the multi-scale Fourier-Laplace transforms, and following Ref. \cite{misra2021} we obtain  
 the following modified KdV equation with nonlinear Landau damping $\propto\Gamma$ (For details, see Ref. \cite{misra2017}).
\begin{equation}
\frac{\partial \phi}{\partial \tau}+A\frac{\partial ^3\phi}{\partial \zeta^3}+B \phi\frac{\partial \phi}{\partial \zeta} +  \Gamma{\cal P}\int_{-\infty}^{\infty} \frac{\partial \phi^2\left(\zeta',\tau\right)}{\partial \zeta'} \frac{1}{\zeta-\zeta'}    d\zeta'=0,\label{K-dV1}
\end{equation}
where $\zeta=\xi-v_p\sigma$ with $v_p$ denoting the phase velocity of EAWs and  the coefficients of the KdV equation are $A=1/P$, $B=Q/P$ and $\Gamma=R/P$, given by,
  \begin{equation}
  \Re P=-\frac{6\pi e^2}{m}\sum_{\alpha=l,h}\frac{n_{\alpha0}T_\alpha e^{\xi_\alpha}}{v_{F\alpha}T_{F\alpha}}\left[\frac{5(v_p^2-v_q^2)+2v_p v_{t\alpha}}{(v_p^2-v_q^2)^2}\right],
    \end{equation}
  \begin{equation}
   \Im P=-6(v_p-v_q)\frac{\pi^2 e^2}{\hbar k}\sum_{\alpha=l,h}\frac{n_{\alpha0}T_\alpha e^{\xi_\alpha}}{v_{F\alpha}T_{F\alpha}v_{t\alpha}^2}\exp\left\lbrace-\left(\frac{v_p-v_q}{v_{t\alpha}}\right)^2\right\rbrace,
   \end{equation}
  \begin{equation}
  Q=-\frac{24\pi me^3}{(\hbar k)^3}\frac{v_p^2+4v_q^2}{\lambda(v_p^2-4v_q^2)}\sum_{\alpha=l,h}\frac{n_{\alpha0}T_\alpha v_{t\alpha}}{v_{F\alpha}T_{F\alpha}}e^{\xi_\alpha},
  \end{equation}
  \begin{equation}
    R=\frac{6\pi^2m e^3}{(\hbar k)^3}\sum_{\alpha=l,h}\frac{n_{\alpha0}T_\alpha }{v_{F\alpha}T_{F\alpha}}e^{\xi_\alpha}\left[\exp\left\lbrace-\left(\frac{v_p-2v_q}{v_{t\alpha}}\right)^2\right\rbrace-2\exp\left\lbrace-\left(\frac{v_p}{v_{t\alpha}}\right)^2\right\rbrace\right]. \label{eq-R-reduced}
   \end{equation}
The expression for the phase velocity $v_p$ can be obtained from the linear dispersion relation [Case IV, Eq. \eqref{eq-EAW}] by considering the limit $k\rightarrow0$. We also note that $P$ becomes complex due to the one plasmon resonance (linear), and so are the dispersive $(\propto A)$, local nonlinear $(\propto B)$, and  the nonlocal nonlinear  $(\propto\Gamma)$ terms. The latter, however, appears due to the phase velocity and multi-plasmon resonances. Such resonances are noted in Ref. \cite{misra2021} with poles  at $v=\omega/k\pm n v_q$, where $n=0,1,2$,  in the integrals appearing in certain expressions for  the second order perturbations, namely,
\begin{equation}
\int_{-\infty}^{\infty} \left[\frac{1}{(v-\omega/k-v_q)^2}-\frac{1}{(v-\omega/k+v_q)^2} \right]f_\alpha^{(0)}  dv, \label{eq-J1}
\end{equation}
 \begin{equation}
 \int_{-\infty}^{\infty} \left(\frac{1}{v-\omega/k-2v_q} +\frac{1}{v-\omega/k+2v_q}-\frac{2}{v-\omega/k}\right) f_\alpha^{(0)} (v)dv. \label{eq-reso}
 \end{equation} 
 It is interesting to note that  although  the form of the KdV equation \eqref{K-dV1} looks similar to that first obtained by Ott and Sudan \cite{ott1969} and later by many authors (See, e.g., Refs. \cite{vandam1973,barman2017}) in classical/semiclassical plasmas,   the   Landau damping term $\propto\Gamma$ appears here as nonlinear    due to the  phase velocity resonance as well as the two-plasmon  resonance processes in the wave-particle interactions. The appearance of such a nonlocal nonlinearity not only modifies the propagation of EAWs but also introduces a new wave damping mechanism.            
From the reduced expression of $R$ [Eq. \eqref{eq-R-reduced}]  it is evident that the contribution of the two-plasmon resonance is higher than that of the phase velocity resonance, implying that the two-plasmon resonance process is the dominant wave damping mechanism for EAWs.    
 \par 
 Meanwhile, the KdV equation   \eqref{K-dV1} conserves the total number of particles,  
  however, the wave energy decays with time   \cite{misra2021},  i.e., 
\begin{equation}
\frac{\partial }{\partial \tau}\int_{-\infty}^{\infty}|\phi(\zeta,\tau)|^2d\zeta\leq0. \label{eq-cons-enr}
\end{equation}    
So, a steady-state solution of Eq. \eqref{K-dV1} with finite wave energy does not exist implying that  the wave amplitude will tend to decay due to the nonlinear resonance. In this context,   an approximate solitary wave solution of Eq. \eqref{K-dV1} can be obtained similar to Ref. \cite{misra2021}  
 \begin{equation}
\phi_0(\tau)=\phi_{00}\left( 1+\frac{\tau}{\tau_0}\right)^{-2/3}, \label{phi-soln}  
\end{equation}
where 
\begin{equation}
\tau_0^{-1}=\frac{3}{4}\Gamma\sqrt{\frac{B}{3A}}\phi_{00}^{3/2}{\cal P}\int_{-\infty}^\infty \int_{-\infty}^\infty \text{sech}^2z\frac{\partial}{\partial z^{\prime}}\left( \text{sech}^4z^{\prime}\right)\frac{dz dz'}{z-z'}\approx 2\Gamma\sqrt{\frac{B}{3A}}\phi_{00}^{3/2}, \label{tau-eq}
\end{equation}
where the   Cauchy principal value  is evaluated as $\approx 2.8$. 
From Eq. \eqref{phi-soln} it is clear that  the nonlinear  Landau damping indeed causes the wave amplitude to decay with time $\sim(\tau+\tau_0)^{-2/3}$ which is   a bit slower than that $\sim(\tau+\tau_0)^{-2}$   predicted by Ott and Sudan \cite{ott1969} in classical plasmas.
 \par 
 Some important points are to be noted. Since the KdV equation accounts for upto the second order perturbations, the lower resonance velocity is due to the two-plasmon processes and it gives the dominant wave-damping mechanism in the description of EAWs in the strong quantum regime.  In the model, the  plasmas  are  not in thermodynamic equilibrium.  However, the theory of EAWs can be studied with the background distribution of electrons in thermodynamic equilibrium (i.e., for a single species with finite thermal velocity). For such a background distribution [Eq. \eqref{eq-distb-fn}] the EAWs tend to have a higher Landau damping rate, making the nonlinear analysis less important.   
  \subsection{Nonlinear Landau damping of Langmuir wave envelopes due to multiplasmon resonances}\label{sec-nld-LW-multi}
We  turn our attention again to the nonlinear evolution of Langmuir wave envelopes, however, in the strong quantum regime. Specifically, we will consider the  wave-particle interaction and the amplitude modulation of Langmuir wave envelopes  in a fully degenerate  plasma, and focus on the regime $k<k_\text{cr}$  where the linear damping is forbidden. Here,   $k_\text{cr}$ is some critical value of the wave number $k$ such that  $\hbar k \sim mv_F$ holds. The basic equations are the Wigner-Moyal equation coupled to the Poisson equation [\textit{cf.} Eqs. \eqref{wigner-1d} and \eqref{poisson-1d}], i.e.,
\begin{equation}
\frac{\partial f}{\partial t}+v\frac{\partial f}{\partial x}-\frac{e m }{2i\pi\hbar^2}\int\int dx_0 dv_0 e^{im_{\alpha}(v-v_0)x_0/\hbar}   
 \left[\phi\left(x+\frac{x_0}{2}\right)-\phi\left(x-\frac{x_0}{2}\right)\right]f(x,v_0, t)=0, \label{eq-wig-mpr}
\end{equation}
 \begin{equation}
 {\frac{\partial^2 \phi}{\partial x^2}}=  \frac{e}{\varepsilon_0} \left(\int f dv-n_0\right), \label{eq-poiss-mpr}
\end{equation}  
where $n_{0}$ is the background number density of electrons and ions. In a fully degenerate plasma,   
 the background distribution of electrons is given by the Fermi-Dirac pressure [\textit{cf.} Eq. \eqref{eq-dist-FD-1d}]
\begin{equation}
 f_0(v)= \left\lbrace\begin{array}{cc} 
\left[{2\pi m^3}/{(2\pi\hbar)^3}\right](v_F^2-v^2),&|{ v}|\leq v_F \\
 0,&|{ v}|> v_F. 
 \end{array}\right.  \label{eq-dist-mpr}
\end{equation}
  Introducing the multiple space-time scales with the stretched coordinates $x\rightarrow x+\epsilon ^{-1} \eta +\epsilon^{-2}\zeta,~ ~t\rightarrow t+\epsilon ^{-1}\sigma$,   the perturbation expansion for  $f(\mathbf{v},x,t)$ and $\phi(x,t)$ with a scaling parameter $\epsilon$, and further expanding the harmonic components of $f$ and $\phi$ in terms of Fourier-Laplace integrals \cite{chatterjee2015,chatterjee2016,ichikawa1974} we obtain the following nonlinear Schr{\"o}dinger (NLS) equation for the evolution of Langmuir wave envelopes in a fully degenerate plasma \cite{misra2017}.
\begin{equation}
i\frac{\partial\phi}{\partial \tau}+P \frac{\partial^2\phi}{\partial \xi^2}%
+Q |\phi|^2 \phi +\frac{R }{\pi}\mathcal{P}\int\frac{ |\phi(\xi^{\prime},\tau)|^2}{%
\xi-\xi^{\prime }} \phi d\xi^{\prime }=0,  \label{eq-nls-mpr}
\end{equation}
Here, $\xi=\eta-v_g \sigma$ with $v_g$ denoting the group velocity of the envelope, given by, 
 $v_g=\lambda_1/\lambda_2$, where
\begin{equation}
\lambda_1=2-\frac{4\pi e^2}{mk^2}\int_C \frac{v_p^2-v^2+v_q^2}{{\left\lbrace (v_p-v)^2-v_q^2
\right\rbrace}^2 } f_0(v) dv,~  
\lambda_2=-\frac{8\pi e^2}{mk^2}\int_C \frac{v_p-v}{{\left\lbrace
(v_p-v)^2-v_q^2\right\rbrace}^2 }
f_0(v) dv.  \label{eq-gr-mpr}
\end{equation}
The coefficients of the dispersion (group velocity), cubic nonlinear
(local), nonlocal nonlinear terms, respectively, are $P,~Q$ and $R$, given
by $P\equiv(1/2)\partial^2\omega/\partial k^2=\beta/ \alpha,~Q= \gamma/
\alpha$ and $R=D/ \alpha$, where 
\begin{equation}
\alpha=-\frac{8\pi e^2}{mk}\int_C \frac{v_p-v}{\left[ (v_p-v)^2-v_q^2\right] ^2} f_0(v)dv,
\label{alpha}
\end{equation}
\begin{equation}
\beta=1+ \frac{4\pi e^2}{\hbar k^3}\int_C \left[\frac{\left(v-v_q
-v_g \right)^2}{\left(v_p-v+v_q\right)^3}
-\frac{\left(v+v_q-v_g \right)^2}{\left(v_p-v- 
v_q\right)^3}\right]f_0(v) dv  \label{beta}
\end{equation}
\begin{equation}
\gamma=\left(\frac{1}{4}\frac{A A_1}{\hbar}-\frac{1}{2\hbar^2}B+C\right)k^2,
\label{gamma}
\end{equation}
\begin{eqnarray}
D=&&-\frac{4\pi e^4}{m \hbar^2k^2} \int_\gamma\left[ \delta\left\lbrace
v-\left(v_p-3v_q \right) \right\rbrace \frac{ 
v-v_g+ v_q}{ \left(v_p-v-v_q
\right)^3 \left( v-v_g + 2v_q\right)}\right.\notag\\
&&\left.+2\delta(v-v_g)v_q \frac{ \left\lbrace \left(
v_p-v\right)^2+v_q^2 \right\rbrace }{\left\lbrace
\left( v_p-v\right)^2-v_q^2 \right\rbrace ^3 } \right]%
f_0(v) dv.  \label{D-expression}
\end{eqnarray}
The expressions for $A,~A_1,~B,$ and $C$ in $\gamma$  are given in 
Appendix \ref{appendix-a}. Also, the reduced expressions for $\alpha,~\beta$, $\gamma$ and $D$  are given in Appendix  \ref{appendix-b}.
\par 
We note that the integrals in $\alpha$ and $ \beta$ do not have any pole except at the linear resonant velocities $v=v^l_{res}\equiv v_p\pm v_q$ which lie  outside the regime of interest $k<k_{cr}$, and accordingly,  the group velocity dispersion $P$ does not have any resonance contribution in the regime. Furthermore, inspecting the denominators of different expressions for $A,~A_1,~B$ and $C$  in $\gamma$,   we find that only   the two- and three-plasmon resonances can occur   at $v_\mathrm{res}^n=v_p-nv_q$ for $n=2,3$.  Thus, in contrast to  classical \cite {chatterjee2015} or semiclassical \cite{chatterjee2016}   plasmas, the nonlinear coefficient $Q$ of the NLS equation is significantly modified by the Landau resonances due to the two- and three-plasmon processes. Also, both the group velocity resonance (the first term of $D$) and the three-plasmon resonance (the second term of $D$) contribute to and modify the nonlocal coefficient $R\propto D$.
\par
\begin{figure*}[tbp]
\includegraphics[scale=0.5]{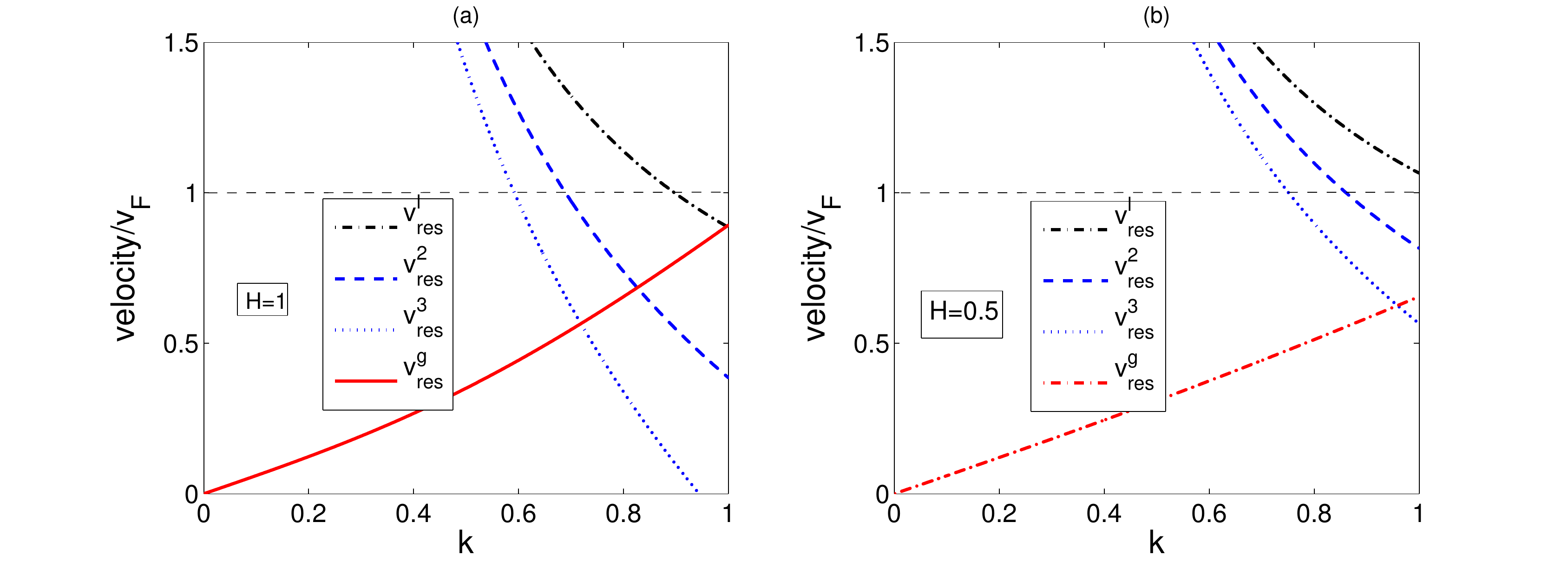}
\caption{The normalized resonant velocities ($\sim v_F$) are plotted against the normalized wave number $k~(\sim \lambda_F^{-1})$ for two different values of the  quantum parameter $H\equiv \hbar\omega_p/k v_F$ to show different parameter regimes, namely semi-classical (e.g., $0<k\ll0.59$ for $H=1$), modest quantum ($0<k\lesssim0.59$ for $H=1$ and $0<k\lesssim0.75$ for $H=0.5$) and strong quantum ($0.591\lesssim k\lesssim0.9$ for $H=1$ and $0.75\lesssim k\lesssim1$ for $H=0.5$) regimes. The figure is reproduced from Ref. \cite{misra2017}. }
\label{fig1-mpr}
\end{figure*}

 
 As noted before in Sec. \ref{sec-nls-sem-cl}, the mass and momentum are conserved for the NLS equation \eqref{eq-nls-mpr}. However,    the wave energy $ 
I_3=\int\left(|\partial_{\xi}\phi|^2-(|Q/2P|)|\phi|^4\right)d\xi$ may not be conserved in presence of the nonlocal nonlinearity  \cite{chatterjee2015,chatterjee2016}. In fact, the time variation of the energy integral
 \begin{equation}
\frac{\partial I_3}{\partial\tau}=-\frac{R}{\pi}\int s^2|\hat{\phi} 
(s,\tau)|^2|\hat{\phi}(-s,\tau)|^2ds.  \label{eq-energ-conser-mpr}
\end{equation}
becomes positive or negative according to when  $R>0$ or $R<0$. A careful exmanination reveals that $R>0$ in the regime     $k<k_\text{cr}$ \cite{misra2017}. In order to explore the regime   in more details,   we   require the dispersion equation   to be reduced in the limit of $k\lambda _{F}\lesssim 1$, i.e.,  one obtains the same    relation as Eq. \eqref{eq-disp-Lang-FD} \cite{eliasson2010,misra2017}.    The expressions for the phase velocity and resonant velocities are obtained from this reduced equation and plotted against the wave number $k$. The results are displayed in Fig. \ref{fig1-mpr}.      
 It is seen that  there are mainly two parameter regimes:  one where the resonant velocities are both the group velocity and  the plasmon resonant velocities  $v_\text{res}^n$, $n=2,3$, and the other where only the group velocity resonance is the damping mechanism.  The detailed analysis is given in Ref. \cite{misra2017}. 
 If $H\sim1$ there is a region of $k$, i.e., $0<k\lesssim0.9$ in which  the linear resonance is forbidden. However, a subregion of it exists, i.e., $0<k\lesssim0.59$ in which    only the group velocity resonance occurs. The two other  subregions exist, namely  $0.59\lesssim k\lesssim0.6953$ and $0.59\lesssim k\lesssim0.9$. In the former, both the three-plasmon and the group velocity resonances can occur, while in the latter,    the group velocity, as well as the two- and three-plasmon resonances can be significant.
In this regime,   the magnitudes of the coefficients $P,~Q$ and $R$ of the NLS equation \eqref{eq-nls-mpr} should be noted. These are, however, useful  for the estimation of frequency shift and the rate of energy transfer in the modulation of Langmuir waves, as well as the  nonlinear evolution of envelope solitons. 
 On the other hand, multi-plasmon resonances can be forbidden, only the group velocity resonance prevails \cite{chatterjee2015,chatterjee2016} if $H$ is reduced  from $H=1$  to $H=0.5$.   
\par 
 Thus,  three   regimes may be of interest: (i)  Semi-classical regime   with   $\hbar k\ll mv_F$  (i.e., $0<k\ll0.59$ for $H\sim1$)  where the quantum effect appears only due to the degeneracy of   background   electrons. The results will be similar to those in Sec. \ref{sec-nls-sem-cl} or in Ref. \cite{chatterjee2016} as  the group velocity is only the resonant velocity;  (ii) Modest quantum regime with  $\hbar k\sim mv_F$ and $v_\text{res}^3>v_F$  (i.e., $0<k\lesssim 0.59$ for $H\sim1$ and  $0<k\lesssim0.75$ for $H\sim0.5$) in which the resonant velocity is still the group velocity.   
The results will be similar to those in semiclassical  plasmas \cite{chatterjee2016}. (iii) Strong quantum regime with $\hbar k\sim mv_F$ and $v_\text{res}^3<v_F$ (i.e., $0.591\lesssim k\lesssim0.9$ for $H=1$ and $0.75\lesssim k\lesssim1$ for $H=0.5$) in which all the resonant velocities come into the picture. However,   the contribution from the three-plasmon resonance becomes higher in  magnitude than the group velocity resonance  until $v_\text{res}^3~(<v_F)\gtrsim v^g_\text{res}$ holds. Thus, in the strong quantum regimes, the three-plasmon resonance plays a decisive role  in the nonlinear Landau damping of Langmuir wave envelopes.
\par
\begin{figure*}[ht]
\centering
\includegraphics[height=2.5in,width=6.5in]{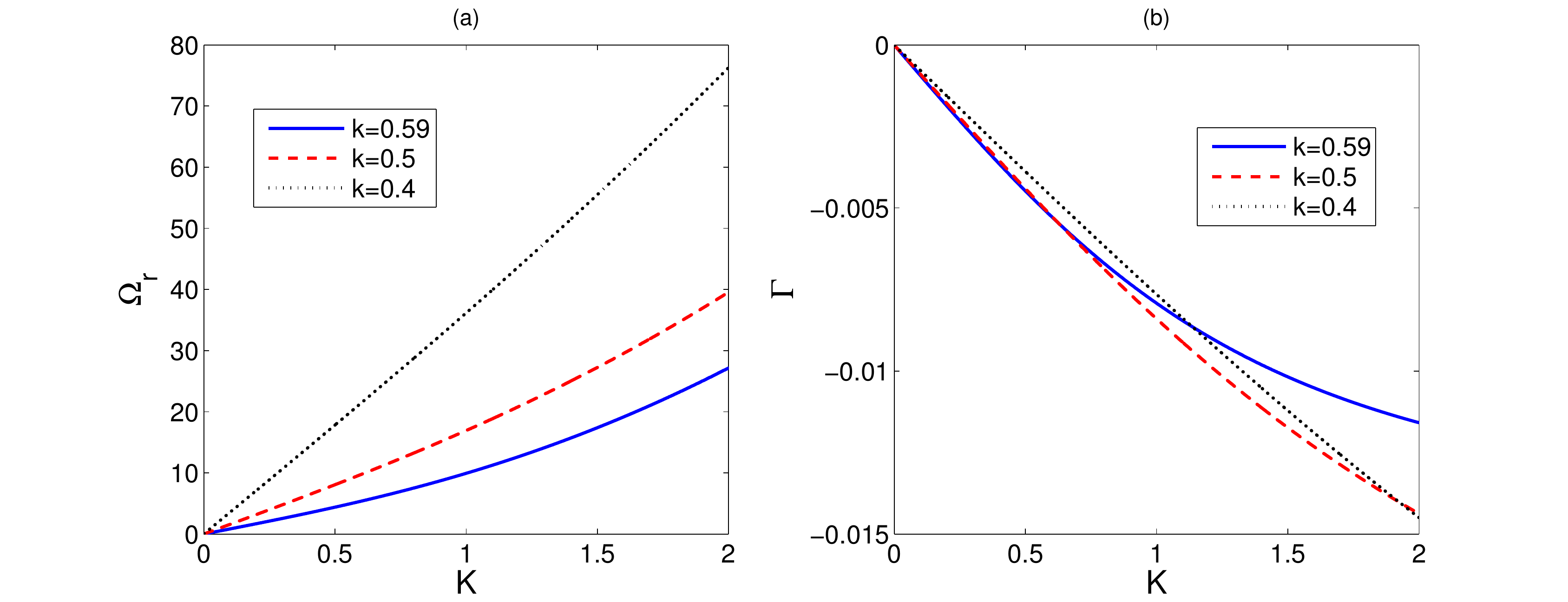}
\caption{The normalized frequency shift $\Omega_r~(\sim\omega_p)$   and the energy transfer rate $\Gamma~(\sim
\omega_p)$   are plotted against the
normalized wave number of modulation $K~(\sim \lambda_F^{-1})$ for different values of the
carrier wave number $k~(\sim \lambda_F^{-1})$ that correspond to semi-classical and modest quantum regimes (where $\Im Q=0$).  The figure is reproduced from Ref. \cite{misra2017}.}
\label{fig2-mpr}
\end{figure*}

\begin{figure*}[ht]
\centering
\includegraphics[height=2.5in,width=6.5in]{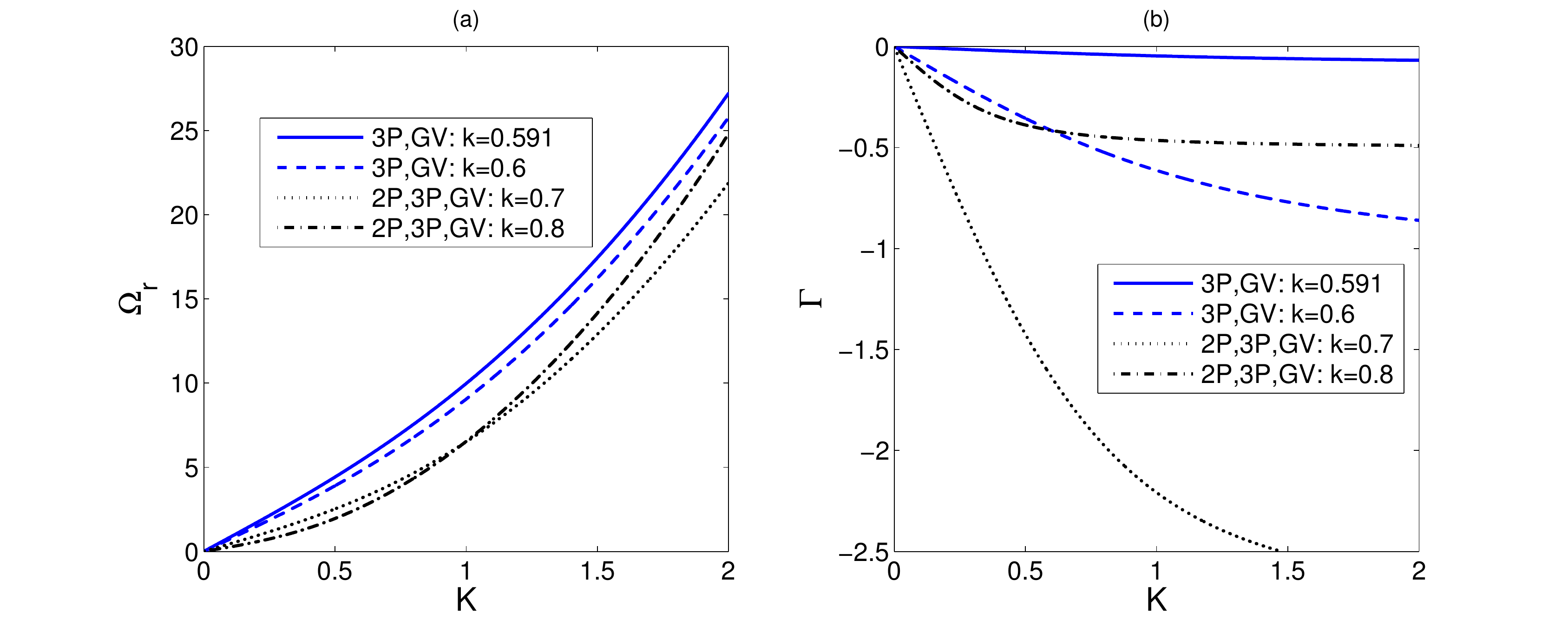}
\caption{ The same as in Fig. \protect\ref{fig3} but in the strong quantum
regime. In the legends, $2P,~3P$ and $GV$, respectively, stand for
two-plasmon, three-plasmon and group velocity resonance effects. The figure is reproduced from Ref. \cite{misra2017}.}
\label{fig3-mpr}
\end{figure*}

 
Similar to Sec. \ref{sec-nls-sem-cl} and   Refs. \cite{chatterjee2016,misra2017}, the  frequency shift $\Omega_r$ and the  energy transfer $\Gamma$ rate can be analyzed in the modulation of Langmuir wave  envelopes.  Although, the forms of the expressions of $\Omega_r$ and $\Gamma$ are the same as in classical \cite{chatterjee2015} or semiclassical \cite{chatterjee2016} plasmas, a significant modification in   both the frequency shift and the  energy transfer  rate is noticed due to  the multi-plasmon resonances. The profiles of $\Omega_r$ and $\Gamma$ are shown Figs. \ref{fig2-mpr} and \ref{fig3-mpr}  especially in the modest quantum and strong quantum regimes (since the semiclassical results are similar to those in Sec. \ref{sec-nls-sem-cl})  for different values of the carrier wave number $k$.      It is found that as  the value  of $k$ decreases  from $k=0.59$ to $k=0.5$ and $k=0.4$, the magnitudes of $P$, $Q$ and $R$ are significantly altered leading to an enhancement of     the frequency shift, however, the values of $|\Gamma|$ increase until   $k=0.5 $, and then  decrease until $k=0.4$.      Thus, in the regime of low wave numbers (below $k=0.5$), although the frequency shift remains high,   the magnitude of $\Gamma$ is greatly reduced, implying that the  energy transfer rate   is relatively low in the semi-classical regimes with $\hbar k/mv_F\ll1$. The energy transfer rate can be maximum near   $k=0.5$ where   $v_{res}^3\gtrsim v_F$ holds.    
 On the other hand,    in the strong quantum regime $0.591\lesssim k\lesssim0.9$,    it is seen from  Fig. \ref{fig3-mpr} that   the frequency shift remains high, however, $|\Gamma|$ attains its minimum value. The effect of the three-plasmon resonance is to decrease  the values of $\Omega_r$ but to increase   $|\Gamma|$.  
 \par 
 It has been established that the nonlocal nonlinearity $(\propto R)$ which appears due to the three-plasmon and group velocity resonances violates the conservation of energy  and the wave damping occurs for $R>0$. It is thus  imperative to study the effects of the Landau resonances on the profile of an envelope soliton solution of Eq. \eqref{eq-nls-mpr}.   Following Refs. \cite{chatterjee2015,chatterjee2016} an approximate soliton
solution of  Eq. \eqref{eq-nls-mpr} with a small effect of the 
nonlinear Landau damping $(\propto R)$ can be obtained whose  amplitude varies as
\begin{equation}
\phi(\xi,\tau)\propto\sqrt{\phi_{0}(\xi,0)}\left(1-i\frac{\tau}{\tau_0}%
\right)^{-1/2},  \label{eq-soliton-mpr}
\end{equation}
where $\tau_0$ is some constant inversely proportional to $R$ and $%
\phi_{0}(\xi,0)$ is the value of $\phi$ at $\tau=0$.
\par
\begin{figure*}[ht]
\centering
\includegraphics[height=2.5in,width=3.8in]{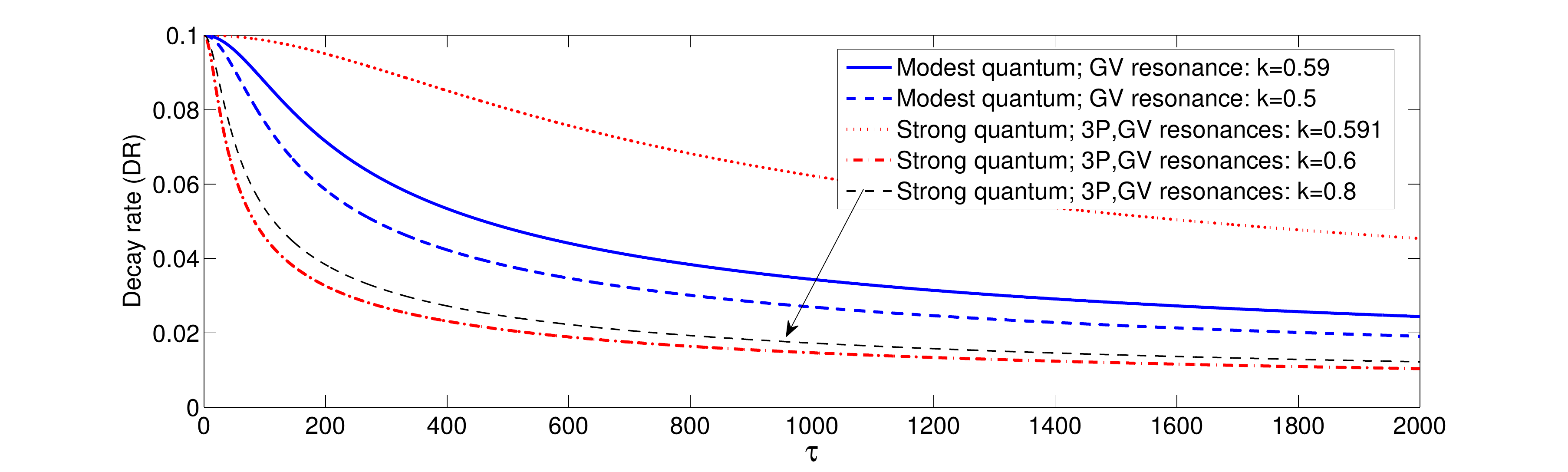}
\caption{The absolute value of the decay rate $DR\equiv|\left(1-i\protect\tau%
/\protect\tau_0\right)^{-1/2}|$ is shown against the normalized time
variable $\tau~(\omega_p^{-1})$ in different parameter regimes as in
the legend. The figure is reproduced from Ref. \cite{misra2017}.}
\label{fig4-mpr}
\end{figure*}

  A qualitative plot of the decay 
rate $DR\equiv|\left(1-i\tau/\tau_0\right)^{-1/2}|$ is shown in Fig. \ref{fig4-mpr} in the modest and strong quantum regimes to show the relative
importance of the group velocity (solid and dashed lines) and three-plasmon
(dotted and dash-dotted lines and as indicated in the figure) resonances. It is found that the decay rate due to the effects of the three-plasmon resonance is faster than that due to the group velocity resonance. 
  
\subsection{Wave-particle interaction including spin dynamics}
While many aspects of wave-particle interaction can be described by the Wigner equation  where spin effects are omitted, 
certain processes must be described by more general equations. 
Kinetic plasma models including spin effects appear in various degrees of
complexity. However, many of the key ingredients in the interaction can be
seen already in relatively simple kinetic models including spin. Thus, as a
starting point, we will make use of the following approximations:

\begin{enumerate}
\item We leave out short-scale (particle dispersive) effects by considering
scale lengths much larger than the characteristic de Broglie wavelength. In the
absence of spin, such an approximation reduces the Wigner-Moyal equation to
the Vlasov equation.

\item We omit weakly relativistic spin effects, i.e., we drop the spin-orbit
interaction. For consistency, this also leaves out the effect of Thompson
precession  and spin polarization currents (only magnetization currents are
retained).

\item We make use of the Hartree approximation.   While many spin effects
are included in the mean-field approximation, the exchange effects  (which are
also dependent on spin) are not included in the model. This is a consistent
approximation  since the scaling of the mean-field terms and the exchange
contribution with the plasma parameters are different.
\end{enumerate}

Two different but equivalent models, derived from first principles, meeting
the above criteria have been derived \cite{zamanian2010,hurst2014}. We will make
use of the governing equations first presented in Ref. \cite{zamanian2010},
which slightly corrects an older somewhat simpler model derived using
semi-classical arguments \cite{brodin2008}. The evolution equation for the
distribution function $f(\mathbf{r,v,}\hat{\mathbf{s}},t)$ reads 
\begin{equation}
\frac{\partial f}{\partial t}+\mathbf{v}\cdot \nabla _{\mathbf{x}}f+\left[ 
-\frac{e}{m}(\mathbf{E}+\mathbf{v}\times \mathbf{B})+\frac{\tilde{\mu}}{%
m}\nabla _{\mathbf{x}}\left( \hat{\mathbf{s}}\cdot \mathbf{B}+\mathbf{B}%
\cdot \nabla _{\hat{\mathbf{s}}}\right) \right] \cdot \nabla _{\mathbf{v}}f+%
\frac{2\tilde{\mu}}{\hbar }(\hat{\mathbf{s}}\times \mathbf{B})\cdot \nabla _{%
\hat{\mathbf{s}}}f=0,  \label{Vlasov}
\end{equation}%
with the current density $\mathbf{J}$ in Maxwell's equations calculated as  
\begin{eqnarray}
\mathbf{J} &=&\mathbf{J}_{f}+\mathbf{J}_{M},  \notag \\
&=&\mathbf{J}_{f}+\nabla _{\mathbf{x}}\times \mathbf{M},  \notag \\
&=&-e\int \mathbf{v}fd^{2}sd^{3}v+\nabla _{\mathbf{x}}\times \left( \tilde{\mu}\int 3\hat{\mathbf{s}}fd^{2}sd^{3}v\right).  \label{Current}
\end{eqnarray}%
where $\mathbf{J}_{f}$ is the free current density, $\mathbf{M}$ is the
magnetization due to the spin, $\tilde{\mu}$ is the electron magnetic moment,
and $\mathbf{J}_{M}$ is the magnetization current. The key features of the
model are:

\begin{enumerate}
\item The phase space is extended with the spin variable $\hat{\mathbf{s}}%
=(\sin \theta _{s}\cos \varphi _{s},\sin \theta _{s}\sin \varphi _{s},\cos
\theta _{s})$, a unit vector on the Bloch sphere that describes the
distribution of spin among the particles at given phase space positions. As
shown above, when calculating sources in Maxwell's equations, an extra
integration  over the
Bloch sphere $d^{2}s=\sin \theta _{s}d\theta _{s}d\varphi _{s}$ is required. The integration is usually made in spherical spin
coordinates  as indicated here and will be used onwards.

\item While the electron magnetic moment is given by the Bohr magneton to a
good approximation  when the wave particle interaction is considered, it is
important to use the QED corrected value $\tilde{\mu}=ge\hbar /4m$, where the
electron g-factor is $g=-2.002319$.

\item Since the spin vector has a fixed length, $\nabla _{\hat{\mathbf{s}}}$
is only a gradient on the surface of the Bloch sphere. Hence it is given by $%
\nabla _{\hat{\mathbf{s}}}=\mathbf{\hat{e}}_{\theta }\partial /\partial
\theta _{s}+(1/\sin \theta _{s})\mathbf{\hat{e}}_{\varphi }\partial
/\partial \varphi _{s}.$
\end{enumerate}
\par 
With the computational aspects of Eq. (\ref{Vlasov}) pointed out  we note
that the second term inside the square bracket (to the right of the Lorentz
force), is the magnetic dipole force, whereas the last term describes the
spin precession. Solving Eq. (\ref{Vlasov}) is similar to solving the
standard Vlasov equation. For the case of homogeneous linearized theory, a
complete description was given in Ref. \cite{lundin2010}. We will not
repeat the technical details here. Rather, we will just point out the main
mechanism affecting wave-particle interaction problems.
\par 
Due to the Larmor gyration, as is well-known, for electromagnetic fields the
perturbed distribution function $f_{1}$ will have an oscillating dependence
on the azimuthal angle in velocity space, i.e., \thinspace $f_{1}\propto \exp
(i\varphi _{v})$. For a constant magnetic field $\mathbf{B}_{0}=B_{0}\mathbf{\hat{z}},$ we have $(e/m)(\mathbf{v}\times \mathbf{B}_{0})\cdot \nabla _{\mathbf{v}}=\omega _{c}\partial /\partial \varphi _{v}$, such that $(e/m)( 
\mathbf{v}\times \mathbf{B}_{0})\cdot \nabla _{\mathbf{v}}f_{1}=i\omega
_{c}f_{1}$, where $\omega _{c}=qB_{0}/m$ is the Larmor gyration frequency.
Similarly, due to the precession of the spin state, the perturbed
distribution function will have the same type of dependence on the spin
azimuthal angle \thinspace $f_{1}\propto \exp (i\varphi _{s}\dot{)}$. As a
result, $(2\tilde{\mu}/\hbar )(\hat{\mathbf{s}}\times \mathbf{B}_{0})\cdot
\nabla _{\hat{\mathbf{s}}}f_{1}=i\omega _{cg}f_{1}$, where $\omega
_{cg}=(g/2)\omega _{c}$. Since the spin $g$-factor is slightly larger than $2$%
, the Larmor gyration and the spin precession will not be exactly in sync,
which has important consequences.
\par 
For the free current contributions, the wave particle interaction terms are
similar to the classical case, i.e., there are only quantum corrections
affecting the magnitudes of the classical terms. Unless the density is very
high and/or the temperature is low, the quantum corrections are relatively
modest. However, the expression for the (spin) magnetization current is much
different, since it allows for new forms of wave-particle interaction. The
reason is that the contribution from $\mathbf{J}_{M}=\nabla _{\mathbf{x}%
}\times \mathbf{M}$ to the linear conductivity tensor gives rise to new
types of resonant denominators in the integrand, related to the slight
mismatch between the Larmor gyration frequency and the spin precession
frequency outlined above. To be specific, the classical denominators giving
raise to wave-particle resonances are modified according to (see Ref. \cite{lundin2010} for details)
\begin{equation}
\frac{1}{\omega -k_{z}v_{z}-n\omega _{c}}\rightarrow \frac{1}{\omega
-k_{z}v_{z}-n\omega _{c}\pm \omega _{cg}}.  \label{Denominator}
\end{equation}%
   As usual in linearized theory for a
magnetized plasma, $n$ is an integer, $\ $and $k_{z}$ is the wavenumber
component along the external magnetic field. Firstly, we note that for $
k_{z}=0$, contributions from $n=\pm 1$ can be much magnified for frequencies 
$\omega \simeq \Delta \omega _{c}$, where $\Delta \omega _{c}=\omega
_{cg}-\omega _{c}=(g/2-1)\omega _{c}$. As a result, the dispersion relation
allows for new wave modes dependent on the spin dynamics. The magnification
due to the frequency resonance compensates for a small prefactor for the
magnetization current (proportional to $\tilde{\mu}B_{0}/k_{B}T$, which is
usually much less than unity), such that $\omega \simeq \Delta \omega _{c}$
applies for most part of the $k-$spectrum. In this sense, these types of spin
modes are similar to the Bernstein waves \cite{swanson2003}, which tend to have
frequencies close to the resonances where $\omega =n\omega _{c}$. The
properties of the spin resonance modes have been studied in more detail in
Refs. \cite{brodin2008,zamanian2010,asenjo2012}.
\par
It is to be noted that even for plasma parameters with relatively small quantum parameters
(with $\tilde{\mu}B_{0}/k_{B}T\ll 1$, $\hbar \omega _{p}/k_{B}T\ll 1$, etc.),
and for $\omega $ not close to $\Delta \omega _{c}$, when the dispersion
relations are classical to a good approximation, the quantum denominators
given in Eq. (\ref{Denominator}) can have important consequences for
wave-particle interactions. The reason is that a strong magnification in the
number of resonant particles can compensate for a small prefactor of the
magnetization current. To study this and for simplicity, we assume a
Maxwell-Boltzmann background distribution. Thus, the Boltzmann factor
proportional to the number of resonant electrons is given by $\exp (-v_{\mathrm{rc}n}^{2}/v_{t}^{2})$, where the classical resonant velocities
making the denominator zero are given by $v_{\mathrm{rc}n}=$ $(\omega
-n\omega _{c})/k_{z}$. By contrast, the spin resonances have (quantum)
resonant velocities $v_{\mathrm{rq}n\mathrm{\pm }}=$ $(\omega -n\omega
_{c}\pm \omega _{cg})/k_{z}$. The most interesting case here is the quantum
resonance $v_{\mathrm{rq}n\mathrm{\pm }}$ with the smallest resonant
velocity, giving the maximum number of resonant particles. If the wave
frequency $\omega $ is of the order of  $\Delta \omega _{c}$ (note that
for an electron-proton plasma $\Delta \omega _{c}\sim \omega _{ci}$), this
occurs when $n=1$ and we use the positive sign for $\omega _{cg}$, in which
case we get 
\begin{equation}
v_{\mathrm{rq1+}}=\frac{(\omega -\Delta \omega _{c})}{k_{z}}.
\end{equation} 
Assuming that $\omega $ is of the same order as $\Delta \omega _{c}$ (but
not necessarily very close), in the long wavelength limit (small wave number
spectrum regime), we will have the strong inequality $\exp [(\omega -\Delta
\omega _{c})^{2}/k_{z}^{2}v_{t}^{2}]\gg \exp \left[-(v_{\mathrm{rc}n})^{2}/v_{t}^{2}\right]$ for all classical resonances (i.e., for any value $n$ for
the classical resonant velocity $v_{\mathrm{rc}n}$). A specific case was
studied in Ref. \cite{zamanian2010}, where the linear damping of ion-cyclotron
modes was considered in the limit of parallel propagation. Here, the dominant
classical damping mechanism was due to ions. It was found that in the
long-wavelength regime, the Boltzmann factor of electrons $\exp [(\omega
-\Delta \omega _{c})^{2}/k_{z}^{2}v_{t}^{2}]$  was considerably larger than
the Boltzmann factor of ions $\exp [(\omega -\omega
_{ci})^{2}/k_{z}^{2}v_{ti}^{2}]$  due to the much lower thermal velocity of
ions $v_{ti}\ll v_{t}$. As a result, the damping mechanism shifted from
quantum spin-resonance damping, in the long wavelength regime, to classical
ion-cyclotron damping in the short wavelength regime.
\par
In the above model (\ref{Vlasov}), the spin resonances only contribute to the
cases when a magnetic wave field is present, and thus a constant background
field $B_{0}\mathbf{\hat{z}}$ is not sufficient. Specifically, if a Langmuir
wave propagates parallel to a constant external magnetic field, there is no
perturbed magnetic field. Hence, when Eq. (\ref{Vlasov}) is applied, the
magnetization current and the corresponding denominators vanish identically.
This is only an approximation,  although, the wave-particle resonance  of the
type (\ref{Denominator}) is possible even for a purely electrostatic wave
field  if Eq. (\ref{Vlasov}) is generalized. Including a weakly relativistic
correction to Eq. (\ref{Vlasov}), in particular, the spin-orbit interaction,
the model is modified according to
\begin{equation}
\hat{\mathbf{s}}\times \mathbf{B\rightarrow }\hat{\mathbf{s}}\times \left( 
\mathbf{B-}\frac{\mathbf{p}\times \mathbf{E}}{2mc^{2}}\right), 
\label{Thomsson}
\end{equation}%
where the extended contribution is referred to as Thomas precession \cite%
{asenjo2012}. We also note that the weakly relativistic theory requires us to
change to a distribution function expressed in terms of momentum, i.e., we
let $f(\mathbf{r,v,}\hat{\mathbf{s}},t)\rightarrow f(\mathbf{r,p,}\hat{%
\mathbf{s}},t)$, although we can still use $\mathbf{p/}m\simeq \mathbf{v}$
to leading order. Moreover, we stress that the factor   $2$ in the
denominator of the second term of the right-hand side of Eq. (\ref{Thomsson}) is
consistent with Lorentz invariance (see e.g., Ref. \cite{jackson1975}). Finally,
since the particles carrying a spin magnetic dipole is moving, they
contribute with a polarization in the laboratory frame \cite{asenjo2012}.
Thus, the total current density is given by $\mathbf{J}=\mathbf{J}_{f}+%
\mathbf{J}_{M}+\mathbf{J}_{P}$, with the polarization current, given by,
 \begin{equation}
\mathbf{J}_{P}=\frac{\partial \mathbf{P}}{\partial t}=\frac{\partial }{\partial t}\left( 3\tilde{\mu}\int \frac{\left( \hat{\mathbf{s}}\times \mathbf{p%
}\right) }{2m}fd^{2}sd^{3}p\right).   \label{Polarization current}
\end{equation}%
With these changes in place (see Ref. \cite{asenjo2012} for the full details
of the extended model), it is possible to study spin-resonance damping of
Langmuir waves propagating parallel to an external magnetic field. As found
in Ref. \cite{ekman2021}, even for parameters where the classical
electrostatic dispersion relation is a good approximation for the real part
of the frequency, for a strong magnetic field (such that $\Delta \omega _{c}$
is of the same order of magnitude as the plasma frequency), the
wave-particle damping can be dominated by the spin resonance effect. While
the physics is much different (electrostatic waves rather than
electromagnetic), the result is similar to the preceding case. That is, in
the long wavelength regime, the number of resonant particles for the spin
resonance is much higher than that for the classical resonance. This leads to
spin resonance damping dominating for long wavelengths, whereas classical
wave-particle damping dominates for shorter wavelengths.
\par
While we have focused on the new wave-particle interaction resonances [as
seen in Eq. (\ref{Denominator}) induced by the spin effects, it should be
noted that spin dynamics can influence wave-particle dynamics through other
mechanisms. For example, when ultra-strong magnetic fields are present,
relativistic Landau quantization will give an increased effective mass to
electrons in the higher energy state, which affect the wave-particle
resonances, see e.g. Ref. \cite{alnaseri2020}. It should be stressed,
however, that most other mechanisms require either very strong magnetic
fields, low temperatures and/or high plasma densities to be significant.   
\section{Summary and discussion} \label{summary}
In this review paper, we have presented a contemporary theoretical knowledge on the wave-particle interactions and Landau damping of electrostatic waves in quantum plasmas.  We have restricted our discussion to the resonant interactions of both the linear and nonlinear   electrostatic waves, especially Langmuir waves and electron/ion-acoustic waves   in semiclassical and relativistic/nonrelativistic quantum plasma regimes.  The characteristics and consequences of wave-particle interaction  including the spin dynamics are also presented.  We have started with the basic concepts of plasma oscillation,   wave-particle interactions and the Landau's linear treatment on wave damping in classical plasmas. Before  moving on to nonlinear regimes, we have demonstrated the  linear theory of Landau damping in non-relativistic and relativistic quantum plasmas with different distributions of background electrons that are relevant to nondegenerate semiclassical plasmas and  degenerate quantum plasmas.  The occurrence of different resonance processes and the wave damping associated with them  are noted and discussed. It is elucidated   that while the phase velocity resonance is still the wave damping mechanism in classical and semiclassical plasmas, the resonant velocity in the quantum  regime is, however,  shifted by the velocity associated with the plasmon quanta (quantum modified Cherenkov resonance).    The  dispersion relations for electrostatic  modes and the Landau damping rates obtained in different  cases  are compared and analyzed.  We have also discussed the significance of spin effects on the wave-particle interactions in  spin plasmas. It is shown that  new wave eigenmodes  exist  as well as new types  of wave-particle resonances can occur  that typically depend   on the anomalous magnetic moment of charged particles.  
\par 
Going beyond the linear theory, we have discussed the wave-particle interactions for homogeneous plasma waves as well as for   localized waves in both the weak and strong quantum regimes.  It is found that in the weak quantum regime, not only   a transition from the     classical to the quantum regime in nonlinear Landau damping occurs, several other new features also take place including the quantum modified bounce frequency  and the  occurrence of bounce-like amplitude oscillations. On the other hand, the linear damping can be suppressed and the nonlinear multi-plasmon resonance can emerge in the strong quantum regime.    We have also considered the evolution of ion-acoustic waves and the modulation of Langmuir wave envelopes in the weak quantum regimes separately. It is shown that similar to classical plasmas, the resonant velocities are still the phase velocity or the group velocity. However, the quantum recoil effect significantly modifies the wave dispersion and nonlinearities and hence the Landau damping rates.   
Also, discussed are the effects of multi-plasmon resonances   on the modulation of Langmuir wave envelopes and low-frequency electron-acoustic waves in the nonlinear regime.  It is found that in contrast to classical and semiclassical plasmas, the multi-plasmon resonance is the dominant wave damping mechanism in the nonlinear evolution of electrostatic solitary waves. 
\par 
The wave-particle interaction is a very vast area of plasma physics. It is extremely rich and vibrant, and it holds   great promise for  various interesting and important applications including laser-based inertial plasma fusion and the laser-based plasma compression  schemes  as well as in high-energy density plasmas such as those in compact astrophysical objects (e.g., interior of white dwarf stars).   Recently,  the quantum kinetic theory of electron plasma waves has been advanced to take into account both electron and photon Landau damping in presence of an arbitrary spectrum of electromagnetic waves \cite{mendonca2016}.   The results could be  relevant in different physical situations such as the early universe and those mentioned before.  So, a possible extension of this study to relativistic quantum plasmas especially in the nonlinear regime could be  interesting  due to a possible overlap between the two  resonance processes.  In spin plasmas, we have limited our discussion to wave-particle interactions in the weak quantum regime where the typical length scale of oscillation is larger than the thermal de Broglie wavelength  and the Zeeman energy is smaller than the particle's thermal energy. Thus, generalizing the theory in the extended quantum regime where the Zeeman energy is comparable or larger than the thermal energy for which the Landau quantization may enter the picture could be a problem of promising research. 
\par 
A large number of wave-particle coupling processes and/or applications have  not been discussed  in this paper although they may be of importance to plasma physics communities. One area where the Landau damping due to the light-matter interactions  (surface plasmons in metal structures) plays a vital role in the dissipation mechanism of surface plasmon polaritons \cite{misra2020}  and provides an intrinsic limitation to plasmonics technology \cite{moghadam2021,abdikian2017}. Such an investigation requires the development of quantum-mechanical theory and is important for understanding of the underlying physical mechanisms for increasing lifetime of surface plasmons, and providing guidelines  in the future design of plasmonic devices \cite{li2013,shahbazyan2016}.  Another area   where the resonant quantum particles interacting with the wave can lead to the bump-on-tail and two-stream instabilities, as well as the particle trapping \cite{daligault2014} and the formation of  phase space structures and wave turbulence to be observed in future experiments \cite{haas2008}. Furthermore, other domains of wave-particle interactions which have been left out include the neutrino Landau damping \cite{mendonca2000} the Landau damping of electron plasma waves in  stimulated Raman and Brillouin scattering \cite{bers2009}, quantum two-stream instability \cite{liang2021}, and  wakefield acceleration \cite{brodin2013} in quantum plasmas.   Also, interesting and related phenomena referred to as anomalous Landau damping  \cite{trott2018} can also occur in non-Fermi Liquids. However, such phenomena is beyond the scope of the present review.  
\par
Finally, we would like to point out that phenomena similar to Landau damping occur in many fields besides plasma physics. Whenever a fundamental wave mode interacts with a continuous spectrum of oscillators, the dynamics will resemble many of the features that have been studied in the manuscript. To appreciate the ubiquitousness of the Landau mechanism, let us point out a few different examples throughout physics as well as in other fields of science. In plasma physics, the same mathematical structure as in Landau damping is seen for example in mode conversion problem  \cite{sedlacek1971}. In other fields of physics, Landau type of damping occurs in contexts such as superfluidity  \cite{fak1990}, Bose-Einstein condensates  \cite{shchedrin2018,mendonca2018}, accelerators  \cite{herr2013}, and quark physics \cite{baier1992}.  Moreover, very similar phenomena can occur in biological systems, e.g., in the flashing of fireflies  \cite{buck1988} and in the periodic firing of the pacemaker cells \cite{winfree1980}.   The above is by no means a complete list, but it should be clear that linear and nonlinear wave-particle mechanisms can be found very broadly in many fields of natural sciences.   
\appendix
The expressions appearing in Sec. \ref{sec-nld-LW-multi}
\section{Expressions for $A,~A_1,~B$ and $C$ in $\gamma$}\label{appendix-a} 
\begin{equation}
A=-\frac{16\pi e^3}{ A_0m^2k^3}{\int_C \frac{(v_p-v)\left[ (v_p-v)^2+%
\frac{v_q^2}{2}\right] }{\left\lbrace (v_p-v)^2-v_q^2\right\rbrace ^2 \left\lbrace \left(v_p-v+v_q%
\right)^2-v_q^2\right\rbrace \left\lbrace \left(v_p-v-%
v_q\right)^2-v_q^2\right\rbrace }%
F^{(0)}(v)dv},  \label{A}
\end{equation}
where
\begin{equation}
A_0={1-\frac{\pi e^2}{k^2m}\int_C \frac{F^{(0)}(v)}{(v_p-v)^2-\left(2v_q \right)^2 }}dv.
\end{equation}
Also, 
\begin{eqnarray}
&&A_1=-{12\pi e^3}\frac{\hbar }{k^2 m^2}\left[ \int_C \frac{1}{\left\lbrace
\left(v_p-v+v_q\right)^2-4v_q^2%
\right\rbrace \left\lbrace \left(v_p-v-v_q\right)^2-%
4v_q^2\right\rbrace dv }\right.  \notag \\
&&\left.+ \frac{3}{2}\int_C \frac{\left[ (v_p-v)^2+v_q^2\right] }{\left\lbrace (v_p-v)^2-4v_q^2\right\rbrace
\left\lbrace \left(v_p-v+2v_q\right)^2-v_q^2\right\rbrace \left\lbrace \left(v_p-v-2v_q\right)^2-%
v_q^2\right\rbrace }\right] F^{(0)}(v)dv,  \notag
\label{A1}
\end{eqnarray}
\begin{eqnarray}
&& B=\frac{4\pi e^4}{k^4 m}\int_C \left[ \frac{1}{\left\lbrace
v_p-v+2v_q\right\rbrace \left\lbrace v_p-v+v_q\right\rbrace \left\lbrace \left(v_p-v+2v_q%
\right)^2-v_q^2\right\rbrace} \right.  \notag \\
&&\left. +\frac{1}{\left\lbrace v_p-v-2v_q\right\rbrace
\left\lbrace v_p-v-v_q\right\rbrace \left\lbrace
\left(v_p-v-2v_q\right)^2-v_q^2%
\right\rbrace}\right.  \notag \\
&& \left. - \frac{2}{\left\lbrace (v_p-v)^2-v_q^2%
\right\rbrace^2 }\right]F^{(0)}(v)dv,  \notag  \label{B}
\end{eqnarray}
\begin{eqnarray}
&& C(k,\omega; v_g)=-\frac{4\pi e^4}{m \hbar^2 k^2}
\int_C \frac{1}{(v_p-v)^2-v_q^2} \frac{I(v)}{v-v_g}dv  \notag \\
&&=-\frac{4\pi e^4}{m \hbar^2 k^4} \int_C \left[ \frac{v-v_g-v_q}{\left\lbrace \left(v_p-v+2v_q\right)^2-v_q^2\right\rbrace \left(v_p-v+v_q\right)^2 \left( v-v_g-2v_q\right)} \right.  \notag \\
&&\left. - \frac{v-v_g+ v_q}{\left\lbrace \left(v_p-v-2v_q\right)^2-v_q^2\right\rbrace
\left(v_p-v-v_q \right)^2 \left( v-v_g + 2v_q\right)} \right.  \notag \\
&&\left. -2v_q \frac{ \left\lbrace \left( v_p-v\right)^2+v_q^2 \right\rbrace }{(v-v_g)\left\lbrace \left(v_p-v\right)^2-v_q^2 \right\rbrace ^3 }-4v_q\frac{v_p-v}{\left\lbrace \left( v_p-v\right)^2-v_q^2 \right\rbrace ^3}\right] F^{(0)}(v)dv,  \notag
\label{C-k-omega}
\end{eqnarray}
where
\begin{equation}
I(v)= \frac{1}{k^2}\left[ \left(v-v_g+v_q \right) \frac{f^{(0)}\left( v+2v_q\right) -f^{(0)}(v)}{{\left\lbrace v_p-\left(v+v_q \right) \right\rbrace }^2} +\left(v-v_g-v_q \right) 
\frac{f^{(0)}(v)-f^{(0)}\left( v-2v_q\right) }{{%
\left\lbrace v_p-\left(v-v_q \right) \right\rbrace }^2}\right].\label{I-v}
\end{equation}
\section{Reduced expressions for $\alpha,~\beta,~\gamma $ and $D$ with the Fermi distribution at zero temperature }\label{appendix-b} 
\begin{equation}
\alpha=-\frac{8m \omega_p^2}{3\hbar k^2 v_F^3} \sum_{j=\pm1} \left(v_p+jv_q
\right) \log \left\vert\frac{v_p+jv_q-v_F}{v_p+jv_q+v_F}\right\vert ,
\label{alpha-eq}
\end{equation}
\begin{equation}
\begin{split}
&\beta=1-\frac{3 m \omega_p^2}{2 \hbar k^3 v_F^2} \sum_{j=\pm1}\left[
\left\lbrace 2\left(v_p+jv_q \right)+\left( v_p-v_g\right)\right\rbrace
\log \left\vert\frac{v_p+jv_q-v_F}{v_p+jv_q+v_F}\right\vert \right. \\
&\left.-\left\lbrace v_F^2-\left(v_p+jv_q \right)^2-2\left(v_p+jv_q
\right)\left(v_p-v_g\right) \right\rbrace \frac{v_F}{ v_F^2-%
\left(v_P+jv_q \right)^2} +\left(v_p-v_g \right)\frac{v_F \left(v_p+jv_q
\right) }{ v_F^2-\left(v_p+jv_q \right)^2}\right],  \label{beta-coeff}
\end{split}%
\end{equation}
\begin{equation}
\gamma=\left(\frac{1}{4}\frac{A A_1}{\hbar}-\frac{1}{2\hbar^2}B+C\right)k^2,
\label{gamma}
\end{equation}
where
\begin{equation}
\begin{split}
&A= - \frac{4em^2 \omega_p^2}{A_0\hbar^3 k^6 v_F^3}\sum_{j=\pm1}\left[
kv_Fv_q +\frac{\omega_p}{6v_q} \left\lbrace \left(v_F^2-\left(v_p+jv_q
\right)^2 \right)\left(4-\frac{jk v_q}{2 \omega_p} \right)\right.\right.
\\
&\left.\left. -6v_q\left(v_p+jv_q \right) \left(1-\frac{j k v_q}{%
2 \omega_p} \right)\right\rbrace \log \left\vert\frac{v_p+jv_q-v_F}{%
v_p+jv_q+v_F}\right\vert\right. \\
&\left.+\frac{\omega_p}{3 v_q} \left\lbrace v_F^2-\left(v_p+j2v_q
\right)^2 \right\rbrace \left(1-\frac{jk v_q}{4\omega_p} \right)\log
\left\vert\frac{v_p+j2v_q-v_F}{v_p+j2v_q+v_F}\right\vert\right. \\
&\left. +i \pi v_q \omega_p \left\lbrace v_F^2-\left(v_p-2v_q
\right)^2 \right\rbrace \left(1+\frac{k v_q}{4\omega_p} \right)\right],
\label{A-reduced}
\end{split}%
\end{equation}
with
\begin{equation}
\begin{split}
& A_0=1-\frac{3\omega_p^2}{16 v_F^2 k^2}\left[2-\sum_{j=\pm1}\frac{jv_q}{%
4 v_F}\left\lbrace v_F^2-\left(v_p+jv_q\right)^2 \right\rbrace \log
\left\vert\frac{v_p+j2v_q-v_F}{v_p+j2v_q+v_F}\right\vert \right] \\
&-i\frac{3\pi  \omega_p^2}{64 v_F^3 v_q k^2}\left\lbrace
v_F^2-\left(v_p-2v_q \right)^2 \right\rbrace.  \label{A-0-reduced}
\end{split}%
\end{equation}
Also, 
\begin{equation}
\begin{split}
&A_1=-\frac{e}{m}\frac{9\omega_p^2 m^3}{4\hbar^2 v_F^3k^5} \sum_{j=\pm1}%
\left[\frac{j}{4}\left\lbrace v_F^2-\left(v_p+jv_q\right)^2 \right\rbrace
\log \left\vert\frac{v_p+jv_q-v_F}{v_p+jv_q+v_F}\right\vert\right. \\
&\left.- \frac{2j}{3}\left\lbrace v_F^2-\left(v_p+j3v_q \right)^2
\right\rbrace \log \left\vert\frac{v_p+j3v_q-v_F}{v_p+j3v_q+v_F}\right\vert
+j\left\lbrace v_F^2-\left(v_p+j2v_q \right)^2 \right\rbrace \log \left\vert%
\frac{v_p+j2v_q-v_F}{v_p+j2v_q+v_F}\right\vert\right] \\
& -i\frac{9\pi \omega_p^2}{16v_F^3k^3}\frac{e }{v_q^2}\left[ \frac{2}{3}%
\left\lbrace v_F^2-\left(v_p-3v_q \right)^2 \right\rbrace-\left\lbrace
v_F^2-\left(v_p-2v_q \right)^2 \right\rbrace\right],
\end{split}%
\end{equation}
\begin{equation}
\begin{split}
&B=-\frac{e^2}{k^7}\frac{3\omega_p^2m^3}{2v_F^3\hbar^3}\sum_{j=\pm1}\left[ 
16 v_F v_q+4\left\lbrace v_F^2-\left(v_p+j2v_q\right)^2
\right\rbrace \log \left\vert\frac{v_p+j2v_q-v_F}{v_p+j2v_q+v_F}%
\right\vert\right. \\
&\left.+\left\lbrace v_F^2-\left(v_p+j3v_q\right)^2 \right\rbrace \log
\left\vert\frac{v_p+j3v_q-v_F}{v_p+j3v_q+v_F}\right\vert -\left\lbrace
v_F^2-\left(v_p+jv_q \right)^2 -8j v_q\left( v_p+jv_q\right)
\right\rbrace \log \left\vert\frac{v_p+jv_q-v_F}{v_p+jv_q+v_F}\right\vert %
\right] \\
& -i \frac{3 \pi e^2 \omega_p^2}{4 k^4 v_F^3}\left[-\frac{1}{ v_q^3}%
\left\lbrace v_F^2-\left(v_p-2v_q\right)^2 \right\rbrace + \frac{1}{%
4 v_q^3}\left\lbrace v_F^2-\left(v_p-3v_q \right)^2 \right\rbrace\right],
\label{B-reduced}
\end{split}%
\end{equation}
\begin{equation}
\begin{split}
& C=-\frac{3}{4} \frac{e^2}{\hbar^2 k^4}\frac{\omega_p^2}{v_F^3}%
\sum_{j=\pm1} \left[ -\frac{1}{8 v_q^3} \frac{v_p+j2v_q-v_g}{%
v_p+jv_q-v_g} \left\lbrace v_F^2-\left(v_p+j3v_q \right)^2 \right\rbrace
\log \left\vert\frac{v_p+j3v_q-v_F}{v_p+j3v_q+v_F}\right\vert +jM_j \log
\left\vert\frac{v_p+jv_q-v_F}{v_p+jv_q+v_F}\right\vert\right. \\
&\left. +jN_j \frac{2v_F}{v_F^2-\left(v_p+jv_q \right)^2 } -\left( \frac{1}{%
2 v_q} \frac{v_p-v_g}{v_p-jv_q-v_g} -\frac{1}{2} \frac{1}{%
v_p+jv_q-v_g}-\frac{1}{2 v_q}\right) \frac{2v_F\left(v_p+jv_q \right) 
}{v_F^2-\left(v_p+jv_q\right)^2 }\right. \\
& \left. -v_q \frac{v_F^2-\left(v_g +jv_q\right)^2 }{\left(v_p-jv_q-%
v_g \right)^3 \left( v_p+jv_q-v_g\right) }\log\left\vert \frac{%
v_g +jv_q-v_F}{v_g+jv_q+v_F}\right\vert +\frac{2v_q}{k^2} \frac{%
\left\lbrace \left( \omega-k v_g\right)^2 +\frac{\hbar^2 k^4}{4m^2}
\right\rbrace\left( v_F^2-v_g^2\right) }{\left(v_p+v_q-v_g \right)^3
\left( v_p-v_q-v_g\right)^3} log \frac{v_g-v_F}{v_g+v_F}\right],
\end{split}
\label{C-reduced}
\end{equation}
with
\begin{equation}
\begin{split}
&M_{1,-1}= \frac{1}{4v_q^2 \left(v_p-v_g\mp 2v_q \right)^2 }\left[%
\left(v_p-v_g\mp 3v_q \right)\left\lbrace v_F^2-3\left( v_p\pm
v_q\right)^2+2\left(v_g\pm v_q \right)\left( v_p\pm
v_q\right)\right\rbrace \right. \\
& \left. \mp4v_q\left(v_p-v_g\mp v_q \right) \left(3v_p-v_g\pm 2v_q
\right) -2 \left(v_p\pm v_q \right)\left( v_p-v_g\right)
\left(v_p-v_g\mp 3v_q \right)\right. \\
&\left.+\left\lbrace 2\left(v_p-v_g\mp v_q \right)+\left(v_p-v_g\mp
3v_q \right)\pm 2\left(v_q-v_g \right)\left(v_q-v_g\mp3v_q
\right)^2\left(\frac{1}{2v_q}\mp\frac{1}{v_p- v_g\mp 2v_q} \right)
\right\rbrace \left\lbrace v_F^2-\left(v_p\pm v_q \right)^2 \right\rbrace %
\right] \\
& +\frac{1}{4v_q^2 \left(v_p- v_g\pm v_q \right)^2} \left[
5\left(v_p-v_g\pm v_q \right) \left\lbrace v_F^2-\left(v_p\pm v_q
\right)^2 \right\rbrace \pm 3v_q \left\lbrace v_F^2-\left(v_p\pm
v_q\right)^2 \right\rbrace \right. \\
&\left. \mp 4v_q^2\left( v_p \pm v_q\right) -\left(3v_p-3v_g\pm 5v_q
\right) \left\lbrace v_F^2\mp 2v_q\left(v_p\pm v_q \right)- \left(v_p\pm v_q
\right)^2 \right\rbrace \right. \\
& \left.+2v_q^2\frac{v_F^2-\left(v_p\pm v_q\right)^2 }{v_p-v_g\pm v_q}%
\right] \pm \frac{v_p}{v_q^2},
\end{split}%
\end{equation}
\begin{equation}
\begin{split}
& N_{1,-1}=\pm \frac{1}{4v_q^2 \left(v_p-v_g\mp v_q \right)^2}\left[%
2v_q\left(v_p-v_g\mp v_q \right)\left\lbrace v_F^2-3\left(v_p\pm
v_q\right)^2 -2\left(v_p-v_g\right) \left(v_p\pm v_q \right)
\right\rbrace \right. \\
&\left.+\left\lbrace v_F^2-\left( v_p\pm v_q \right)^2 \right\rbrace
\left(v_p-v_g\right)\left(v_p-v_g\mp 3v_q \right) \right] \\
&\mp\frac{1}{4v_q{\left(v_p-v_g\mp v_q \right)^2}}\left[
\left(v_p-v_g\mp 3v_q \right)\left\lbrace v_F^2-\left(v_p\pm v_q
\right)^2 \right\rbrace \pm 4v_q\left(v_p\pm v_q \right)\left(v_p-v_g\pm
v_q \right)\right] \\
&\pm \frac{1}{4v_q^2}\left[v_p\pm v_q- \left\lbrace v_F^2-\left(v_p\pm v_q
\right)^2 \right\rbrace \right],
\end{split}%
\end{equation}
\begin{equation}
D=\frac{3 e^2 \pi \omega_p^2}{4 k\hbar m v_F^3}\left[(v_F^2-v_g^2) \frac{(v_p-v_g)^2+v_q^2}{{\left\lbrace(v_p-%
v_g)^2-v_q^2 \right\rbrace^3 }} +\frac{1}{8 v_q^4}\left\lbrace
v_F^2 -\left(v_p-3v_q \right)^2 \right\rbrace \frac{ v_p-v_g-2v_q}{%
v_p-v_g-v_q}\right].  \label{D-reduced}
\end{equation}
This expression \eqref{D-reduced} of $D$ is obtained by using
the following relations. 
\begin{eqnarray}
&& \lim_{\nu _g\rightarrow 0} \frac{1}{\Omega-Kv+i\nu_g}=\frac{1}{\Omega-Kv}%
-i\pi\frac{1}{|K|}\delta \left(v-\frac{\Omega}{K} \right),  \notag \\
&& \lim_{\nu _3\rightarrow 0} \frac{1}{\omega-kv-3kv_q+i\nu_3%
}= \frac{1}{\omega-kv-3kv_q}  \notag \\
&&-i\pi\frac{1}{|K|} \delta\left(v-v_p+3v_q
\right),
\end{eqnarray}
and we have made use of $\Omega/K\rightarrow v_g$. The infinitesimal
quantities $|\nu_g|$ and $|\nu_3|$ are taken to anticipate the Landau
damping terms associated with the group velocity and three-plasmon
resonances. 
\par
Thus, the reduced expressions of $P,~Q$ and $R$ can be obtained from the relations $P=\beta/\alpha$, $Q=\gamma/\alpha$ and $R=D/\alpha$.
  \begin{acknowledgements}
   One of us (APM) acknowledges  support from Science and Engineering Research Board (SERB), Government of India, for a research project (under Core Research Grant) with sanction order no. CRG/2018/004475. The authors thank the anonymous referees for their useful comments. 
 \end{acknowledgements}
 
\section*{Conflict of interest statement} 
 The authors have no actual or potential conflicts of interest to declare in relation to this article.

\bibliographystyle{spphys}       
\bibliography{Reference-RMPP}   

%
%

\end{document}